%Superconformal Ward Identities and their Solution
%Plain TeX file using harvmac

\input harvmac
%\draft
\input amssym.def
\input amssym
\baselineskip 14pt
\magnification\magstep1

\parskip 6pt
%For smaller font footnotes
\newdimen\itemindent \itemindent=32pt
\def\textindent#1{\parindent=\itemindent\let\par=\resetpar%
\indent\llap{#1\enspace}\ignorespaces}

\let\oldpar=\par
\def\resetpar{\oldpar\parindent=20pt\let\par=\oldpar}

\font\ninerm=cmr9 \font\ninesy=cmsy9
\font\eightrm=cmr8 \font\sixrm=cmr6
\font\eighti=cmmi8 \font\sixi=cmmi6
\font\eightsy=cmsy8 \font\sixsy=cmsy6
\font\eightbf=cmbx8 \font\sixbf=cmbx6
\font\eightit=cmti8
\def\eightpoint{\def\rm{\fam0\eightrm}
  \textfont0=\eightrm \scriptfont0=\sixrm \scriptscriptfont0=\fiverm
  \textfont1=\eighti  \scriptfont1=\sixi  \scriptscriptfont1=\fivei
  \textfont2=\eightsy \scriptfont2=\sixsy \scriptscriptfont2=\fivesy
  \textfont3=\tenex   \scriptfont3=\tenex \scriptscriptfont3=\tenex
  \textfont\itfam=\eightit  \def\it{\fam\itfam\eightit}%
  \textfont\bffam=\eightbf  \scriptfont\bffam=\sixbf
  \scriptscriptfont\bffam=\fivebf  \def\bf{\fam\bffam\eightbf}%
  \normalbaselineskip=9pt
  \setbox\strutbox=\hbox{\vrule height7pt depth2pt width0pt}%
  \let\big=\eightbig  \normalbaselines\rm}
\catcode`@=11 %
\def\eightbig#1{{\hbox{$\textfont0=\ninerm\textfont2=\ninesy
  \left#1\vbox to6.5pt{}\right.\n@@space$}}}
\def\vfootnote#1{\insert\footins\bgroup\eightpoint
  \interlinepenalty=\interfootnotelinepenalty
  \splittopskip=\ht\strutbox %
  \splitmaxdepth=\dp\strutbox %
  \leftskip=0pt \rightskip=0pt \spaceskip=0pt \xspaceskip=0pt
  \textindent{#1}\footstrut\futurelet\next\fo@t}
\catcode`@=12 %
\def \de{\delta}

\def \si{\sigma}

\def \ga{\gamma}

\def \pr{\partial}

\def \ha{{\hat a}}

\def \hf{{\hat f}}

\def \J{{\rm J}}

\def \X{{\rm X}}
\def \hD{{\hat D}}
\def \cD{{\hat {\cal D}}}
\def \hG{{\hat {\cal G}}}
\def \hV{{\hat V}}
\def \zz{{\bar x}}
\def \yz{{\bar y}}
\def \bz{{\bar z}}
\def \bta{{\bar \eta}}

\def \l{\big \langle}
\def \r{\big \rangle}
\def \ep{\epsilon}
\def \vep{\varepsilon}
\def \half{{\textstyle {1 \over 2}}}
\def \thir{{\textstyle {1 \over 3}}}
\def \quar{{\textstyle {1 \over 4}}}
\def \ts{\textstyle}

\def \d{{\rm d}}

\def \x{{\rm x}}
\def \y{{\rm y}}

\def \tx{{\tilde {\rm x}}}

\def \A{{\cal A}}
\def \B{{\cal B}}
\def \C{{\cal C}}
\def \D{{\cal D}}
\def \E{{\cal E}}
\def \F{{\cal F}}
\def \G{{\cal G}}
\def \H{{\cal H}}
\def \I{{\cal I}}
\def \J{{\cal J}}
\def \K{{\cal K}}

\def \N{{\cal N}}
\def \O{{\cal O}}
\def \P{{\cal P}}
\def \Q{{\cal Q}}
\def \S{{\cal S}}
\def \U{{\cal U}}
\def \V{{\cal V}}
\def \W{{\cal W}}
\def \Y{{\cal Y}}

\def \hD{{\hat D}}
\def \tQ{{\tilde Q}{}}
\def \dal{{\dot \alpha}}
\def \dbe{{\dot \beta}}

\def \bQ{{\bar Q}}
\def \bS{{\bar S}}
\def \bsi{\bar \sigma}
\def \bj{{\bar \jmath}}
\def \bJ{{\bar J}}
\def \hF{{\hat \F}}
\def \hU{{\hat \U}}

\def \tx{{\tilde {\rm x}}}

\def \tsi{{\tilde \sigma}}

\def \x{{\rm x}}
\def \y{{\rm y}}

\def \vphi{{\varphi}}
\def \bpsi{{\overline \psi}}
\def \bep{{\bar \epsilon}}
\def \hep{{\hat {\epsilon}}}
\def \hbep{{\hat {\bep}}}
\def \bet{{\bar \alpha}}

\def \dal{{\dot \alpha}}
\def \dbe{{\dot \beta}}

\def \bga{{\bar \gamma}}
\def \tx{{\tilde {\rm x}}}
\def \bsi{\bar \sigma}
\def \lam{\sigma}
\def \mun{\tau}
\def \oD{{\overline D}}
\def \scs{\scriptstyle}
\font \bigbf=cmbx10 scaled \magstep1
\def\toinf#1{\mathrel{\mathop{\longrightarrow}\limits_{\scriptstyle{#1}}}}

%References
\lref\DZ{E. D'Hoker and D.Z. Freedman, Supersymmetric Gauge Theories
and the AdS/CFT Correspondence, Lectures given at Theoretical Advanced 
Study Institute in Elementary Particle Physics (TASI 2001), hep-th/0201253.}
\lref\hughtwo{J. Erdmenger and H. Osborn, {Conserved Currents and the 
Energy Momentum Tensor in Conformally Invariant Theories for General 
Dimensions}, Nucl. Phys. B483 (1997) 431, hep-th/9605009.}
\lref\hughone{H. Osborn and A. Petkou, {Implications of Conformal 
Invariance for Quantum Field Theories in $d>2$}
Ann. Phys. (N.Y.) {231} (1994) 311, hep-th/9307010.}
\lref\HO{H. Osborn, {$\N=1$ Superconformal Symmetry in Four-Dimensional
Quantum Field Theory}, Ann. Phys. (N.Y.) 272 (1999) 243, hep-th/9808041.}
\lref\Sei{S. Lee, S. Minwalla, M. Rangamani and N. Seiberg, {Three-Point
Functions of Chiral Operators in $D=4$, $\N=4$ SYM at Large $N$},
Adv. Theor. Math. Phys.  2 (1998) 697, hep-th/9806074.}
\lref\LW{K. Lang and W. R\"uhl, Nucl. Phys. {B402} (1993) 573.}
\lref\Pet{A.C. Petkou, Ann. Phys. (N.Y.) 249 (1996) 180, hep-th/9410093.}
\lref\Fone{S. Ferrara, A.F. Grillo, R. Gatto and G. Parisi, Nucl. Phys. 
B49 (1972) 77\semi
S. Ferrara, A.F. Grillo, R. Gatto and G. Parisi, Nuovo Cimento 19A
(1974) 667.}
\lref\Ftwo{S. Ferrara, A.F. Grillo and R. Gatto, Ann. Phys. 76 (1973) 161.}
\lref\Dob{V.K. Dobrev, V.B. Petkova, S.G. Petrova and I.T. Todorov,
Phys. Rev. D13 (1976) 887.}
\lref\Ext{E. D'Hoker, D.Z. Freedman, S.D. Mathur, A. Matusis and
L. Rastelli, Extremal Correlators in the AdS/CFT correspondence,
{\it in} The Many Faces of the Superworld, ed. M.A. Shifman,
hep-th/9908160\semi
B. Eden, P.S. Howe, C. Schubert, E. Sokatchev and P.C. West,
Extremal Correlators in Four-dimensional SCFT, Phys. Lett. B472 (2000) 323,
hep-th/9910150\semi
B. Eden, P.S. Howe, E. Sokatchev and P.C. West, Extremal and
Next-to-Extremal N Point Correlators in Four-dimensional SCFT,
Phys. Lett. B494 (2000), hep-th/0004102\semi
M. Bianchi and S. Kovacs, Nonrenormalization of Extremal Correlators in
N=4 SYM Theory, Phys. Lett. B468 (1999) 102, hep-th/9910016\semi
J. Erdmenger and M. P\'erez-Victoria, Nonrenormalization of 
Next-to-Extremal Correlators in N=4 SYM and the AdS/CFT correspondence,
Phys. Rev. D62 (2000) 045008, hep-th/9912250\semi
E. D'Hoker, J. Erdmenger, D.Z. Freedman and M. P\'erez-Victoria,
Near Extremal Correlators and Vanishing Supergravity Couplings in 
AdS/CFT, Nucl. Phys. B589 (2000) 3, hep-th/0003218.}
\lref\Free{D.Z. Freedman, S.D. Mathur, A. Matusis and L. Rastelli, 
Nucl. Phys. B546 (1999) 96, hep-th/9804058\semi
D.Z. Freedman, S.D. Mathur, A. Matusis and L. Rastelli, 
Phys. Lett. B452 (1999) 61, hep-th/9808006\semi
E. D'Hoker and D.Z. Freedman, Nucl. Phys. B550 (1999) 261,
hep-th/9811257\semi
E. D'Hoker and D.Z. Freedman, Nucl. Phys. B544 (1999) 612, hep-th/9809179.}
\lref\FreeI{E. D'Hoker, D.Z. Freedman and L. Rastelli, 
Nucl. Phys. B562 (1999) 395, hep-th/9905049.}
\lref\FreeD{E. D'Hoker, D.Z. Freedman, S.D. Mathur, A. Matusis and 
L. Rastelli, Nucl. Phys. B562 (1999) 353, hep-th/9903196.}
\lref\AF{L. Andrianopoli and S. Ferrara, {On short and long $SU(2,2/4)$
multiplets in the AdS/CFT correspondence}, Lett. Math. Phys. 48 (1999) 145, 
hep-th/9812067.}
\lref\Short{S. Ferrara and A. Zaffaroni, {Superconformal Field Theories,
Multiplet Shortening, and the AdS${}_5$/SCFT${}_4$ Correspondence},
Proceedings of the Conf\'erence Mosh\'e Flato 1999, vol. 1, ed. G. Dito and 
D. Sternheimer, Kluwer Academic Publishers (2000), hep-th/9908163.}
\lref\short{F.A. Dolan and H. Osborn, {On short and semi-short
representations for four-dimensional superconformal symmetry}, Ann. Phys.
307 (2003) 41, hep-th/0209056.} 
\lref\pert{F. Gonzalez-Rey, I. Park and K. Schalm, {A note on four-point
functions of conformal operators in $N=4$ Super-Yang Mills},
Phys. Lett. B448 (1999) 37, hep-th/9811155\semi
B. Eden, P.S. Howe, C. Schubert, E. Sokatchev and P.C. West,
{Four-point functions in $N=4$ supersymmetric Yang-Mills theory at
two loops}, Nucl. Phys. B557 (1999) 355, hep-th/9811172; {Simplifications
of four-point functions in $N=4$ supersymmetric Yang-Mills theory at
two loops}, Phys. Lett. B466 (1999) 20, hep-th/9906051\semi
M. Bianchi, S. Kovacs, G. Rossi and Y.S. Stanev, {On the
logarithmic behaviour in $\N=4$ SYM theory}, JHEP 9908 (1999) 020, 
hep-th/9906188\semi
M. Bianchi, S. Kovacs, G. Rossi and Y.S. Stanev, Anomalous dimensions
in $\N$=4 SYM theory at order $g^4$, Nucl. Phys. B584 (2000) 216, 
hep-th/0003203\semi
B. Eden, C. Schubert and E. Sokatchev, Three loop four point
correlator in N=4 SYM, Phys. Lett. B482 (2000) 309, hep-th/0003096.}
\lref\hpert{G. Arutyunov, S. Penati, A. Santambrogio and E.Sokatchev,
Four-point correlators of BPS operators in $\N=4$ SYM at order $g^4$,
Nucl. Phys. B670 (2003) 103, hep-th/0305060.}
\lref\Arut{G. Arutyunov and S. Frolov, {Four-point Functions of Lowest
Weight CPOs in $\N=4$ SYM${}_4$ in Supergravity Approximation},
Phys. Rev. D62 (2000) 064016, hep-th/0002170.}
\lref\Three{G. Arutyunov and S. Frolov, {Three-point function of the 
stress-tensor in the AdS/CFT correspondence}, Phys. Rev. D60 (1999) 026004,
hep-th/9901121.}
\lref\ADHS{G. Arutyunov, F.A. Dolan, H. Osborn and E. Sokatchev,
Correlation Functions and Massive Kaluza-Klein Modes in the AdS/CFT
Correspondence, Nucl. Phys. B665 (2003) 273, hep-th/0212116.}
\lref\Degen{G. Arutyunov and E. Sokatchev, On a Large N Degeneracy
in $\N=4$ SYM and the AdS/CFT Correspondence, Nucl. Phys. B663 (2003) 163,
hep-th/0301058.}
\lref\ASok{F.A. Dolan, L. Gallot and E. Sokatchev, On Four-point Functions 
of $\half$-BPS operators in General Dimensions, hep-th/0405180.}
\lref\Hokone{E. D'Hoker, S.D. Mathur, A. Matusis and L. Rastelli, {The 
Operator Product Expansion of $N=4$ SYM and the 4-point Functions of 
Supergravity}, Nucl. Phys. B589 (2000) 38, hep-th/9911222.}
\lref\OPEN{G. Arutyunov, S. Frolov and A.C. Petkou, {Operator Product
Expansion of the Lowest Weight CPOs in $\N=4$ SYM${}_4$ at Strong Coupling},
Nucl. Phys. B586 (2000) 547, hep-th/0005182; 
(E) Nucl. Phys. B609 (2001) 539.}
\lref\OPEW{G. Arutyunov, S. Frolov and A.C. Petkou, {Perturbative and
instanton corrections to the OPE of CPOs in $\N=4$ SYM${}_4$}, Nucl. 
Phys. B602 (2001) 238, hep-th/0010137; (E) Nucl. Phys. B609 (2001) 540.}
\lref\Edent{B. Eden, A.C. Petkou, C. Schubert and E. Sokatchev, {Partial
non-renormalisation of the stress-tensor four-point function in $N=4$
SYM and AdS/CFT}, Nucl. Phys. B607 (2001) 191, hep-th/0009106.}
\lref\Except{G. Arutyunov, B. Eden, A.C. Petkou and E. Sokatchev, 
{Exceptional non-renorm-alization properties and OPE analysis of
chiral four-point functions in $\N=4$ SYM${}_4$}, 
Nucl. Phys. B620 (2002) 380, hep-th/0103230.}
\lref\BKRS{M. Bianchi, S. Kovacs, G. Rossi and Y.S. Stanev, {Properties
of the Konishi multiplet in  $\N=4$ SYM theory}, JHEP 0105 (2001) 042,
hep-th/0104016.}
\lref\HMR{L. Hoffmann, L. Mesref and W. R\"uhl, {Conformal partial wave
analysis of AdS amplitudes for dilaton-axion four-point functions},
Nucl. Phys. B608 (2001) 177, hep-th/0012153\semi
%L. Hoffmann, T. Leonhardt, L. Mesref and W. R\"uhl, Composite fields,
%generalized hypergeometric functions and $U(1)_Y$ symmetry in the
%AdS/CFT correspondence, in Karpacz 2001, New developments in fundamental
%interaction theories, hep-th/0102161\semi
T. Leonhardt, A. Meziane and W. R\"uhl, Fractional BPS Multi-Trace
Fields of $\N=4$ SYM${}_4$ from AdS/CFT, Phys. Lett. B552 (2003) 87, 
hep-th/0209184.}

\lref\one{F.A. Dolan and H. Osborn, {Implications of $\N=1$ 
Superconformal Symmetry for Chiral Fields}, Nucl. Phys. B593 (2001) 599, 
hep-th/0006098.}
\lref\Dos{F.A. Dolan and H. Osborn, {Conformal four point functions
and the operator product expansion}, Nucl. Phys. B599 (2001) 459, 
hep-th/0011040.}
\lref\scft{F.A. Dolan and H. Osborn, {Superconformal symmetry,
correlation functions and the operator product expansion}, Nucl. Phys. B629
(2002) 3, hep-th/0112251.}

\lref\Hoff{L.C. Hoffmann, A.C. Petkou and W. R\"uhl, 
Phys. Lett. B478 (2000) 320, hep-th/0002025\semi
L.C. Hoffmann, A.C. Petkou and W. R\"uhl, hep-th/0002154.}
\lref\HoffR{L. Hoffmann, L. Mesref and W. R\"uhl, AdS box graphs, unitarity
and operator product expansions,
Nucl. Phys. B589 (2000) 337,  hep-th/0006165.}
\lref\Witt{E. Witten, Adv. Theor. Math. Phys. 2 (1998) 253, hep-th/9802150.}
\lref\HSW{P.S. Howe, E. Sokatchev and P.C. West, {3-Point Functions in 
$N=4$ Yang-Mills}, Phys. Lett. B444 (1998) 341, hep-th/9808162.}
\lref\Eden{B. Eden, P.S. Howe, A. Pickering, E. Sokatchev and P.C. West,
{Four-point functions in $N=2$ superconformal field theories},
Nucl. Phys. B581 (2000) 523, hep-th/0001138.}
\lref\Intr{K. Intriligator, {Bonus Symmetries of ${\cal N} =4$ 
Super-Yang-Mills Correlation Functions via AdS Duality}, 
Nucl. Phys. B551 (1999) 575, hep-th/9811047.}
\lref\Bon{K. Intriligator and W. Skiba, {Bonus Symmetry and the Operator
Product Expansion of ${\cal N} =4$ Super-Yang-Mills},
Nucl. Phys. B559 (1999) 165, hep-th/9905020.}
\lref\Non{G. Arutyunov, B. Eden and E. Sokatchev, {On 
Non-renormalization and OPE in Superconformal Field Theories}, Nucl. Phys. 
B619 (2001) 359, hep-th/0105254.}
\lref\bpsN{B. Eden and E. Sokatchev, {On the OPE of 1/2 BPS Short
Operators in $N=4$ SCFT${}_4$}, Nucl. Phys. B618 (2001) 259, hep-th/0106249.}
\lref\Hes{P.J. Heslop and P.S. Howe, {OPEs and 3-point correlators of
protected operators in $N=4$ SYM}, Nucl. Phys. B626 (2002) 265, 
hep-th/0107212.}
\lref\Pen{S. Penati and A. Santambrogio, {Superspace approach to 
anomalous dimensions in ${\N}=4$ SYM}, Nucl. Phys. B614 (2001) 367, 
hep-th/0107071.}
\lref\Hok{A.V. Ryzhov, {Quarter BPS Operators in $\N=4$ SYM}, 
JHEP 0111 (2001) 046, hep-th/0109064\semi
E. D'Hoker and A.V. Ryzhov, {Three Point Functions of Quarter BPS 
Operators in $\N=4$ SYM}, JHEP 0202 (2002) 047, hep-th/0109065.}
\lref\HH{P.J. Heslop and P.S. Howe, {A note on composite operators in
$N=4$ SYM}, Phys. Lett. 516B (2001) 367, hep-th/0106238.}
\lref\HesHo{P.J. Heslop and P.S. Howe, Aspects of $\N=4$ SYM, hep-th/0307210.}
\lref\Howe{P.J. Heslop and P.S. Howe, Four-point functions in $N=4$ SYM,
JHEP 0301 (2003) 043, hep-th/0211252.}

\lref\Vretare{L. Vretare, Formulas for Elementary Spherical Functions and
Generalized Jacobi Polynomials, Siam J. on Mathematical Analysis, 15 (1984)
805.}
\lref\Koo{T. Koornwinder, Two-variable Analogues of the Classical Orthogonal
Polynomials, in Theory and Applications of Special Functions, ed. R. Askey,
Academic Press, New York (1975).}
\lref\ortho{C.F. Dunkl and Y. Xu, Orthogonal Polynomials of Several Variables,
Cambridge University Press, Cambridge (2001).}
\lref\Class{M. G\"unaydin and N. Marcus, {The spectrum of the $S^5$
compactification of the chiral $N=2$, $D=10$ supergravity and the unitary
supermultiplets of $U(2,2/4)$}, Class. and Quantum Gravity, 2 (1985) L11.}
\lref\phd{F.A. Dolan, Aspects of Superconformal Quantum Field Theory,
University of Cambridge PhD thesis (2003).}

\lref\Dirac{P.A.M. Dirac, Wave Equations in Conformal Space,
Annals of Math. 37 (1936) 823; {\it in} ``The
Collected Works of P.A.M. Dirac 1924-1948'', edited by R.H. Dalitz, CUP
(Cambridge) 1995.}
\lref\CFT{F.A. Dolan and H. Osborn, {Conformal Partial Waves and the Operator 
Product Expansion}, Nucl. Phys. B678 (2004) 491, hep-th/0309180.}

\lref\Barg{V. Bargmann and I.T. Todorov, Spaces of analytic functions on a complex
cone as carriers for the symmetric tensor representations of SO(n), J. of Math.
Phys. 18 (1977) 1141.}
\lref\Dob{V.K. Dobrev, G. Mack, V.B. Petkova, S.G. Petrova and I.T. Todorov,
Harmonic Analysis on the n-dimensional Lorentz Group and its Application to
Conformal Quantum Field Theory, Springer Lecture Notes in Physics, 63, 
Springer-Verlag (Berlin - Heidelberg) 1977.}

{\nopagenumbers
\rightline{DAMTP/04-51}
\rightline{hep-th/0407060}
\vskip 1.5truecm
\centerline {\bigbf Superconformal Ward Identities and their Solution}
\vskip 1.5 true cm
\centerline {M. Nirschl and H. Osborn${}^\dagger$}

\vskip 12pt
\centerline {\ Department of Applied Mathematics and Theoretical Physics,}
\centerline {Wilberforce Road, Cambridge CB3 0WA, England}
\vskip 1.5 true cm

{\eightpoint
\parindent 1.5cm{

{\narrower\smallskip\parindent 0pt

Superconformal Ward identities are derived for the the four point
functions of chiral primary BPS operators for $\N=2,4$ superconformal
symmetry in four dimensions. Manipulations of arbitrary tensorial fields
are simplified by introducing a null vector so that the four point functions
depend on two internal $R$-symmetry invariants as well as two conformal 
invariants. The solutions of these identities are 
interpreted in terms of the operator product expansion and are shown
to accommodate long supermultiplets with free scale dimensions and also
short and semi-short multiplets with protected dimensions. The decomposition
into $R$-symmetry representations is achieved by an expansion in terms of
two variable harmonic polynomials which can be expressed also in terms
of Legendre polynomials. Crossing symmetry conditions on the four point
functions are also discussed.



\narrower}}

\vfill
\line{${}^\dagger$ 
address for correspondence: Trinity College, Cambridge, CB2 1TQ, 
England\hfill}
\line{\hskip0.2cm emails:
{{\tt mn244@damtp.cam.ac.uk} and \tt ho@damtp.cam.ac.uk}\hfill}
}
%\vskip0.5cm

%PACS: 11.10-z; 11.25.Hf; 11.10Kk; 04.62+v
\eject}
\pageno=1

\newsec{Introduction}

Since the discovery of the AdS/CFT correspondence there has been a
huge resurgence of interest in superconformal theories in
four dimensions, for a review see \DZ. In particular for the $\N=4$ 
superconformal $SU(N)$ gauge theory, to which the AdS/CFT correspondence 
is most directly applicable, many new and exciting results have been
obtained. Much has been discovered concerning the spectrum of operators
and their scale dimensions both in the large $N$ limit, through the 
supergravity approximation to the AdS/CFT correspondence, and also 
perturbatively as an expansion in the coupling $g$.

Of the operators present in the the theory the simplest are the chiral
primary operators belonging to $SU(4)_R$ $R$-symmetry representations, 
with Dynkin labels $[0,p,0]$. 
These are represented by symmetric traceless rank $p$
tensors formed by gauge invariant traces of the elementary scalar
fields and satisfy BPS like constraints so that they belong to short
supermultiplets of the superconformal group $PSU(2,2|4)$. They are
therefore protected against renormalisation effects and have scale
dimension $\Delta=p$. Their three point functions have been fully
analysed \Sei. For the case of $p=2$, when the supermultiplet contains
the energy momentum tensor, the four point functions have been
found both perturbatively \pert\ and in the large $N$ limit \Arut.
Such results have also been extended more recently to chiral primary
operators with $p=3,4$ \refs{\hpert,\ADHS,\Degen}. The explicit results
for the four point correlation functions has then allowed an analysis
of those operators which contribute to the operator product expansion 
for two chiral primary operators 
\refs{\Hokone,\OPEN,\OPEW,\Edent,\Except,\BKRS,\HMR}.

To take the analysis of the operator product expansion of correlation
functions beyond the lowest scale dimension operators for each 
$SU(4)_R$ representation it is necessary to have an explicit form
for the conformal partial waves which give the contribution of
a quasi-primary operator of arbitrary scale dimension and spin  and
all its conformal descendants to conformally covariant four point
functions. In four dimensions a simple expression was found in \Dos.
In addition since all operators in a superconformal multiplet must
have the same anomalous dimensions it is desirable to have a 
procedure for analysing the operator product expansion for each 
supermultiplet as a single contribution. This depends on a solution
of all superconformal Ward identities since this should allow all
possible operator product expansion contributions to be found in
a form compatible with the superconformal symmetry. For the simplest
case of the four point function for $[0,2,0]$ chiral primary operators
this was undertaken in \scft\ and applied to determine the one
loop anomalous dimensions for all operators with lowest order twist two.

The procedure adopted in \scft\ is somewhat involved and does not
simply generalise to correlation functions of more general chiral
primary operators. As was shown in \scft\ the superconformal Ward
identities are simplified if they are expressed in terms of new
variables $x,\zz$ rather than the usual conformal invariants. In
terms of the standard correspondence for the space-time coordinates
$x^a \to \x = x^a\si_a $, where $\x$ is a $2\times 2$ spinorial matrix such 
that $\det \x = - x^2$, then, for four points $x_1,x_2,x_3,x_4$ and 
$x_{ij} = x_i - x_j$, $x,\zz$ may be defined, as shown in \Howe, as the 
eigenvalues of $\x_{12}\,\x_{42}{}^{\!\!-1} \x_{43} \, \x_{13}{}^{\!\! -1}$. 
By conformal transformations we may choose a frame such that
$\x_2=0, \, \x_3=\infty, \, \x_4 = 1$ and $\x_1 = \pmatrix{x & 0 \cr 0& \zz}$.
The two conformal invariants are then given in terms of $x,\zz$ by\foot{Since
$1+u-v = x+ \zz$ and $1+u^2+v^2-2uv-2u-2v=(x-\zz)^2$ it is easy to invert 
these results to obtain  $x,\zz$ in terms of $u,v$ up to the
arbitrary sign of the square root $\sqrt{(x-\zz)^2}$.  For any $f(u,v)$
there is a corresponding symmetric function ${\hat f}(x,\zz)={\hat f}(\zz,x)$
such that ${\hat f}(x,\zz)=f(u,v)$.},
\eqn\eiguv{\eqalign{
u={}&\det( \x_{12}\,\x_{42}{}^{\!\!-1} \x_{43}\, \x_{13}{}^{\!\! -1})= x\zz \, , \cr
v= {}& \det(1- \x_{12}\,\x_{42}{}^{\!\!-1}\x_{43} \, \x_{13}{}^{\!\! -1}) = 
\det( \x_{14}\,\x_{24}{}^{\!\!-1} \x_{23} \, \x_{13}{}^{\!\! -1}) =
(1-x)(1-\zz) \, .\cr}
}
For a Euclidean metric on space-time $x,\zz$ are complex conjugates.
In our analysis we find that the linear equations which follow 
from superconformal invariance naturally separate into ones involving just $x$ 
and conjugate equations with $x \to \zz$.

A technical complication in dealing with arbitrary chiral primary
operators represented by symmetric traceless tensorial fields
$\vphi_{r_1 \dots r_p}$ is that the four point function for four
chiral primary fields, for arbitrary $p_1,p_2,p_3,p_4$, involves in general
a proliferation of independent tensorial invariants as the $p_i$
are increased. The construction of projection operators corresponding to
different $R$-symmetry representations also becomes a non trivial exercise.
Such tensorial complications in the analysis of superconformal Ward identities 
and also in applying the operator product expansion are avoided here
by taking
\eqn\tphi{
\vphi_{r_1 \dots r_p}(x) \to
\vphi^{(p)}(x,t) = \vphi_{r_1\dots r_p}(x)\, t_{r_1} \dots t_{r_p} \, ,
}
where $t$ is an arbitrary complex vector satisfying\foot{Such null vectors may also
be motivated by considering the harmonic superspace approach and were used
similarly for instance in \refs{\ADHS,\hpert}. Our application is independent 
of the harmonic superspace formalism and is essentially
motivated just by the requirement of simplifying the treatment of arbitrary rank
symmetric traceless tensors, we do not anywhere consider the conjugate of $t$.}
\eqn\nul{
t^2=0 \, .
}
(For a more mathematical discussion of using such vectors for the treatment of
representations of $SO(n)$ see \Barg, see also appendix A in \Dob).
Clearly $\vphi_{r_1\dots r_p}$ can be recovered from $\vphi^{(p)}$.
The four point function then becomes a homogeneous polynomial in 
$t_1,t_2,t_3,t_4$, of respective degree $p_1,p_2,p_3,p_4$, invariant under
simultaneous rotations on all $t_i$'s. Due to the condition \nul\ for each 
$t_i$ the conformally covariant four point function is reducible to an
invariant function $\F(u,v;\lam,\mun)$ with $\lam,\mun$ the two independent
invariants, homogeneous of degree zero in each $t_i$, which are analogous to the 
conformal invariants $u,v$,
\eqn\deflu{
\lam = {t_1{\cdot t_3} \, t_2{\cdot t_4} \over
t_1{\cdot t_2} \, t_3{\cdot t_4}} \, , \qquad
\mun = {t_1{\cdot t_4} \, t_2{\cdot t_3} \over
t_1{\cdot t_2} \, t_3{\cdot t_4}} \, .
}
In general $\F(u,v;\lam,\mun)$ is a polynomial in $\lam,\mun$, with  degree 
determined by the $p_i$, where the number of independent terms match exactly 
the number of tensorial invariants necessary for the general decomposition 
of the four  point function for the corresponding symmetric traceless 
tensorial fields, for $p_i=p$ there are $\half (p+1)(p+2)$ terms.

Just as  the invariants 
$u,v$ are expressed in terms of $x,\zz$ it is convenient to write
$\lam,\mun$ in a similar  form involving new variables $\alpha,\bet$,
\eqn\defab{
\lam = \alpha \bet \, , \qquad \mun = (1-\alpha)(1-\bet) \, .
}
For the $\N=2$ case further restrictions impose $\alpha=\bet$. The
superconformal Ward identities are then simply expressed in terms of
$\hF(x,\zz;\alpha,\bet) = \F(u,v;\lam,\mun)$, with $\hF$ symmetric in
$x,\zz$ and also a symmetric polynomial in $\alpha,\bet$. 
The superconformal identities constrain $\hF(x,\zz;\alpha,\bet)$
for $\alpha = 1/x$ to be expressible only in terms of a function involving
$\zz$ and $\bet$ (for $\N=2$ just a single variable function of $\zz$ 
appears). Taking into account the symmetry conditions the unconstrained 
or dynamical part of $\F$ is therefore of the form
\eqn\uncon{
\F(u,v;\lam,\mun)_{\rm dynamical} =
( \alpha x -1  ) ( \alpha \zz - 1 ) ( \bet x -1  ) ( \bet \zz - 1 ) \,
\H(u,v;\lam,\mun) \, ,
}
with $\H(u,v;\lam,\mun)= {\hat \H}(x,\zz;\alpha,\bet)$ a polynomial in $\lam,\mun$,
or $\alpha,\bet$, of reduced degree. In the $\N=2$ the corresponding result is
\eqn\uncont{
\F(u,v;\lam,\mun)_{\rm dynamical} = ( \alpha x -1  ) ( \alpha \zz - 1 ) \, 
\H(u,v;\lam,\mun) \, , \qquad \H(u,v;\lam,\mun) = {\hat \H}(x,\zz;\alpha) \, .
}
Similar results were previously obtained by Heslop and Howe \Howe\ based on 
expansions in terms of Schur polynomials for $SU(2,2|2)$ and 
$PSU(2,2|4)$\foot{In eq. (49) of \Howe\ 
$S_{020}(Z) = (X_1-Y_1)(X_1-Y_2)(X_2-Y_1)(X_2-Y_2)$ which appears
as an overall factor in the Schur polynomial for long representations.}.
Using the formalism of harmonic superspace \ASok\ also provides a method
for deriving superconformal identities which are equivalent to those obtained
here.

These results have a natural interpretation in terms of the operator
product expansion when the four point function is expanded in terms
of conformal partial waves corresponding to operators
with various scale dimensions $\Delta$ and spins $\ell$ belonging to
the various possible representations of the $R$-symmetry group. The
conformal partial waves are explicit functions of $u,v$, more simply
given in terms of $x,\zz$ \Dos. To disentangle the different
$R$-symmetry representations the correlation functions are also expanded
in terms of two variable harmonic polynomials  corresponding to
the possible $R$-symmetry representations which may be formed. Explicit
simple expressions are found here for these harmonic polynomials in
the $\N=4$ case using the variables $\alpha,\bet$
(for $\N=2$ they reduce to a single variable Legendre polynomial). If
$\H$ in \uncon\ is simultaneously expanded in such harmonic polynomials and 
conformal partial waves then the factors multiplying $\H$, for each term in 
the expansion, through various recurrence relations generate contributions 
corresponding to all operators belonging to a single superconformal long 
multiplet. For such a long multiplet the scale dimension has only a lower 
bound due to unitarity and, in a perturbative expansion, $\H$
therefore includes dynamical renormalisation effects leading to anomalous
scale dimensions. The remaining parts in the solution of the superconformal
identities which involve functions of $x$ or $\zz$ are also analysed here.
For $\N=2$ there is a single variable function $f(x)$ whereas for $\N=4$
the superconformal identities allow for $f(x,\alpha)$, which is polynomial
in $\alpha$ and satisfies the constraint that it is a constant $k$ when
$\alpha=1/x$. These functions correspond to semi-short multiplets with 
protected scale dimensions, determined by $\ell$ and the $R$-symmetry 
representation, and also in special cases to short multiplets. The full
set of possible semi-short supermultiplets are obtained by decomposing
long multiplets at the unitarity threshold and the short multiplet
contributions are realised by extending the semi-short results to 
$\ell=-1,-2$.

The superconformal Ward identities are most powerful for four point
functions which are extremal, so that there is only one possible $SU(4)_R$ 
invariant coupling, or next-to-extremal when there are just three
invariant couplings. Various calculations \Ext\ have shown that the
correlation functions are identical with the results obtained in free
field theory. The superconformal Ward identities  for the extremal case
show that the correlation function depends only on the constant $k$
whereas the next-to-extremal case is given just by the function $f(x,\alpha)$,
without any dynamical contribution of the form exhibited in \uncont.

For four point functions of identical primary operators there are further
constraints arising from crossing symmetry which corresponds to the 
permutation group $\S_3$. In such four point functions $\S_3$ acts on $u,v$
and also $\lam,\mun$ so that $\F(u,v;\lam,\mun)$ is invariant up to an 
explicit overall factor. All invariant contributions to $\F$ may be formed 
by combining $\S_3$ irreducible representations constructed from 
functions of $u,v$ and also $\lam,\mun$. 
Crossing symmetry further constrains the single variable functions
$f(x)$ or $f(x,\alpha)$ that arise in solving the superconformal Ward
identities. We argue here, based on superconformal representation
theory and analyticity requirements, that these can be extended to
a fully crossing symmetric contribution to $\F(u,v;\lam,\mun)$ in
terms of two variable crossing symmetric polynomials which are
constructed here. These essentially correspond to generalised free 
field contributions to the correlation function. A similar discussion
is also applicable to the next-to-extremal correlation function in the
case of thee identical operators and there is still a $\S_3$ symmetry.

In detail the structure of this paper is then as follows. In section 2
we derive the superconformal Ward identities for $\N=2$ superconformal
symmetry and these are applied in section 3 by analysing the contributions
of different supermultiplets in the operator product expansion. The
discussion is extended to the $\N=4$ case in sections 4 and 5. For
the operator product expansion it is shown how there are potential
contributions from non-unitary semi-short supermultiplets although
they may be cancelled so that only unitary multiplets remain.
In section 6 we take into account the restrictions imposed by 
crossing symmetry making use of $\S_3$ representations. In section 7 we
summarise some results obtained previously for large $N$ in the framework
of this paper and a few comments are made in a conclusion.
Various technical issues are addressed in four appendices. In
appendix A we discuss how derivatives involving the null vector $t_r$
are compatible with $t^2=0$. In appendix B we consider two variable harmonic
polynomials, depending on $\lam,\mun$ given in \deflu, which are used
in the expansion of general four point correlation functions. Appendix C
describes some differential operators which play an essential role
in our analysis whereas in appendix D we consider non unitary semi-short
representations for $PSU(2,2|4)$ which are important in our operator
product analysis.

\newsec{Superconformal Ward Identities, $\N=2$}

The algebraic complications involved in the analysis of Ward identities
are much simpler for $\N=2$ superconformal symmetry. In this case the
$R$-symmetry group is just $U(2)$ and discussion of the representations
is much easier. In order to facilitate the comparison with the $\N=4$
case later we consider BPS chiral primary operators which belong
to representations of  $SU(2)_R$ symmetry for $R=n$, an integer. The
BPS condition requires that the scale dimension $\Delta =2n$. Such fields
form superconformal primary states for a short supermultiplet with necessarily
unrenormalised scale dimensions. The fields in this case are represented by
symmetric traceless tensors $\vphi_{r_1 \dots r_n}$ with $r_i=1,2,3$.
To derive the Ward identities we need to consider just the superconformal 
transformations at the lowest levels of the multiplet. First
\eqn\ptph{
\de \vphi_{r_1\dots r_n} = \hep \, \tau_{(r_n} \psi_{r_1\dots r_{n-1})} +
\bpsi{}_{(r_1\dots r_{n-1}} \tau_{r_n)} \, \hbep \, ,
}
where $\psi_{i r_1\dots r_{n-1}\alpha}, \, \bpsi{}_{i r_1\dots r_{n-1}\dal}$
are spinor fields, traceless and symmetric on the indices $r_1\dots r_{n-1}$,
 satisfying, with $i=1,2$ and $\tau_r$ the usual Pauli matrices,
\eqn\tpsi{
\tau_r \, \psi_{r r_1\dots r_{n-2}} = 0 \, , \qquad
\bpsi{}_{r_1\dots r_{n-2}r} \, \tau_r = 0 \, .
}
Thus both $\psi$ and $\bpsi$ belong to $SU(2)_R$ representations with 
$R=n-\half$. In \ptph\ we have\foot{Thus $4$-vectors
are identified with $2\times 2$ matrices using the hermitian $\si$-matrices
$\si_a, \, \tsi_a, \, \si_{(a} \tsi_{b)} = - \eta_{ab}1$,
$x^a \to \x_{\alpha\dal} =
x^a (\si_a)_{\alpha\dal}, \ \tx^{\dal\alpha} = x^a (\tsi_a)^{\dal\alpha}
= \ep^{\alpha\beta}\ep^{\smash {\dal \dbe}} \x_{\smash {\beta \dbe}}$,
with inverse $x^a = - {1\over 2}{\rm tr}(\si^a \tx)$. We have $x{\cdot y}
= x^a y_a = - {1\over 2}{\rm tr}(\tx \y)$, $\det \x = -x^2$, 
$\x^{-1} = -\tx/x^2$.}
\eqn\scep{
\hep_i{}^{\! \alpha}(x)= \ep_i{}^{\! \alpha} - i \, \bta{}_{i\dal}
\tx^{\dal\alpha} \, , \qquad \hbep{}^{\, i \dal}(x) = \bep{}^{\, i \dal}
+ i \, \tx^{\dal\alpha} \eta^i{}_{\! \alpha} \, .
}
where $\ep_i{}^{\! \alpha}, \bta{}_{i\dal}, \bep{}^{\, i \dal},
\eta^i{}_{\! \alpha}$ are the $R=\half$ anticommuting parameters for an 
$\N=2$ superconformal transformation. In addition to \ptph\ we use
\eqn\ptps{\eqalign{
\de \psi_{r_1\dots r_{n-1}\alpha} ={}&
i \pr_{\alpha\dal}\, \vphi_{r_1\dots r_{n-1}s} \, \tau_s \hbep{}^{\, \dal}
+ 4n \, \vphi_{r_1\dots r_{n-1}s} \, \tau_s \eta_\alpha \cr
&{}+ J_{r_1 \dots r_{n-1}\alpha\dal} \, \hbep{}^{\, \dal} - {n-1\over 2n-1}\,
\tau_{(r_1} J_{r_2 \dots r_{n-1})s\alpha\dal }\, \tau_s \hbep{}^{\, \dal}\, ,
\cr}
}
where $J_{r_1 \dots r_{n-1}\alpha\dal}$, a symmetric traceless rank $n-1$
tensor, is a $R=n-1$ current. Using \ptps\ together with its conjugate 
we may verify closure of the superconformal algebra acting on 
$\vphi_{r_1\dots r_n}$,
\eqn\clos{
[\de_2 , \de_1] \vphi_{r_1\dots r_n} = - v{\cdot \pr} \vphi_{r_1\dots r_n}
- n(\si + \bsi)\, \vphi_{r_1\dots r_n} + n \, t_{(r_n|s}\,
\vphi_{r_1\dots r_{n-1})s} \, ,
}
where $v^a$, which is quadratic in $x$, and $\si , \bsi, t_{rs}=-t_{sr}$,
which are linear in $x$, are constructed from 
$\hep_1, \hbep_1, \hep_2,\hbep_2$.

For the general analysis here we define $\vphi^{(n)}(x,t)$ as in \tphi,
where $t_r$ is here a $3$-vector, and in a similar fashion also
$\psi^{(n-1)}_\alpha (x,t) = \psi_{r_1\dots r_{n-1}\alpha}(x)\, 
t_{r_1} \dots t_{r_{n-1}}$ and $\bpsi{}^{\,(n-1)}_\dal (x,t) =
\bpsi{}_{r_1\dots r_{n-1}\dal}(x)\, t_{r_1} \dots t_{r_{n-1}}$ while
$J^{(n-1)}_{\alpha\dal} (x,t) = J_{r_1 \dots r_{n-1}\alpha\dal}(x) \,
t_{r_1} \dots t_{r_{n-1}}$. With this notation \ptph\ may be rewritten as
\eqn\varp{
\de \vphi^{(n)}(t) = \hep \,\tau{\cdot t} \, \psi^{(n-1)}(t) + 
\bpsi{}^{\,(n-1)}(t)\, \tau{\cdot t}\, \hbep \, ,
}
and \ptps\ becomes
\eqn\vars{\eqalign{
\de \psi^{(n-1)}_\alpha (t) = {}& {1\over n} \,
\tau{\cdot {\pr \over \pr t}}\, i \pr_{\alpha\dal} \vphi^{(n)}(t) \, 
\hbep{}^{\, \dal}
+ 4 \, \tau{\cdot {\pr \over \pr t}} \vphi^{(n)}(t) \eta_\alpha \cr
&{}+ \bigg ( 1 - {1\over 2n-1}\,
\tau{\cdot t} \, \tau{\cdot {\pr \over \pr t}} \bigg ) 
J^{(n-1)}_{\alpha\dal}(t) \, \hbep{}^{\, \dal} \, . \cr}
}
A precise form for differentiation with respect to $t_r$ satisfying \nul\
is given in appendix A. The conditions \tpsi\ are now
\eqn\condit{
\tau{\cdot {\pr \over \pr t}} \, \psi^{(n-1)}_\alpha (t) = 0 \, , \qquad
\bpsi{}^{\,(n-1)}(t)\, \tau{\cdot {\overleftarrow{\pr \over \pr t}}} =0 \, .
}

The four point correlation functions of interest here then have the form
\eqn\Fourp{\eqalign{
\l \vphi^{(n_1)}& (x_1,t_1) \, \vphi^{(n_2)}(x_2,t_2)\, 
\vphi^{(n_3)}(x_3,t_3)\, \vphi^{(n_4)}(x_4,t_4 )\r \cr
&{} = { r_{23}^{\, \, \Sigma - 2n_2 - 2n_3}  \,
r_{34}^{\, \, \Sigma - 2n_3 - 2n_4}\over
r_{13}^{\, \, 2n_1}\,  r_{24}^{\, \, \Sigma - 2n_3} }\, F(u,v;t) \, ,
\quad \Sigma = n_1+n_2+n_3+n_4 \, , \cr}
}
where
\eqn\defrij{
r_{ij} = (x_i-x_j)^2 \, ,
}
and $u,v$ are the two independent conformal invariants, 
\eqn\defuv{
u= {r_{12} \, r_{34} \over r_{13} \, r_{24}} \, , \qquad \qquad
v= {r_{14} \, r_{23} \over r_{13} \, r_{24}} \, ,
}
which is equivalent to \eiguv.
$F(u,v;t)$ is also a $SU(2)_R$ scalar which is specified further later, clearly
there is a freedom to modify it by suitable powers of $u$ or $v$ at the expense
of changing the terms involving $r_{ij}$ in \Fourp. The choice made on \Fourp\
has some convenience in the later discussion.

The fundamental superconformal Ward identities arise from expanding
\eqn\ward{
\de \l \psi^{(n_1-1)}_\alpha (x_1,t_1) \, \vphi^{(n_2)}(x_2,t_2)\, 
\vphi^{(n_3)}(x_3,t_3)\, \vphi^{(n_4)}(x_4,t_4 )\r =0 \, ,
}
using \varp\ and \vars. This gives, suppressing the arguments $t_i$ for
the time being,
\eqn\exw{\eqalign{
& {1\over n_1}\, i \pr_{1\alpha\dal}\, \tau{\cdot {\pr \over \pr t_1}}
\l \vphi^{(n_1)} (x_1) \, \vphi^{(n_2)}(x_2)\, 
\vphi^{(n_3)}(x_3)\, \vphi^{(n_4)}(x_4)\r \, \hbep{}^{\, \dal}(x_1) \cr
& \qquad {} + 4 \, \tau{\cdot {\pr \over \pr t_1}}
\l \vphi^{(n_1)} (x_1) \, \vphi^{(n_2)}(x_2)\, 
\vphi^{(n_3)}(x_3)\, \vphi^{(n_4)}(x_4)\r \, \eta_\alpha \cr
& {}+ \bigg ( 1  - {1\over 2n_1-1}\,
\tau{\cdot t_1} \, \tau{\cdot {\pr \over \pr t_1}} \bigg ) 
\l J^{(n_1-1)}_{\alpha\dal}(x_1)  \, \vphi^{(n_2)}(x_2)\,
\vphi^{(n_3)}(x_3)\, \vphi^{(n_4)}(x_4)\r \, \hbep{}^{\, \dal}(x_1) \cr 
&{}+ \l \psi^{(n_1-1)}_\alpha (x_1) \, \bpsi{}^{\,(n_2-1)}_\dal (x_2) \, 
\vphi^{(n_3)}(x_3)\, \vphi^{(n_4)}(x_4)\r \, \tau{\cdot t_2} \,
\hbep{}^{\, \dal}(x_2) \cr
&{}+ \l \psi^{(n_1-1)}_\alpha (x_1) \, \vphi^{(n_2)}(x_2) \, 
\bpsi{}^{\,(n_3-1)}_\dal (x_3) \, \vphi^{(n_4)}(x_4)\r \,\tau{\cdot t_3} \,
\hbep{}^{\,\dal}(x_3)\cr
&{}+ \l \psi^{(n_1-1)}_\alpha (x_1) \, \vphi^{(n_2)}(x_2)\,
\vphi^{(n_3)}(x_3)\, \bpsi{}^{\, (n_4-1)}_\dal(x_4)\r \, \tau{\cdot t_4} \,
\hbep{}^{\, \dal}(x_4)  = 0 \, . \cr }
}
To apply this we make use of general expressions compatible with conformal
invariance for each four point function which appears. Thus
\eqn\fourpsi{\eqalign{
& \l \psi^{(n_1-1)}_\alpha (x_1) \, \bpsi{}^{\,(n_2-1)}_\dal (x_2) \, 
\vphi^{(n_3)}(x_3)\, \vphi^{(n_4)}(x_4)\r \cr
&\quad {}= 2i \, { r_{23}^{\, \, \Sigma - 2n_2 - 2n_3}  \,
r_{34}^{\, \, \Sigma - 2n_3 - 2n_4}\over
r_{13}^{\, \, 2n_1}\,  r_{24}^{\, \, \Sigma - 2n_3} }\, \bigg ( 
{1\over r_{12}} \x_{12\alpha\dal}\, R_2 + {1\over r_{13}r_{24}}
(\x_{13} \tx_{34} \x_{42})_{\alpha\dal}\, S_2 \bigg ) \, , \cr
& \l \psi^{(n_1-1)}_\alpha (x_1) \, \vphi^{(n_2)}(x_2) \,
\bpsi{}^{\,(n_3-1)}_\dal (x_3) \, \vphi^{(n_4)}(x_4)\r \cr
&\quad {}= 2i \, { r_{23}^{\, \, \Sigma - 2n_2 - 2n_3}  \,
r_{34}^{\, \, \Sigma - 2n_3 - 2n_4}\over
r_{13}^{\, \, 2n_1}\,  r_{24}^{\, \, \Sigma - 2n_3} }\, \bigg (
{1\over r_{13}} \x_{13\alpha\dal}\, R_3 + {1\over r_{14}r_{23}}
(\x_{14} \tx_{42} \x_{23})_{\alpha\dal}\, S_3 \bigg ) \, , \cr
& \l \psi^{(n_1-1)}_\alpha (x_1) \, \vphi^{(n_2)}(x_2)\,
\vphi^{(n_3)}(x_3)\, \bpsi{}^{\, (n_4-1)}_\dal(x_4)\r \cr
&\quad {}= 2i \, { r_{23}^{\, \, \Sigma - 2n_2 - 2n_3}  \,
r_{34}^{\, \, \Sigma - 2n_3 - 2n_4}\over
r_{13}^{\, \, 2n_1}\,  r_{24}^{\, \, \Sigma - 2n_3} }\, \bigg (
{1\over r_{14}} \x_{14\alpha\dal}\, R_4 + {1\over r_{13}r_{24}}
(\x_{13} \tx_{32} \x_{24})_{\alpha\dal}\, S_4 \bigg ) \, , \cr }
}
where $R_n,S_n$ are functions of $u,v$ and also scalars formed from $t_i$
(to verify completeness of the basis chosen in \fourpsi\ we use relations 
such as $\x_{13} \tx_{34} \x_{42} + \x_{14} \tx_{43} \x_{32} 
= r_{34} \, \x_{12}$). In addition we have
\eqn\fourJ{\eqalign{
& \l J^{(n_1-1)}_{\alpha\dal}(x_1)  \, \vphi^{(n_2)}(x_2)\,
\vphi^{(n_3)}(x_3)\, \vphi^{(n_4)}(x_4)\r \cr
&\quad {}= 2i \, { r_{23}^{\, \, \Sigma - 2n_2 - 2n_3}  \,
r_{34}^{\, \, \Sigma - 2n_3 - 2n_4}\over
r_{13}^{\, \, 2n_1}\,  r_{24}^{\, \, \Sigma - 2n_3} }\, \Big (
\X_{1[23]\alpha\dal} \, I + \X_{1[43]\alpha\dal} \, J \Big ) \, , \cr}
}
for 
\eqn\defX{
\X_{i[jk]} = {\x_{ij}\tx_{jk}\x_{ki}\over r_{ij}\, r_{ik}} =
{1\over r_{ij}} \x_{ij} - {1\over r_{ik}} \x_{ik} \, ,
}
which transforms under conformal transformations as a vector at $x_i$ and
is antisymmetric in $jk$. 

Using \fourpsi\ and \fourJ\ in \exw, noting that
\eqn\deep{
i \pr_{1\alpha\dal}\, {1\over r_{13}^{\, \, 2n_1}} \, \hbep{}^{\, \dal}(x_1)
+ 4n_1 \, {1\over r_{13}^{\, \, 2n_1}} \, \eta_\alpha = 
- 4n_1 i \, {1\over r_{13}^{\, \, 2n_1+1}} \, \hbep{}^{\, \dal}(x_3) \, ,
}
and $\pr_{1\alpha\dal} u = 2u \, \X_{1[23]\alpha\dal}, \,
\pr_{1\alpha\dal} v = 2v \, \X_{1[43]\alpha\dal}$, we may decompose \exw\
into independent contributions involving $\hbep(x_1)$ and
$\hbep(x_3)$ (note that $\x_{12} \hbep(x_2)/r_{12} = \x_{13}\hbep(x_2)/r_{13}
+ \X_{1[23]}  \hbep(x_1)$ and also for $x_2\to x_4$)
giving two linear relations,
\eqna\Wardrel
$$\eqalignno{
& { 1\over n_1} \, \tau{\cdot {\pr \over \pr t_1}} \big (
\X_{1[23]} \,  u\pr_u  + \X_{1[43]} \, v\pr_v \big ) F
+ \bigg ( 1  - {1\over 2n_1-1}\,
\tau{\cdot t_1} \, \tau{\cdot {\pr \over \pr t_1}} \bigg ) \big (
\X_{1[23]} \, I  + \X_{1[43]} \, J \big ) \cr
&{}+ \X_{1[23]} \, R_2 \, \tau{\cdot t_2} + \X_{1[43]} \, R_4 \,\tau{\cdot t_4}
+  \big ( u \X_{1[23]} -v \X_{1[43]} \big ) (  S_2 \, \tau{\cdot t_2}
-  S_4 \, \tau{\cdot t_4} ) = 0 \, , & \Wardrel{a} \cr
& \bigg ( - 2 \, \tau{\cdot {\pr \over \pr t_1}} F 
+ \big ( R_2 +(u-v) S_2 \big ) \tau{\cdot t_2} + R_3 \, \tau{\cdot t_3} + 
\big ( R_4 +(1-u+ v) S_4 \big ) \tau{\cdot t_4} \bigg ) 
{1\over r_{13}}\x_{13} \cr
&{}+  \big ( v S_2 \, \tau{\cdot t_2} + S_3 \, \tau{\cdot t_3}
- v S_4 \, \tau{\cdot t_4} \big ) {1\over r_{14}r_{23}} \, 
\x_{14} \tx_{42} \x_{23} = 0 \, . & \Wardrel{b} \cr}
$$
It is easy to decompose \Wardrel{a,b}\ into independent equations but crucial
simplifications are obtained essentially by diagonalising each $2\times 2$
spinorial equation in terms  of new variables $x,\zz$
which, as mentioned in the introduction, are the eigenvalues of 
$\x_{12}\,\x_{42}{}^{\!\!-1} \x_{43} \, \x_{13}{}^{\!\! -1}$. These are 
related to the conformal invariants $u,v$ defined in \defuv\ by \eiguv.
In \Wardrel{b} the spinorial matrix
$\tx_{41}{}^{\! -1} \tx_{42} \,\tx_{32}{}^{\! -1} \tx_{31}$ may be replaced
by $1/(1-x)$ and in \Wardrel{a} we may effectively replace $ \X_{1[23]} \to
1/x$ and $\X_{1[43]} \to - 1/(1-x)$ and in each case also for $x\to \zz$.
Using 
\eqn\part{
{\pr \over \pr x} = \zz \, {\pr \over \pr u} - (1-\zz) \, {\pr \over \pr v} \, ,
}
and the definitions
\eqn\hatRJ{
{T_2} = R_2 + x S_2 \, , \quad {T_3} = R_3 + {1\over 1-x} S_3 \, ,
\quad {T_4} = R_4 + (1-x) S_4 \, , \quad
K = {1\over x} I - {1\over 1-x} J \, ,
}
we may then obtain from \Wardrel{a,b}
\eqna\redeq
$$\eqalignno{
{ 1\over n_1} \, \tau{\cdot {\pr \over \pr t_1}} \, {\pr \over \pr x} F
= {}& - \bigg ( 1  - {1\over 2n_1-1}\,
\tau{\cdot t_1} \, \tau{\cdot {\pr \over \pr t_1}} \bigg ) K
- {1\over x} \, {T_2}\, \tau{\cdot t_2} 
+ {1\over 1-x } \, {T_4}\, \tau{\cdot t_4} \, , & \redeq{a} \cr
2 \, \tau{\cdot {\pr \over \pr t_1}} F = {}& {T_2}\, \tau{\cdot t_2}
+ {T_3}\, \tau{\cdot t_3} + {T_4}\, \tau{\cdot t_4}  \, . 
& \redeq{b} \cr}
$$
together with the corresponding equations obtained by $x\to \zz$ in 
\redeq{a,b}\ with also $T_i\to {\bar T}_i,\, K\to {\bar K}$, which are defined
just as in \hatRJ\ for $x\to \zz$.

The equations in \redeq{a,b}\ are equations for $\pr F/\pr t_1$. The 
integrability conditions, which are required by virtue of
$(\tau{\cdot \pr_{t_1}})^2=0$, are satisfied since we have, for $i=2,3,4$,
\eqn\cons{
\tau{\cdot {\pr \over \pr t_1}} \, {T_i} = 0 \, , \qquad
{T_i} \, \tau{\cdot {\overleftarrow{\pr \over \pr t_i}}} =0 \, ,
}
as a consequence of \condit. To reduce \redeq{a,b} into separate equations
we first write,
\eqn\UVS{
T_i \, \tau{\cdot t_i} = \tau {\cdot V_i} + W_i \, 1 \, , 
}
where $W_i$ and $V_{i,r}$ are respectively a scalar and a vector. From 
the results of appendix A we further decompose $V_i$ uniquely in the form
\eqn\VU{
V_i = {1\over n_1}\, {\pr \over \pr t_1} U_i + \hV_i \, , \qquad
t_1 {\cdot V_i} = U_i \, , \quad t_1 {\cdot \hV_i} = 0 \, .
}
The first equation in \cons\ then gives
\eqn\VS{
{\pr \over \pr t_1} {\cdot V_i} = 0 \, , \qquad i 
{\pr \over \pr t_1} \times V_i + {\pr \over \pr t_1} W_i = 0 \, ,
}
where we may let $V_i \to \hV_i$ without change. From \VS\ we may then find
\eqn\VSt{
L_1 W_i = i n_1 \hV_i \, ,
}
where we define the $SU(2)_R$ generators by
\eqn\genL{
L_i = t_i \times {\pr \over \pr t_i} \, .
}

Substituting \UVS\ into \redeq{a} gives
\eqn\Fx{
{\pr \over \pr x} F = - {1\over x} \, U_2  + {1\over 1-x} \, U_4 \, ,
}
and
\eqn\eqJ{
{n_1 \over 2n_1-1}\, K = - {1\over x} \, W_2 + {1\over 1-x} \, W_4 \, ,
}
which is just an equation giving $K$, and also
\eqn\eqJV{
- {1 \over 2n_1-1}\, iL_1 K = - {1\over x} \, \hV_2 + 
{1\over 1-x} \, \hV_4 \, .
}
It is easy to see that this follows from \eqJ\ as a consequence of \VSt.
Similarly substituting \UVS\ into \redeq{b} gives three equations
\eqn\FYn{
2n_1 \, F = \sum_{i=2}^4 \, U_{i} \, , 
}
and
\eqn\Vz{
\sum_{i=2}^4 \, \hV_{i} =0 \, , 
}
as well as
\eqn\Sz{
\qquad \sum_{i=2}^4 \, W_{i} =0 \, .
}
Clearly \Vz\ follows from \Sz. 

An essential constraint may also be obtained
from the second equation in \cons\ which gives
\eqn\contw{
(n_i + 1 ) T_i \, \tau {\cdot t_i} = - ( T_i \, \tau {\cdot t_i} )
\, i \tau {\cdot \overleftarrow L}_i \, .
}
With the decomposition \UVS\ this leads to
\eqna\relSU
$$\eqalignno{
(n_i +1) W_i = {}& -i L_i {\cdot V_i} \, ,  & \relSU{a} \cr
(n_i +1) V_i ={}& - L_i \times V_i - i L_i W_i \, . & \relSU{b} \cr}
$$
Contracting  \relSU{b} with $L_i$, and using $L_i \times L_i = - L_i,
\ L_i {\!}^2 W_i = n_i(n_i+1) W_i$, gives \relSU{a}. In addition we have from
$T_i(\tau{\cdot t_i})^2 = 0 $
\eqn\ttT{
t_i{\cdot V_i}=0 \, , \qquad i \, t_i \times V_i = t_i \, W_i \, .
}
With the aid of the results in appendix A we may obtain $(2n_1+1)\, \pr_i {\cdot (}
t_i \, W_i - i \, t_i \times V_i ) = (2n_i+3) \big ( (n_i+1) W_i + i L_i {\cdot V_i}
\big )$ so that \ttT\ implies \relSU{a}. Similarly, since $\pr_i \times (t_i\times V_i)
= (\pr_i \times t_i) \times V_i) + \pr_i (t_i {\cdot V_i}) -
\pr_i {\cdot (}t_i \, V_i)$, we have from appendix A 
$(2n_i+1) \pr_i \times (t_i\times V_i) =
- (2n_i+3) ( L_i \times V_i + (n_i+4) V_i )$ and
$(2n_i+1) \pr_i \times (t_i W_i) = - (2n_i+3) L_i W_i $. Hence it is clear that \ttT\ 
also implies \relSU{b}\foot{The converse follows using $t_i {\cdot L_i} =0 , \
t_i \times L_i = t_i \, t_i{\cdot \pr_i}$ and $t_i \times (L_i \times V_i)
= - t_i \times V_i$.}.

Using \VU\ and \VSt\ for $\hV_i$ in \relSU{a} we obtain
\eqn\LLU{
\big ( L_1 \! \cdot L_i + n_1 ( n_i +1) \big ) W_i =
\half \big ( (L_1 + L_i)^2 + (n_1+n_i)(n_1+n_i+1) \big ) W_i = - i
{\pr \over \pr t_1} {\cdot L_i } \, U_i \, .
}
$U_i(u,v;t)$, which is defined by \UVS, is a homogeneous polynomial
in $t_1,t_i$ of ${\rm O}(t_1^{n_1}, t_i^{n_i})$ such that the
$SU(2)_R$ representation with $R_{(1i)} = n_1+n_i$ is absent. In 
consequence the operator $(L_1 + L_i)^2 + (n_1+n_i)(n_1+n_i+1)$, which
commutes with $\pr_1{\cdot L_i}$, in \LLU\ may be inverted to give $W_i$ in 
terms of $U_i$. Alternatively we may obtain from \ttT
\eqn\LUW{
i \, t_i \times \pr_1 U_i = - t_1 \times L_1 W_i + n_1 \, t_i W_i \, .
}

To analyse these equations further we now consider the decomposition of $F$ 
and also $U_i$ in terms of $SU(2)_R$ scalars.
We first assume the $n_i$ are ordered so that
\eqn\order{
n_1 \le n_2 \le n_3 \le n_4 \, , 
}
and further assume
\eqn\inn{
\qquad n_4 = n_1 +n_2 + n_3 - 2E \, ,
}
for integer $E=0,1,2\dots$, where $E$ is a measure of how close the
correlation function is to the extremal case. With \order\ and \inn\ 
$F$, which is ${\rm O}(t_1^{n_1},t_2^{n_2},
t_3^{n_3},t_4^{n_4})$, can in general be written in the form
\eqn\exF{
F(u,v;t) = \big ( t_1 {\cdot t}_4 \big )^{n_1-E} \big ( t_2 {\cdot t}_4 
\big )^{n_2-E} \big ( t_1 {\cdot t}_2 \big )^{E} 
\big ( t_3 {\cdot t}_4 \big )^{n_3} \, \F(u,v;\lam,\mun) \, ,
}
where $\F$ is a polynomial in $\lam,\mun$, defined in \deflu,
with all terms $\lam^p\mun^q$ satisfying $p+q\le E$.
If $E> n_1$ then all terms in $\F$ must contain a factor $\mu^{E-n_1}$ to
cancel negative powers of $t_1 {\cdot t}_4$ in \exF. Since $t_i$ 
are three dimensional vectors $t_{1[r} t_{2s} t_{3t} t_{4u]} =0$ so that
$\lam,\mun$ are not independent but obey the relation
\eqn\conlm{
\Lambda \equiv \lam^2 + \mun^2  + 1 - 2\lam \mun - 2\lam - 2 \mun = 0 \, .
}
This may be solved in terms of a single variable $\alpha$ by
\eqn\defa{
\lam = \alpha^2 \, , \qquad \mun = (1-\alpha)^2 \, ,
}
so that 
\eqn\FhF{
\F(u,v;\lam,\mun) = {\hF}(x,\zz;\alpha) \, . 
}
${\hat \F}(x,\zz;\alpha)$ is symmetric in $x,\zz$ and, for $E \le n_1$, 
is a polynomial in $\alpha$ of degree $2E$, so 
that there are ${2E+1}$ independent coefficients, while if $E > n_1$ then it
must be of the form $(1-\alpha)^{2(E-n_1)}p(\alpha)$  with $p$ a polynomial
of degree $n_1$, so that the number of coefficients is $2n_1+1$.
These results correspond exactly of course to the number of 
$SU(2)_R$ invariants which can be formed in the four point function,
subject to \inn, together with \order, that can be found using standard 
$SU(2)$ representation multiplication rules. 

A similar expansion to \exF\ can be given for each $U_i$ 
\eqn\exF{\eqalign{
U_i(x,\zz;t)  = {}& \big ( t_1 {\cdot t}_4 \big )^{n_1-E} 
\big ( t_2 {\cdot t}_4 \big )^{n_2-E}
\big ( t_1 {\cdot t}_2 \big )^{E} \big ( t_3 {\cdot t}_4 \big )^{n_3} \, 
\U_i(x,\zz;\lam,\mun) \, ,\cr
& \U_i(x,\zz;\lam,\mun) = {\hU}_i(x,\zz;\alpha) \, . \cr}
}
The analysis of \eqJ\ and \Sz\ depends on using \LLU, or \LUW, as shown in 
appendix C, to relate $W_i$ and $U_i$. Defining
\eqn\WW{
W_i = i \,
t_2 {\cdot (t_3 \times t_4)} \, \big ( t_1 {\cdot t}_4 \big )^{n_1-E} \!
\big ( t_2 {\cdot t}_4 \big )^{n_2-E} \! \big ( t_1 {\cdot t}_2 \big )^{E-1} 
\big ( t_3 {\cdot t}_4 \big )^{n_3-1} \W_i \, ,
}
then we obtain
\eqn\crit{
2(2n_1-1) \W_i = \hD_i \hU{}_{i}  \, ,
}
where $\hD_i$ are linear operators given by
\eqn\DDD{
\hD_2 = {\d \over \d \alpha} + {2(E - n_1)\over 1-\alpha} \, , \quad
\hD_3 = {\d \over \d \alpha} - {2n_1\over \alpha} +
{2(E-n_1)\over 1-\alpha} \, , \quad
\hD_4 = {\d \over \d \alpha} + {2 E\over 1-\alpha} \, .
}

The superconformal identities \Fx, \FYn\ and \Sz\ then become
\eqna\superid
$$\eqalignno{
{\pr \over \pr x} \hF = {}& - {1\over x} \, \hU_2 + 
{1\over 1-x}\, \hU_{4} \, , & \superid{a} \cr
2n_1 \, \hF = {}& \hU_{2} + \hU_{3} + \hU_{4} \, , 
& \superid{b} \cr
\hD_2 \, \hU_{2} + {}& \hD_3 \,\hU_{3} + \hD_4 \, \hU_{4} = 0 \, .
& \superid{c} \cr}
$$
By acting on \superid{b} with $\hD_3$ and using \superid{c} we may obtain
\eqn\FDD{
\hD_3 \hF = - {1\over \alpha}\, \hU_{2} - 
{1\over \alpha(1-\alpha)} \, \hU_{4} \, ,
}
and substituting in \superid{a} gives
\eqn\diF{
\Big ( x {\pr \over \pr x} - \alpha \hD_3 \Big ) \hF =
\Big ( {x\over 1-x} + {1\over 1-\alpha} \Big ) \hU_{4} \, .
}      
The right hand side of \diF\ vanishes when $\alpha=1/x$ leaving an
equation for $\hF$ alone. With the explicit
form for $\hD_3$ in \DDD\ we have
\eqn\resF{
{\pr \over \pr x} \bigg ( x^{2n_1} \Big ( 1 - {1\over x} \Big )^{2(n_1-E)}
\hF \Big ( x,\zz; {1\over x} \Big ) \bigg ) = 0 \, .
}
Together with its partner or conjugate equation involving $\pr/\pr \zz$ 
\resF\ provides the final result for the constraints due to superconformal 
identities for the four point function when $\N=2$.

For the $\N=2$ case we may also require instead of \inn
\eqn\inc{
n_4 = n_1 +n_2 + n_3 - 2E - 1 \, ,
}
since $F$ can then be written as
\eqn\exFc{
F(u,v;t) = \big ( t_1 {\cdot t}_4 \big )^{n_1-E}
\big ( t_2 {\cdot t}_4 \big )^{n_2-E-1}\big ( t_1 {\cdot t}_2 \big )^{E} 
\big ( t_3 {\cdot t}_4 \big )^{n_3-1} \, t_2 \, {\cdot \, t_3 \times t_4}\
{\hat \F}(x,\zz;\alpha) \, .
}
There is an essentially unique expression in \exFc, with a single function 
${\hat \F}$ as a consequence of identities for the various possible 
vector cross products for null vectors  which take the form
\eqn\idthree{\eqalign{
t_1 \, {\cdot \, t_2 \times t_3} \ t_2{\cdot t_4} = {}& \half
(\lam - \mun + 1) \, t_2 \,{\cdot \, t_3 \times t_4} \ 
t_1{\cdot t_2} \, ,\cr
t_1 \, {\cdot \, t_2 \times t_4} \ t_2{\cdot t_3} = {}& \half
(\lam - \mun - 1) \, t_2 \,{\cdot \, t_3 \times t_4} \ 
t_1{\cdot t_2}\,,\cr 
t_1 \, {\cdot \, t_3 \times t_4} \ t_2{\cdot t_4} \, \mun = {}& \half
(\lam + \mun - 1) \, t_2 \,{\cdot \, t_3 \times t_4} \
t_1{\cdot t_4}\, . \cr} 
}
Since, as shown in appendix B, effectively $t_1 {\cdot t}_4 \, 
t_2 \, {\cdot \, t_3 \times t_4} = {\rm O}(1-\alpha)$ we can take
in \exFc, if $n_1-E\ge1$, $(1-\alpha){\hat \F}(x,\zz;\alpha)$ to be a polynomial 
of degree $2E+1$. If $n_1-E<1$ then ${\hat \F}(x,\zz;\alpha)$ must contain
a factor $(1-\alpha)^{2(E-n_1)-1}$. It is easy to see that the number
of independent coefficients matches with the number of independent
terms in the four point function obtained by counting possible
representations in each case.

There is a similar expansion as \exFc\ for $U_i$.
Instead of \WW\ and \crit\ we now have
\eqn\crittwo{
W_i = {i \over 2n_1-1} \, 
\big ( t_1 {\cdot t}_4 \big )^{n_1-E-1} \!
\big ( t_2 {\cdot t}_4 \big )^{n_2-E} \!
\big ( t_1 {\cdot t}_2 \big )^{E} \big ( t_3 {\cdot t}_4 \big )^{n_3} \mun
\, \hD_i \hU_{i} \, ,
}
with $\hD_i$ exactly as in \DDD. In consequence the superconformal
identities reduce to \superid{a,b,c} and we may derive the final result
\resF, albeit with $E$ given by \inc.

\newsec{Solution of Identities, $\N=2$}

Although in the $\N=2$ case the identities can be solved rather trivially
we show here how they may be put in a form which makes the connection
with the operator product expansion, and the possible supermultiplets
which may contribute to it, rather obvious. For the purposes of analysing
the operator product expansion for $x_1\sim x_2$ an alternative form to 
\Fourp\ is more convenient so we write
\eqn\fourp{\eqalign{
\l \vphi^{(n_1)} (x_1,t_1)& \, \vphi^{(n_2)}(x_2,t_2)\,
\vphi^{(n_3)}(x_3,t_3)\, \vphi^{(n_4)}(x_4,t_4 )\r \cr
&{} = {1 \over r_{12}^{\, \, n_1+n_2}\,  r_{34}^{\, \, n_3+n_4} }
\bigg ( {r_{24}\over r_{14}} \bigg )^{\! n_1-n_2} 
\bigg ( {r_{14}\over r_{13}} \bigg )^{\! n_3-n_4} G(u,v;t) \, ,\cr}
}
where
\eqn\GF{
G(u,v;t) = u^{n_1+n_2} \, v^{n_1+n_4-n_2-n_3} F(u,v;t) \, .
}

For application of the superconformal Ward identities here it is
convenient here to replace the variable $\alpha$ by $y$ where
\eqn\defy{
y = 2\alpha -1 \, ,
}
and $x,\zz$ by $z,\bz$ given by
\eqn\defz{
z= {2\over x} -1 \, , \qquad \bz= {2\over \zz} -1 \, .
}
Assuming now
\eqn\exG{
G(u,v;t) = \big ( t_1 {\cdot t}_4 \big )^{n_1-E} \big ( t_2 {\cdot t}_4 
\big )^{n_2-E} \big ( t_1 {\cdot t}_2 \big )^{E} 
\big ( t_3 {\cdot t}_4 \big )^{n_3} \, \G(u,v;y) \, ,
}
the solution of \resF\ and its conjugate equation, maintaining the
symmetry under $z \leftrightarrow \bz$, becomes
\eqn\Gf{
\G ( u,v; z) = u^{n_1+ n_2 -2E} f(\bz) \, , \qquad
\G ( u,v; \bz ) = u^{n_1+ n_2 -2E} f(z) \, ,
}
where $f$ is an unknown single variable function. Since $\G(u,v;y)$ 
is just a polynomial
in $y$ \Gf\ requires
\eqn\Gsol{
\G(u,v;y) = u^{n_1+n_2-2E}\, {(y-\bz)f(\bz) - (y-z) f(z) \over z-\bz}
+ (y-z)(y-\bz)\, \K(u,v;y) \, ,
}
where $\K(u,v;y)$ is undetermined, if $\G(u,v;y)$ is a polynomial of
degree $2E$ in $y$ then clearly $\K$ is a polynomial of degree $2E-2$.

The operator product expansion applied to this correlation function is realised
by expanding it in terms of conformal partial waves 
$\G_\Delta^{(\ell)}(u,v;\Delta_{21}, \Delta_{43})$, $\Delta_{ij} = \Delta_i -
\Delta_j$, which represent the contribution to a four point function for
four scalar fields, with scale dimensions $\Delta_i$, from an operator of
scale dimension $\Delta$ and spin $\ell$, and all its conformal descendants.
Explicit expressions, in four dimensions, were found in \Dos\
which are simple in terms of the variables
$x,\zz$ defined in \eiguv\foot{$$\eqalign{
\G_\Delta^{(\ell)}(u,v;\Delta_{21},\Delta_{43}) ={}&  {u^{{1\over 2}(\Delta-\ell)}
\over x - \zz}\Big ( ( - \half x)^\ell x \, F ( \half (\Delta + \Delta_{21}+\ell),
\half (\Delta - \Delta_{43}+\ell);\Delta+\ell;x )\cr
&\quad {} \times F ( \half (\Delta + \Delta_{21}-\ell-2),
\half (\Delta - \Delta_{43}-\ell-2);\Delta-\ell-2 ;\zz ) -
x\leftrightarrow \zz \Big ) \, . \cr}
$$}, which satisfy
\eqn\Gsym{\eqalign{
\G_\Delta^{(\ell)}(u,v;\Delta_{21}, \Delta_{43}) = {}& (-1)^\ell v^{{1\over 2}
\Delta_{43}} \, \G_\Delta^{(\ell)}(u/v,1/v;-\Delta_{21}, \Delta_{43}) \cr
= {}&  v^{{1\over 2}( \Delta_{43} - \Delta_{21})} \,
\G_\Delta^{(\ell)}(u,v;-\Delta_{21}, - \Delta_{43}) \, . \cr}
}
For this case the expansion is also over the contributions for differing $SU(2)_R$
$R$-representations and has the form, if $n_1 \ge E$,
\eqn\OPEG{
\G(u,v;y) = \sum_{R=n_4-n_3}^{n_1+n_2} \sum_{\Delta,\ell} a_{R,\Delta,\ell} \, 
P_{R+n_3-n_4}^{(2n_1-2E, 2n_2-2E)}(y)
\ \G_\Delta^{(\ell)}(u,v;2(n_2-n_1), 2(n_4-n_3) ) \, ,
}
with $P_n^{(a,b)}$ a Jacobi polynomial. For $a$ a negative integer
$P_n^{(a,b)}(y) \propto (1-y)^{-a}$ and $n+a \ge 0$. Hence when $n_1<E$ we
require a similar expansion to \OPEG\ but with $R=n_2-n_1, \dots , n_1+n_2$
and then $\G(u,v;y) \propto \mun^{E- n_1}$ as required in \exG\ to avoid negative
powers of $t_1 {\cdot t}_4$. The different terms appearing in the sum in \OPEG\ 
then determine the necessary spectrum of operators required by this correlation 
function. The symmetry properties of this operator product expansion follow
from \OPEG\ and $P_n^{(a,b)}(y)=(-1)^n P^{(b,a)}_n(-y)$.

We first consider the case when $n_1=n_2=n_3=n_4=n$, so that $E=n$.
To apply \Gf\ we first consider the expansion in terms of Legendre polynomials
(to which the Jacobi polynomial reduce in this case),
\eqn\Gexp{
\G(u,v;y) = \sum_{R=0}^{2n} a_R(u,v)  P_R(y) \, , \qquad 
\K(u,v;y) = \sum_{R=0}^{2n-2} A_R(u,v)  P_R(y) \, .
}
The $P_R(y)$ in \Gexp\ correspond to the $2n+1$ possible $SU(2)_R$ 
invariants for the four point function \fourp\ and, as a consequence of 
results in appendix B, the coefficients  $a_R$ represent the contribution
to the correlation function from operators belonging just to the $SU(2)_R$ 
$R$-representation in the operator product  expansion for 
$\vphi^{(n)} (x_1,t_1) \, \vphi^{(n)}(x_2,t_2)$. From \Gsol\ it is easy to
see that the single variable function $f$ involves terms linear in $y$ and
so contributes only  for $R=0,1$  giving
\eqn\af{
a^f_0 = { z f(z) - \bz f(\bz ) \over z - \bz} \, ,
\qquad a^f_1 = - { f(z) - f(\bz ) \over z - \bz} \, .
}
Using the expansion in \Gexp\ for $\K$ in \Gsol\ and standard recurrence
relations for Legendre polynomials gives corresponding expressions for 
$a_R$. For the terms involving $A_R$ we have
\eqn\asol{\eqalign{
a^{A_R}_{R+2} = {}& {(R+1)(R+2) \over (2R+1)(2R+3)}\, A_{R} \, , \qquad
a^{A_R}_{R-2} =  {(R-1)R \over (2R-1)(2R+1)}\, A_{R} \, , \cr
a^{A_R}_{R+1} = {}& -{2(R+1) \over 2R+1}\, {1-v \over u} \, A_{R} \, , \qquad
a^{A_R}_{R-1} = -{2R \over 2R+1}\, {1-v \over u} \, A_{R} \, , \cr
a^{A_R}_{R} = {}& \bigg ( 2{1+v\over u} - {1\over 2} 
+ {1 \over 2(2R-1)(2R+3)}\bigg )  A_{R} \, . \cr}
}
For $R\ge 2$ $a_R$ is therefore given in terms $A_{R\pm 2},A_{R\pm 1},A_R$
while for $R=0,1$, with \af, we have
\eqn\asoln{\eqalign{
a_0 ={}&  a_0^f + \Big ( 2\, {1+v\over u} - 
{2\over 3}\Big ) A_0 - {{2\over 3}}\, {1-v\over u}\, A_1 
+ {{2\over 15}}\,  A_2 \, , \cr 
a_1 = {}&  a_1^f  - 2{1-v\over u}\, A_0 +
\Big (2\, {1+v\over u} - {2\over 5}\Big ) A_1 - {{4\over 5}}\, {1-v \over u}\, 
A_2 + {{6\over 35}} \, A_3 \, . \cr}
}
In \asol\ and \asoln\ any contributions involving $A_R$ for $R>2n-2$
should be dropped.

The significance of the results given by \asol\ and \asoln\ is that they
correspond exactly to the $\N=2$ supermultiplet structure of operators
appearing in the operator product expansion. Each $a_R(u,v)$ may then be expanded
in terms of $\G_\Delta^{(\ell)}(u,v)\equiv \G_\Delta^{(\ell)}(u,v;0,0)$
\eqn\pwave{
a_R(u,v) = \sum_{\Delta,\ell} b_{R,\Delta,\ell} \, 
\G_\Delta^{(\ell)}(u,v) \, .
}
The conformal partial waves $\G_\Delta^{(\ell)}(u,v)$ satisfy crucial recurrence 
relations \scft,
\eqnn\recurG
$$\eqalignno{
- 2\,{1-v \over u} \, \G_{\Delta}^{(\ell)}(u,v) = {}& 
4 \, \G_{\Delta-1}^{(\ell+1)}(u,v) + \G_{\Delta-1}^{(\ell-1)}(u,v)
+ a_s \, \G_{\Delta+1}^{(\ell+1)}(u,v) 
+ \quar a_{t-1} \, \G_{\Delta-1}^{(\ell+1)}(u,v) \, , \cr
\Big ( 2\, {1+v \over u} - 1 \Big ) \, \G_{\Delta}^{(\ell)}(u,v) 
= {}&  4 \, \G_{\Delta-2}^{(\ell)}(u,v) + 4a_s \, \G_{\Delta}^{(\ell+2)}(u,v)
+ \quar a_{t-1} \, \G_{\Delta}^{(\ell-2)}(u,v)  \cr
\noalign{\vskip -4pt}
&\qquad {} + \quar a_s a_{t-1} \, \G_{\Delta+2}^{(\ell)}(u,v) \, , &\recurG \cr}
$$
where
\eqn\defst{
s = \half ( \Delta + \ell ) \, , \quad t = \half ( \Delta - \ell ) \, ,
\qquad a_s = {s^2 \over (2s-1)(2s+1)} \, .
}
In \recurG\ $a_{t-1}>0$ if $\Delta > \ell +3$.

If $A_R$ is restricted to a single partial wave so that
\eqn\azG{
A_R \to \G_{\Delta+2}^{(\ell)} \, .
}
then, using \recurG\ with \asol, 
\eqnn\longR
$$\eqalignno{
& \hskip 3cma^{A_R}_{R'} \to a_{R'}^{\vphantom g}
\big (\A^\Delta_{R,\ell} \big)  = \sum_{(\Delta',\ell')} b_{(\Delta';\ell')}
\, \G^{(\ell')}_{\Delta'}\, ,\cr & |R'-R|=2 , \  (\Delta';\ell') =
(\Delta+2;\ell) \, , \ \ |R'-R|=1 , \  (\Delta';\ell') =
(\Delta+3,\Delta+1;\ell\pm 1 ) \, , \cr & R'-R=0 , \  (\Delta';\ell') =
(\Delta+4,\Delta;\ell), \,  (\Delta';\ell') = (\Delta+2;\ell\pm2, \ell) \,
. &\longR \cr} 
$$ 
This gives exactly the expected
contributions corresponding to those operators present in 
a long $\N=2$ supermultiplet, which we may denote $\A^\Delta_{R,\ell}$, 
whose lowest dimension operator has dimension $\Delta$, spin $\ell$ belonging 
to the $SU(2)_R$ $R$-representation. From \asol\ and the positivity 
constraints for \recurG\ we may then easily see  that in \pwave\
$b_{(\Delta',\ell')} >0$ for $\Delta>\ell+1$.
For a unitary representation, so that all states in
$\A^\Delta_{R,\ell}$ have positive norm, (we consider here multiplets whose
$U(1)_R$ charge is zero) the requirement is
\eqn\ineq{
\Delta \ge  2 R + \ell + 2\, .
}
Since $\G_{\Delta}^{(\ell)}(u,v) = u^{{1\over 2}(\Delta-\ell)}F(u,v)$
with $F(u,v)$ expressible as a power series in $u,1-v$ we must have
from \azG\ for $u\sim 0$,
\eqn\unit{
A_R(u,v) \sim u^{R+2+\ep} \, , \qquad \ep \ge 0 \, .
}

The contribution of the single variable function $f$ \af\ represents 
operators just with twist $\Delta- \ell =2$. From the results in \Dos\
we have
\eqn\ttwo{
\G_{\ell+2}^{(\ell)}(u,v) = u \, { g_{\ell+1}(x) - g_{\ell+1}(\zz) \over x-\zz} 
= - 2 \, { g_{\ell+1}(x) - g_{\ell+1}(\zz) \over z-\bz}\, ,
}
for
\eqn\gell{
g_\ell(x)= (-\half x)^{\ell-1} x F(\ell,\ell;2\ell;x) = -{2 \over z^{\ell}}\, 
F \Big ( \half \ell, \half \ell+\half ; \ell+\half ;{1\over z^2} \Big ) \, , 
}
where $F$ is just an ordinary hypergeometric function\foot{$g_\ell(x)
\propto Q_{\ell-1}(z)$ with $Q_\nu$ an associated Legendre function.}. As shown 
in \scft\ $g_\ell$ satisfies
\eqn\recurg{
z \, g_\ell(x) = - g_{\ell-1} (x) - a_\ell \, g_{\ell+1}(x) \, .
}
In general we therefore expand the single variable function $f$ in \af\ in the
form
\eqn\expaf{
f(z) = \sum_{\ell=0}^\infty b_\ell \, g_\ell(x) \, .
}
For this to be possible $f(z)$ must be analytic in $1/z$, or equivalently in $x$.
If we consider just $f\to 2 g_{\ell+2}$ and use \recurg\ in \af\
then $a_R^f \to a_R(\C_{0,\ell})$ where 
\eqn\afl{
a_1(\C_{0,\ell}) = \G_{\ell+3}^{(\ell+1)} \, , \qquad
a_0(\C_{0,\ell}) = \G_{\ell+2}^{(\ell)} + 
a_{\ell+2} \, \G_{\ell+4}^{(\ell+2)} \, .
}
These results for $a_0,a_1$ then correspond to the contributions of operators
belonging to a semi-short $\N=2$ supermultiplet $\C_{0,\ell}$ whose lowest
dimension operator is a $SU(2)_R$ singlet with spin $\ell$ and
$\Delta = \ell+2$, i.e. at the unitarity threshold \ineq.

In general we denote by $\C_{R,\ell}$ the semi-short multiplet
whose lowest dimension operator has spin $\ell$, belongs to the 
representation $R$, and has $\Delta=2R+\ell$, so that the bound \ineq\
is saturated. At the unitarity threshold given by \ineq\ a long 
multiplet $\A^\Delta_{R,\ell}$ may be decomposed into
two semi-short supermultiplets $\C_{R,\ell}$ and $\C_{R+1,\ell-1}$, 
\short. This is reflected in the contributions to the four point function
since, with $a_{R'}^{\vphantom g}(\A^\Delta_{R,\ell})$ defined by \azG\ 
and \longR,
\eqn\ACC{
a_{R'}(\A^{2 R + \ell + 2}_{R,\ell}) = 
4 \, a_{R'}(\C_{R,\ell}) + {R+1 \over 2R+1} \,a_{R'}(\C_{R+1,\ell-1}) \, .
}
where we take
\eqn\sem{\eqalign{
a_R({\C_{R,\ell}}) = {}& \G_{2R+\ell+2}^{(\ell)}
+ \quar a_R\, \G_{2R+\ell+4}^{(\ell)} 
+ a_{R+\ell+2} \,\G_{2R+\ell+4}^{(\ell+2)} \, , \cr
a_{R-1}({\C_{R,\ell}}) = {}& {R\over 2R+1} \Big \{ \G_{2R+\ell+3}^{(\ell+1)} 
+ \quar \, \G_{2R+\ell+3}^{(\ell- 1)} +
\quar a_{R+\ell+2} \, \G_{2R+\ell+5}^{(\ell+1)} \Big \} \, ,\cr
a_{R-2}({\C_{R,\ell}}) = {}& 
{(R-1)R \over 4(2R-1)(2R+1)}\, \G_{2R+\ell+4}^{(\ell)} \, , \cr
a_{R+1}({\C_{R,\ell}}) = {}& 
{ R+1\over 2R+1}\, \G_{2R+\ell+3}^{(\ell+1)} \, .  \cr}
}
For $R=0$ \sem\  coincides with \afl. Thus the contribution of any
semi-short supermultiplet $\C_{R,\ell}$, $R=0,1,\dots ,2n-1$, to the 
four point function may be obtained by combining the results for 
long supermultiplets at unitarity threshold with \afl. There is no
reason why any particular  $\C_{R,\ell}$, except $\C_{0,0}$ which 
contains the energy momentum tensor and the conserved $SU(2)_R$ current,
should be present but if $f(z)$ is non zero it is necessary for there 
to be at least one semi-short contribution involving operators with 
protected dimensions.

A special case arises if we set $\ell=-1$. Formally, as shown in \short,
$\C_{R,-1} \simeq \B_{R+1}$ where $\B_R$ denotes the short supermultiplet
whose lowest dimension operator belongs to the $R$-representation with
$\Delta=2R, \, \ell=0$, obeying the full $\N=2$ shortening conditions.
The conformal partial waves as shown in \scft\ satisfy
\eqn\Gell{
(\quar )^{\ell-1} \G^{(-\ell)}_\Delta = - \G^{(\ell-2)}_\Delta \, , \qquad
\G^{(-1)}_\Delta = 0 \, ,
}
and hence from \sem\ we have
\eqn\CB{
a_{R'}(\C_{R,-1}) = {R+1 \over 2R+1} \, a_{R'} ( \B_{R+1} ) \, ,
}
where
\eqn\aBR{\eqalign{
a_R(\B_R) = {}& \G_{2R}^{(0)} \, , \qquad
a_{R-1}(\B_R) = {R\over 2R+1} \, \G_{2R+1}^{(1)} \, , \cr
a_{R-2}(\B_R) = {}& {(R-1)R\over 4(2R-1)(2R+1)} \, \G_{2R+2}^{(0)} \, . \cr}
}
The operators whose contributions appear in \aBR\ are just those
expected for the short supermultiplet $\B_R$ and there are possible 
contributions to the four point function for $R=1,2,\dots ,2n$. Since
\eqn\Gzero{
\G^{(0)}_0 (u,v) = 1 \, ,
}
then it is easy to see from \aBR\ that
\eqn\azero{
a_R(\B_0) = a_R(\I) = \de_{R0} \, ,
}
where $ a_R(\I)$ denotes the contribution of the identity in the operator
product expansion. Besides \CB\ we may also note that
\eqn\CBtwo{
a_{R'}(\C_{R,-2}) = -4 \, a_{R'} ( \B_R ) \, .
}
For $R=0$ this is in accord with \azero\ since $\G^{(-2)}_0 = -4$.

Apart from the case of the correlation function for four identical operators
as considered  above there are other solutions of the superconformal Ward 
identities which are of interest corresponding to extremal and next-to-extremal
correlation functions \Ext. The extremal case corresponds to taking $E=0$ in \inn.
There is then a unique $SU(2)_R$  invariant coupling which also follows from the
requirement that $\F$ in \exF, or $\G$ where in \GF\
$G(u,v;t) = (t_1 {\cdot t_4})^{n_1}(t_2 {\cdot t_4})^{n_2} 
(t_3 {\cdot t_4})^{n_3} \G(u,v)$, must be independent of $\lam,\mun$ 
and hence equivalently also of $\alpha$. In this case the result 
\resF\ for $\pr_x$ and its conjugate for $\pr_\zz$ simply imply
\eqn\Gext{
\G(u,v) = C \, u^{n_1+n_2} \, ,
}
where $C$ is independent of both $x,\zz$ and thus a constant. 
To interpret this in terms of the operator
product expansion for $\vphi^{(n_1)} (x_1,t_1) \vphi^{(n_2)}(x_2,t_2)$
we may use the result from \Dos,
\eqn\Gextr{
\G_{\Delta_1+\Delta_2}^{(0)}(u,v; \Delta_{21}, \Delta_1 + \Delta_2 ) 
= u^{{1\over 2}(\Delta_1+\Delta_2)} \, .
}
The result \Gext\ then shows that the only operators contributing
to the operator product expansion in the extremal case
have $\Delta=2(n_1+n_2), \ \ell=0$ and necessarily $R=n_1+n_2$. Such
operators can only be found as the lowest dimension operator
in the short supermultiplet $\B_{n_1+n_2}$.

For the next-to-extremal case we set $E=0$ in \inc. The solution 
of \resF\ can be conveniently expressed as
\eqn\next{
( 1 - z  )\, \G  ( u,v;z  ) = u^{n_1+n_2-1} \, f(\bz) \, ,
}
where in \GF\ we have
$G(u,v;t) = (t_1 {\cdot t_4})^{n_1}(t_2 {\cdot t_4})^{n_2-1}
(t_3 {\cdot t_4})^{n_3-1}\, t_2 \, {\cdot \, t_3 \times t_4} \,
\G(u,v;y)$. For $E=0$, $(1-y)\G(u,v;y)$ is linear in $y$ and from
\next\ and its conjugate we may find
\eqn\nextG{
(1-y)\G(u,v;y) = u^{n_1+n_2-1} \, {(y-\bz)f(\bz) - (y-z)f(z)
\over z - \bz} \, ,
}
so that $\G$ is determined just by the single variable function $g$
in this case.

For the next-to-extremal correlation function there are just two 
independent $SU(2)_R$ invariant couplings.
In a similar fashion to \Gexp, we have an expansion, from appendix B,
in terms of two Jacobi polynomials
\eqn\Gextexp{
(1-y)\, \G(u,v;y) = a_{n_1+n_2-1}(u,v)\, P_0^{(2n_1-1,2n_2-1)}(y)
+  a_{n_1+n_2}(u,v)\, P_1^{(2n_1-1,2n_2-1)}(y) \, ,
}
where $a_R$, $R=n_1+n_2-1, \, n_1+n_2$ represent the contribution of 
the two possible $R$-representations of $SU(2)_R$ in this case. From
\nextG\ we obtain
\eqn\Solg{\eqalign{
a_{n_1+n_2-1} = {}& - {1\over n_1+n_2} \, u^{n_1+n_2} \, 
{ f(z) - f(\bz) \over z - \bz} \, , \cr
a_{n_1+n_2} = {}& u^{n_1+n_2} \bigg ( { z f(z)
- \bz f(\bz ) \over z - \bz } 
+ {n_1-n_2 \over n_1+n_2} \, { f(z) - f(\bz) \over z - \bz} \bigg )
\, . \cr}
}
To interpret this in terms of the operator product expansion we may
use, extending \ttwo,
\eqn\text{\eqalign{
& \G_{\Delta_1+\Delta_2+\ell}^{(\ell)}(u,v;\Delta_{21}, \Delta_1+\Delta_2+2 )
= u^{{1\over 2}(\Delta_1+\Delta_2)}\, {g_{\ell+1}(x;\Delta_1,\Delta_2) - 
g_{\ell+1}(\zz;\Delta_1,\Delta_2) \over x-\zz} \, ,\cr
& g_\ell(x;\Delta_1,\Delta_2 )= (-\half x)^{\ell-1} x F(\ell+\Delta_2 -1 ,\ell;
2\ell+ \Delta_1+\Delta_2 - 2 ;x) \, . \cr}
}
In consequence only operators with twist $\Delta_1+\Delta_2$ can contribute for 
the solution for $a_R$ given by \Solg. If in \Solg\ let we $f(z) \to 2 
g_{\ell+2}(x;2n_1,2n_2)$ then 
$a_R \to a_R(\C_{n_1+n_2-1,\ell})$ where
\eqn\semin{\eqalign{
a_{n_1+n_2}& (\C_{n_1+n_2-1,\ell}) = {1\over n_1+n_2} \,
\G^{(\ell+1)}_{2n_1+2n_2+\ell+1} \, , \cr
a_{n_1+n_2-1}& (\C_{n_1+n_2-1,\ell}) \cr
= {}& \G^{(\ell)}_{2n_1+2n_2+\ell}
+ (n_2-n_1){(\ell+1)(2n_1+2n_2+\ell)\over (n_1+n_2+\ell)
(n_1+n_2+\ell+1)(n_1+n_2)}\, \G^{(\ell+1)}_{2n_1+2n_2+\ell+1} \cr
&{}+ {(\ell+2)(2n_1+\ell+1)(2n_2+\ell+1)(2n_1+2n_2+\ell)\over 
(n_1+n_2+\ell+1)^2(2n_1+2n_2+2\ell+1)(2n_1+2n_2+2\ell+3)}\, 
\G^{(\ell+2)}_{2n_1+2n_2+\ell+2} \, , \cr}
}
where $\G^{(\ell')}_{2n_1+2n_2+\ell'}$ are as in \text\ with $\Delta_i \to 2n_i$.
The contributions appearing in \semin\ correspond to those expected  from the
semi-short supermultiplet $\C_{n_1+n_2-1,\ell}$. Using $\C_{R,-1} \simeq \B_{R+1}$
again we may obtain the contribution for the short multiplet $\B_{n_1+n_2}$,
\eqn\ss{
a_{R}(\C_{n_1+n_2-1,{-1}}) = {1\over n_1+n_2} \, a_{R}(\B_{n_1+n_2})\,,
}
giving
\eqn\ash{\eqalign{
a_{n_1+n_2}(\B_{n_1+n_2}) = {}& \G^{(0)}_{2n_1+2n_2} \, , \cr
a_{n_1+n_2-1}(\B_{n_1+n_2}) = {}& {4n_1n_2 \over (n_1+n_2)(2n_1+2n_2+1)}\,
\G^{(1)}_{2n_1+2n_2+1} \, . \cr}
}
For the next-to-extremal correlation function therefore only the protected
short and semi-short supermultiplets $\B_R$ and $\C_{R-1,\ell}$ can
contribute to the operator product expansion.

By analysis \refs{\Non,\bpsN} of three point functions the 
possible $\N=2$ supermultiplets which may appear in the operator product
expansion of two $\N=2$ short supermultiplets is determined by the 
decomposition, for $n_2\ge n_1$,
\eqn\decomp{
\B_{n_1} \otimes \B_{n_2} \simeq \bigoplus_{n=n_2-n_1}^{n_2+n_1} \B_n
\oplus \bigoplus_{\ell\ge 0} \bigg (\bigoplus_{n=n_2-n_1}^{n_2+n_1-1} 
\C_{n,\ell} 
\oplus \bigoplus_{n=n_2-n_1}^{n_2+n_1-2} \A^\Delta_{n,\ell}\bigg ) \, ,
}
where for $\A^\Delta_{n,\ell}$ all $\Delta > 2n+\ell+2$ is allowed.
By considering also the corresponding result for $\B_{n_3}\otimes\B_{n_4}$
in all cases discussed above the general solution of the 
$\N=2$ superconformal identities accommodates all possible $\N=2$ 
supermultiplets which may contribute to the four point function in the
operator product expansion according to \decomp. In the
extremal case it is clear that only $\B_{n_1+n_2}$ contributes while
for the next-to-extremal case long multiplets which undergo
renormalisation are also excluded.

\newsec{Superconformal Ward Identities, $\N=4$}

We here describe an analysis of the superconformal Ward identities
for the four point function of $\N=4$ chiral primary operators
belonging to the $SU(4)_R$ $[0,p,0]$ representation with $\Delta=p$
represented by symmetric traceless fields
$\vphi_{r_1, \dots , r_p}(x)$, $r_i=1,\dots, 6$. As in \tphi\ we 
define $\vphi^{(p)}(x,t)$,  homogeneous of degree $p$ in $t$, in terms
of a six dimensional null vector $t_r$. The superconformal transformation
of $\vphi^{(p)}(x,t)$ is then expressible in the form
\eqn\susp{
\de \vphi^{(p)}(x,t) = - \hep(x) \, \gamma {\cdot t} \, 
\psi^{(p-1)}(x,t) + \bpsi {}^{(p-1)}(x,t)\, \bga {\cdot t} \,\hbep(x) \, ,
}
where the conformal Killing spinors $\hep^\alpha_i(x), \,
\hbep{}^{i\dal}(x)$ are as in \scep, with $i=1,2,3,4$ and 
$\ga_r{}^{\! ij} = - \ga_r{}^{\! ji}, \, \bga_{rij} = \half \vep_{ijkl} 
\ga_r{}^{\! kl}$ are $SU(4)$ gamma matrices, 
$\ga_r \bga_s + \ga_s \bga_r = - 2 \de_{rs} 1$, $\gamma_r{\!}^\dagger =
- \bga_r$. In \susp\ $\psi^{(p-1)}{}_{i\alpha}(x,t)$, 
$\bpsi{}^{(p-1)}{}^i{}_\dal(x,t)$ are homogeneous spinor fields of degree 
$p-1$ in $t$ and satisfy constraints similar to \condit
\eqn\conp{
\gamma {\cdot {\pr \over \pr t}} \, \psi^{(p-1)}_\alpha (x,t) = 0 \, , 
\qquad \bpsi{}^{\,(p-1)}(x,t)\, 
\bga {\cdot {\overleftarrow{\pr \over \pr t}}} =0 \, ,
}
which are necessary for them to belong to $SU(4)_R$ representations
$[0,p-1,1], \, [1,p-1,0]$.
At the next level the superconformal transformations involve a current
belonging to the $[1,p-1,1]$ representation which corresponds to
a homogeneous field of degree $p-1$ with one $SU(4)_R$ vector index
$J^{(p-1)}{}_{r\alpha\dal}(x,t)$ satisfying
\eqn\conJ{
t_r J^{(p-1)}{}_{r\alpha\dal}(x,t) = 0 \, , \qquad
{\pr \over \pr t_r } J^{(p-1)}{}_{r\alpha\dal}(x,t) = 0 \, .
}
The superconformal transformation of $\psi^{(p-1)}(x,t)$, neglecting
$\hep$ terms, is then
\eqn\varp{\eqalign{
\de \psi^{(p-1)}_\alpha (x,t) = {}& {1\over p} \,
\bga {\cdot {\pr \over \pr t}}\, i \pr_{\alpha\dal} \vphi^{(p)}(x,t) \,
\hbep{}^{\, \dal} (x)
+ 2 \, \bga {\cdot {\pr \over \pr t}} \vphi^{(p)}(x,t)\, \eta_\alpha \cr
&{}+ \bigg ( 1 + {1\over 2p+2}\,
\bga{\cdot t} \, \gamma{\cdot {\pr \over \pr t}} \bigg )
J^{(p-1)}{}_{r \alpha\dal}(x,t) \, \bga_r \hbep{}^{\, \dal}(x) \, . \cr}
}
Superconformal transformations which generate the full BPS multiplet
listed in \Class\ can be obtained similarly to \scft\ but the superconformal
Ward identities depend only on \susp\ and \varp. 

The general four point function of chiral primary operators can be
written in an identical form to \Fourp,
\eqn\Fourpf{\eqalign{
\l \vphi^{(p_1)}& (x_1,t_1) \, \vphi^{(p_2)}(x_2,t_2)\,
\vphi^{(p_3)}(x_3,t_3)\, \vphi^{(p_4)}(x_4,t_4 )\r \cr
&{} = { r_{23}^{\, \, \Sigma - p_2 - p_3}  \,
r_{34}^{\, \, \Sigma - p_3 - p_4}\over
r_{13}^{\, \, p_1}\,  r_{24}^{\, \, \Sigma - p_3} }\, F(u,v;t) \, ,
\quad \Sigma = \half ( p_1+p_2+p_3+p_4 )  \, . \cr}
}
The derivation of superconformal Ward identities initially follows an
almost identical path as that in section 2 leading to \redeq{a,b}.
With similar definitions to \fourpsi, \fourJ, taking $2n_i\to p_i$, and 
\hatRJ\ we find
\eqna\redeqf
$$\eqalignno{
{ 1\over p_1} \, \bga{\cdot {\pr \over \pr t_1}} \, {\pr \over \pr x} F
= {}& - \bigg ( 1  + {1\over 2p_1+2 }\,
\bga{\cdot t_1} \, \gamma {\cdot {\pr \over \pr t_1}} \bigg ) \bga{\cdot K}
- {1\over x} \, {T_2}\, \bga{\cdot t_2}
+ {1\over 1-x } \, {T_4}\, \bga{\cdot t_4} \, , & \redeqf{a} \cr
\bga{\cdot {\pr \over \pr t_1}} F = {}& {T_2}\, \bga{\cdot t_2}
+ {T_3}\, \bga{\cdot t_3} + {T_4}\, \bga{\cdot t_4}  \, .
& \redeqf{b} \cr}
$$
Instead of \cons\ we have the constraints, which follow from \conp\ and \conJ,
\eqn\consa{
\gamma{\cdot {\pr \over \pr t_1}} \, {T_i} = 0 \, , \quad
{T_i} \, \bga{\cdot {\overleftarrow{\pr \over \pr t_i}}} =0 \, , \qquad
t_1 {\cdot K} = {\pr \over \pr t_1}  {\cdot K} = 0 \, . 
}

As with \UVS\ we exhibit the dependence on $SU(4)$ gamma matrices by
writing
\eqn\TV{
T_i \, \bga{\cdot t_i} = \bga {\cdot V_i} + {\ts {1\over 6}} \,
\bga_{[r} \gamma_s \bga_{u]} W_{i,rsu} \, .
}
Since we take\foot{Note that $(\gamma_1 \bga_2 \gamma_3 \bga_4
\gamma_5 \bga_6)^\dagger = - \gamma_1 \bga_2 \gamma_3 \bga_4
\gamma_5 \bga_6$.} $\gamma_{[r}\bga_s\gamma_u\bga_v\gamma_w \bga_{z]} =
i \vep_{rsuvwz}$,
\eqn\seld{
 \bga_{[r} \gamma_s \bga_{u]} = - {\ts {1\over 6}} \, 
i \vep_{rsuvwz} \bga_v \gamma_w \bga_z \, ,
}
so that we must require the self-duality condition
\eqn\sdual{
W_{i,rsu} = {\ts {1\over 6}} \, i \vep_{rsuvwz} W_{i,vwz} \, .
}
Imposing the first equation in \consa\ we have
\eqn\consis{
{\pr \over \pr t_1} {\cdot V_i} = 0 \, , \qquad
\pr_{1[r} V_{i,s]} = \pr_{1u} W_{i,rsu} \, .
}
Just as in \VU\ we write,
\eqn\VUp{
V_i = {1\over p_1}\, {\pr \over \pr t_1} U_i + \hV_i \, , \qquad
t_1 {\cdot V_i} = U_i \, , \quad t_1 {\cdot \hV_i} = 0 \, .
}
so that in \consis\ we may let $V_i \to \hV_i$.

Using \TV\ with \VUp, and $\bga{\cdot t_1} \gamma {\cdot \pr_1} 
\bga{\cdot K} = - p_1 \, \bga{\cdot K} + \half 
\bga_{[r} \gamma_s \bga_{u]} L_{1,rs} K_u$,
\redeqf{a,b} may be decomposed into three pairs of equations,
\eqn\FU{
{\pr \over \pr x} F = -{1\over x}\, U_2 + {1\over 1-x}\, U_4 \, , \qquad
\qquad \  p_1 F = \sum_{i=2}^4 U_i  \, ,
}
and
\eqn\KV{
{p_1+2 \over 2p_1 +2} \, K_r = -{1\over x}\, \hV_{2,r} + {1\over 1-x}\, 
\hV_{4,r}\, , \qquad \sum_{i=2}^4 \hV_{i,r} = 0 \, ,
}
and
\eqn\KS{
{3 \over 2p_1 +2} \, \big ( L_{1,[rs} K_{u]} \big )_{\rm sd} 
= -{1\over x}\, W_{2,rsu} + {1\over 1-x}\, W_{4,rsu} \, , \qquad
\sum_{i=2}^4 W_{i,rsu} = 0 \, ,
}
where we define for $i=1,2,3,4$
\eqn\defL{
L_{i,rs} = t_{ir} \pr_{is} - t_{is} \pr_{ir} \, ,
}
and for any $X_{rsu} = X_{[rsu]}$ the self dual part, satisfying \sdual, is given by
\eqn\Xsd{
\big ( X_{rsu} \big )_{\rm sd} = \half \, X_{rsu} + {\ts {1\over 12}} \,
i \vep_{rsuvwz} X_{vwz} \, .
}

Since $2t_{1s} \pr_{1[r} V_{i,s]} = - p_1 \hV_{i,r}$ we may obtain from 
\consis\ 
\eqn\VT{
p_1 \hV_{i,r} = -L_{1,su}W_{i,rsu} \, ,
}
which gives $\hV_{i,r}$ in terms of $W_{i,rsu}$.
Furthermore from \consa\ $L_{1,rs} K_s = K_r$ and using also, as a 
consequence of the commutation relations for $L_1$, $[L_{1,rs},L_{1,ru}] 
= - 4 L_{1,su}$ we have $L_{1,rs}L_{1,ru} K_s = 3 K_u$. With, in addition,  
$\half L_{1,rs}L_{1,rs}\, K_u = - (p_1-1)(p_1+3) K_u$ we may then obtain
\eqn\LLK{
3L_{1,rs}L_{1,[rs} K_{u]} = - 2p_1(p_1+2) K_u \, .
}
Since also $\vep_{rsuvwz} L_{1,su}L_{1,vw} K_z = 0$ it is clear from \VT\
and \LLK\ that eqs. \KS\ imply \KV. However, if we define 
\eqn\Wbar{
{\bar W}_{i,rsu} = 3\big ( L_{1,[rs} \hV_{i,u]} \big )_{\rm sd} - (p_1+2)
\, W_{i,rsu} \, ,
}
with $\hV_{i,u}$ determined by \VT, then as a consequence of
\KV\ and \KS\ we must also require
\eqn\KW{
{1\over x}\, {\bar W}_{2,rsu} = {1\over 1- x} \, {\bar W}_{4,rsu} \, .
}

{}From the second equation in \consa\ we may obtain
$\half L_{i,rs} (T_i \bga{\cdot t_i}) \gamma_r \bga_s 
= (p_i+4) T_i \bga{\cdot t_i}$ which leads to the relations
\eqna\LVW
$$\eqalignno{
& L_{i,rs} V_{i,s}- L_{i,su} W_{i,rsu} = (p_i+4) V_{i,r} \, , &\LVW{a} \cr
& 3 \big( L_{i,[rs} V_{i,u]} \big )_{\rm sd} 
+ 3L_{i,[r|v}W_{i,su]v} = (p_i+4) W_{i,rsu} \, , &\LVW{b} \cr}
$$
where $L_{i,[u|v} W_{i,rs]v}$ is self dual as a consequence
of \sdual. We also have from $T_i \bga{\cdot t_i} \, \gamma{\cdot t_i} = 0$
\eqn\Ttt{
t_i {\cdot V_i} = 0 \, , \qquad t_{i[r} V_{i,s]} + W_{i,rsu} t_{iu} = 0 \, .
}
For consistency we note that $\pr_{is} ( t_{i[r} V_{i,s]} + W_{i,rsu} t_{iu})
= 0$ is identical with \LVW{a}. Furthermore using \sdual\ we have
$(\pr_{i[r} \, W_{i,su]v} t_{iv})_{\rm sd} = \half \big (
\pr_{i[r} \, W_{i,su]v} t_{iv} - \pr_{i v} \, t_{i[r} W_{i,su]v} \big ) + \thir
\pr_{iv} ( W_{i,rsu} t_{iv} )$ and, from appendix A, $(p_i+2)\pr_{iv} 
( W_{i,rsu} t_{iv} ) = {(p_i+3)(p_i+4)}\, W_{i,rsu}$ while acting on $W_{i,suv}$
similarly $(p_i+2)\pr_{i[r} t_{iv]} = - (p_i+3)\, \half L_{i,rs}$. Hence we have
demonstrated that 
$\big (\pr_{i[r} ( t_{i s} V_{i,u]} + W_{i,su]v} t_{iv} \big )_{\rm sd}=0$
is identical to \LVW{b} so that this equation is also implied by \Ttt.

Combining \Ttt\ with \VUp\ gives the essential equation
\eqn\WUp{
t_{i[r} \pr_{1s]} U_i + p_1 \, t_{i[r} \hV_{i,s]} + p_1 \, W_{i,rsu} t_{iu} = 0 \, ,
}
where $p_1 \hV_{i,s}$ is determined by \VT.

As in \exF\ we may expand the correlation function $F$, as defined in
\Fourpf, in terms of $SU(4)$ invariants
\eqn\exFp{
F(u,v;t) = \big ( t_1 {\cdot t}_4 \big )^{p_1-E} \big ( t_2 {\cdot t}_4
\big )^{p_2-E} \big ( t_1 {\cdot t}_2 \big )^{E}
\big ( t_3 {\cdot t}_4 \big )^{p_3} \, \F(u,v;\lam,\mun) \, ,
}
where we assume
\eqn\ppp{
p_1 \le p_2 \le p_3 \le p_4 \, , \qquad 2E = p_1 +p_2 + p_3 - p_4 \, .
}
In \exFp\ $\F(u,v;\lam,\mun)$ is a polynomial in $\lam,\mun$ consistent
with $F(u,v;t) = {\rm O}(t_1^{p_1},t_2^{p_2},t_3^{p_3},t_4^{p_4})$ and 
hence $E\ge 0$ is a integer. For $p_1 \ge E$ then 
$\F$ is expressible as a polynomial of degree $E$ in $\lam,\mun$, i.e.
a linear expansion in the $\half (E+1)(E+2)$ independent monomials 
$\lam^p\mun^q$ with $p+q \le E$. 
For $p_1 < E$ it is necessary also that $ q \ge E - p_1$ giving
only $\half (p_1 +1)(p_1 + 2)$ independent terms. It is easy to see that 
this matches the number of invariants that may be constructed
by finding common representations in $[0,p_1,0] \otimes [0,p_2,0]$ and
$[0,p_3,0] \otimes [0,p_4,0]$ using the tensor product result
\eqn\Oprod{
[0,p_1,0] \otimes [0,p_2,0] \simeq \bigoplus_{r=0}^{p_1} 
\bigoplus_{s=0}^{p_1-r} [r,p_2-p_1+2s,r] \, .
}
Hence representations $[r,p_2-p_1+2s,r]$ may contribute for
$s=0,\dots , n-r , \ r =0, \dots , n$ with $n=E$ if $p_1 \ge E$, otherwise 
$n=p_1$.

In an exactly similar fashion to \exFp\ we may express $U_i(x,\zz;t)$
in terms of $\U_i(x,\zz;\lam,\mun)$ so that \FU\ becomes
\eqn\FUc{
{\pr \over \pr x} \F = -{1\over x}\, \U_2 + {1\over 1-x}\, \U_4 \, , \qquad
\qquad \  p_1 F = \U_2 + \U_3 + \U_4   \, .
}

Furthermore we may also decompose $W_{i,rsu}(x,\zz;t)$ for $i=2,3,4$
in terms of four independent self dual tensors,
\eqn\exWp{\eqalign{
W_{i,rsu}= - & \big ( t_1 {\cdot t}_4 \big )^{p_1-E} 
\big ( t_2 {\cdot t}_4 \big )^{p_2-E} \big ( t_1 {\cdot t}_2 \big )^{E-2}
\big ( t_3 {\cdot t}_4 \big )^{p_3-1} \, \cr
&{} \times \Big ( \big ( t_{1[r} t_{2s} t_{3u]} \big )_{\rm sd} \,
t_2 {\cdot t_4} \, \A_i
+ \big ( t_{1[r} t_{4s} t_{2u]} \big )_{\rm sd} \,
t_2 {\cdot t_3} \, \B_i \cr
&\quad {} + \big ( t_{1[r} t_{3s} t_{4u]} \big )_{\rm sd} \,
t_2 {\cdot t_3} \, t_2 {\cdot t_4} \, {1\over t_3{\cdot t_4}} \, \C_i + 
\big ( t_{2[r} t_{3s} t_{4u]} \big )_{\rm sd} \,
t_1 {\cdot t_2} \, \W_i\Big ) \, , \cr}
}
with $\A_i,\B_i,\C_i$ and $\W_i$ polynomials in $\lam,\mun$ of degree $E-2$
and $E-1$, if $p_1 \ge E$. From its definition in \TV\ we must have
\eqn\ABC{
\C_2 = \B_3 = \A_4 = 0 \, .
}
The result \KS\ then requires
\eqn\sumW{
\A_2 + \A_3 = 0 \, , \quad \B_2 + \B_4 = 0 \, , \quad \C_3 + \C_4 = 0 \, ,
\quad \W_2 + \W_3 + \W_4 = 0 \, .
}

We may similarly decompose $\hV_{i,r}$ in the form
\eqn\exVp{\eqalign{
\hV_{i,r}= {} & \big ( t_1 {\cdot t}_4 \big )^{p_1-E-1}
\big ( t_2 {\cdot t}_4 \big )^{p_2-E} \big ( t_1 {\cdot t}_2 \big )^{E-1}
\big ( t_3 {\cdot t}_4 \big )^{p_3-1} \, \cr
&{} \times \big ( ( t_{2r} \, t_1 {\cdot t_4} -  t_{4r} \, t_1 {\cdot t_2})
t_3 {\cdot t_4} \, \I_i
+ ( t_{3r} \, t_1 {\cdot t_4} -  t_{4r} \, t_1 {\cdot t_3})
t_2 {\cdot t_4} \, \J_i
+ t_{1r}\, t_2 {\cdot t_4} t_3 {\cdot t_4} \, \V_i \big ) \, , \cr}
}
where we impose $t_1{\cdot \hV_i}=0$. The coefficient of $t_{1r}$ is determined 
by the requirement $\pr_1{\cdot \hV_i}=0$,
\eqn\OV{
(p_i+2)\V_i =  - \O_\lam \I_i + \I_i  - 
(\lam \O_\lam - \mun \O_\mun)  \J_i + \J_i \, ,
}
with differential operators
\eqn\defO{\eqalign{
\O_\lam = {}& ( \lam + \mun -1) \, {\pr \over \pr \lam} + 2
\mun {\pr \over \pr \mun} + p_1 - 2E +1 \, , \cr
\O_\mun = {}& 2 \lam {\pr \over \pr \lam} + ( \lam + \mun - 1) \bigg (
{\pr \over \pr \mun} + { p_1 - E \over \mun} \bigg )  - p_1 +1 \, . \cr}
}
Using \VT\ we get
\eqn\solIJ{\eqalign{
6p_1 \, \I_i = {}& (p_1+2) ( \lam\, \A_i - \mun\, \B_i ) - 
(\lam \O_\lam - \mun \O_\mun) \W_i \, , \cr
6p_1 \, \J_i = {}& - (p_1+2) ( \A_i - \mun\, \C_i ) + \O_\lam \W_i  \, . \cr}
}
{}From \OV\ we then obtain
\eqn\VABC{
6p_1 \, \V_i = \tau \big ( (\O_\lam + 1 ) \B_i - (\O_\mun + 1 ) \A_i
- ( \lam (\O_\lam + 1 ) - \mun (\O_\mun + 1 ) ) \C_i \big ) \, .
}

As a consequence of \WUp\ the coefficients in \exFp\ are not independent but
we have  relations which determine $\A_i, \B_i , \C_i$ for each $i$,
\eqn\relABC{\eqalign{
(p_1+1)\, \A_2 = {}& 3\, {\pr \over \pr \lam} \U_2 + \half ( \O_\lam - p_1) \W_2 \, , \cr
(p_1+1)\, \B_2 = {}& - 3 \bigg ( {\pr \over \pr \mun} +{p_1 -E \over \mun} \bigg )\U_2 
+ \half ( \O_\mun - p_1) \W_2 \, , \cr
(p_1+1)\,\lam  \A_3 = {}&  3 \bigg ( \lam {\pr \over \pr \lam} + 
\mun {\pr \over \pr \mun}
- E  \bigg )\U_3 + \half ( \lam \O_\lam - \mun \O_\mun - p_1) \W_3 \, , \cr
(p_1+1)\, \lam \C_3 = {}& 3 \bigg ( {\pr \over \pr \mun} +{p_1 -E \over \mun} \bigg )\U_3 
-  \half ( \O_\mun +  p_1) \W_3 \, , \cr
(p_1+1)\,\mun  \B_4 = {}&- 3 \bigg ( \lam {\pr \over \pr \lam} + 
\mun {\pr \over \pr \mun}
- E  \bigg )\U_4 -  \half ( \lam \O_\lam - \mun \O_\mun + p_1) \W_4 \, , \cr
(p_1+1)\, \tau \C_4 = {}& - 3 \, {\pr \over \pr \lam} \U_4 - \half 
( \O_\lam + p_1) \W_4 \, . \cr}
}
Combining this with \sumW\ and also the result in \FUc\ for $p_1 \F$ leads to
\eqn\FUW{\eqalign{
\bigg ( \mun {\pr \over \pr \mun} + p_1 -E  \bigg ) \F = {}& \U_4 + {\ts {1\over 6}}
(\lam - \mun -1 ) \W_4 - {\ts {1\over 3}} \, \mun \W_2 \, , \cr
\bigg ( \lam {\pr \over \pr \lam} + \mun {\pr \over \pr \mun} - E  \bigg ) \F = {}&
- \U_2 + {\ts {1\over 6}} (\lam - \mun -1 ) \W_2 - {\ts {1\over 3}} \, \W_4 \, . \cr}
}
If ${\bar W}_{i,rsu}$ given by \Wbar\ is defined in terms of 
${\bar \A}_i, {\bar \B}_i, {\bar \C_i}$ and ${\bar \W}_i$ as in \exWp\ then
it is easy to see that ${\bar \W}_i = -(p_1+2) \W_i$ and, as a consequence of
\solIJ\ and \relABC,
\eqnn\ABCW
$$\eqalignno{
2(p_1+1){\bar \A}_i ={}&  \bigg ( {\pr \over \pr \lam} 
\big ( \lam \O_\lam - \mun \O_\mun + p_1 + 2 \big ) 
+ \Big ( \lam {\pr \over \pr \lam} + \mun {\pr \over \pr \mun} - E +1 \Big )
\big ( \O_\lam - p_1 - 2 \big ) \bigg ) \W_i \, , \cr
&{} \mun {\bar \C}_i - {\bar \A}_i = \O_\lam \W_i \, , \qquad 
\mun {\bar \B}_i - \lam {\bar \A}_i = (\lam \O_\lam - \mun \O_\mun) \W_i \, . 
& \ABCW \cr}
$$
Hence \KW\ reduces to just
\eqn\WW{
{1\over x} \, \W_2 = {1\over 1-x} \, \W_4 \, .
}

In terms of the variables $\alpha,\bet$, defined in \defab,  \FUW\ becomes
\eqn\relF{
\alpha(1-\alpha) {\pr \over \pr \alpha} \F + E \, \alpha \F - p_1 \, \F =
- (1-\alpha) \U_2 - \U_4 + {\ts {1\over 6}} (\alpha - \bet ) \big ( (1-\alpha) \W_2
+ \W_4 \big ) \, , 
}
together with the conjugate equation obtained for $\alpha \leftrightarrow \bet$.
If this is used together with \FUc\ for $\pr_x \F$ we may eliminate $\U_2$ to
obtain
\eqn\WarF{\eqalign{
\bigg ( x {\pr \over \pr x} - \alpha {\pr \over \pr \alpha}& - E \, 
{\alpha \over 1-\alpha} + p_1 \, {1\over 1-\alpha} \bigg ) \F \cr 
= {}& \bigg ( {x \over 1-x} + {1\over 1-\alpha} \bigg ) \U_4 - {\ts {1\over 6}} 
(\alpha - \bet ) \bigg ( \W_2 + {1\over 1-\alpha} \W_4 \bigg ) \cr
= {}& {1-\alpha x \over (1-\alpha)(1-x) } \Big ( \U_4 - {\ts {1\over 6}} 
(\alpha - \bet ) \W_4 \Big ) \, , \cr}
}
where we have used \WW. Writing $\F(u,v;\lam,\mun)={\hat \F}(x,\zz;\alpha,\bet)$
evidently
\eqn\Wwp{
\bigg ( x {\pr \over \pr x} - \alpha {\pr \over \pr \alpha} + E \, + 
(p_1 - E) \, {1\over 1-\alpha} \bigg ) {\hat \F}(x,\zz;\alpha,\bet) \Big |_{
\alpha= {1\over x}}  = 0 \, ,
}
which is solved by writing
\eqn\solW{
u^E ( 1-v)^{p_1 - E} {\hat \F}\big (x,\zz;{1\over x},\bet \big ) =
f(\zz,\bet ) \, .
}
Together with the conjugate equation in which $\alpha \to \bet$ \solW\ is the
basic solution of the superconformal Ward identities in this context.

\newsec{Solution of Identities, $\N=4$}

We here extend the results of section 3 to the $\N=4$ case. As previously
it is more convenient for consideration of the operator product expansion
to change from $F(u,v;t)$ to $G(u,v;t)$, defined in a similar fashion to \fourp\
with $2n_i \to p_i$. Writing $G(u,v;t)$ in a similar fashion to \exFp\ then 
the corresponding function $\G$ is given in terms of $\F(u,v;\lam,\mun)$ by
\eqn\GFp{
\G(u,v;\lam,\mun) = u^{{1\over 2}(p_1+p_2)}\, v^{p_1-E}\F(u,v;\lam,\mun) \, .
}
For the applications in the section it is convenient to write
\eqn\Gyy{
\G(u,v;\lam,\mun) = \hG(u,v;y,\yz)=\hG(u,v;\yz,y) \, ,
}
where $\hG$ depends on the variables
\eqn\yy{
y = 2\alpha-1 \, , \qquad \yz=2\bet-1 \, .
}
The solution \solW\ then gives, with $z,\bz$ defined in \defz,
\eqn\solid{
\hG(u,v;z,\yz) = u^{{1\over 2}(p_1+p_2)-E}f(\bz,\yz) \, , \qquad 
\hG(u,v;\bz,\yz) = u^{{1\over 2}(p_1+p_2)-E} f(z,\yz) \, . 
}
For consistency, since $f(z,z)=f(\bz,\bz)$,  we must have
\eqn\real{
f(z,z) = k \, .
}

In general the conformal partial wave expansion and the decomposition into
contributions for differing $SU(4)_R$ representations further into conformal
partial waves is realised by writing
for $p_1 \ge E$.
\eqn\OPEf{ \eqalign{
\hG(u,v;y,\yz) = {}& \sum_{0\le m \le n \le E} a_{nm}(u,v) \,
P_{nm}^{(p_1-E, p_2-E )}(y,\yz) \cr
= {}& \sum_{0\le m \le n \le E}\sum_{\Delta,\ell} a_{nm,\Delta,\ell} \,
P_{nm}^{(p_1-E, p_2-E )}(y,\yz) \ \G_\Delta^{(\ell)}(u,v;p_2-p_1,p_4-p_3) \, ,\cr}
}
where $\G_\Delta^{(\ell)}$ are described in section 3 and $P_{nm}^{(a,b)}(y,\yz)$
are symmetric polynomials of degree $n$ (i.e. for an expansion in
terms of the form $(y\yz)^s(y^t+\yz^t)$, $s+t\le n$) which are discussed in 
appendix B and which are given in terms of Jacobi polynomials
\eqn\poly{
P^{(a,b)}_{nm}(y,\yz) =
{ P^{(a,b)}_{n+1}(y) P^{(a,b)}_{m\vphantom{1}}(\yz) - 
P^{(a,b)}_{m\vphantom{1}} (y) P^{(a,b)}_{n+1} (\yz) 
\over y - \yz} = - P^{(a,b)}_{m-1\, n+1}(y,\yz)  \, ,
}
In \OPEf\  $a_{nm,\Delta,\ell}$ then corresponds to the presence of an operator 
in the operator product expansion ifor $\vphi^{(p_1)}$ and  $\vphi^{(p_2)}$ 
belonging to the $SU(4)_R$ representation with Dynkin labels 
$[n-m,p_1+p_2-2E+2m,n-m]$ and  with scale dimension $\Delta$, spin $\ell$. 
The expansion \OPEf\ also extends to $p_1 < E$ save that then $m \ge E - p_1$ 
and $P^{(a,b)}_{nm}(y,\yz) \propto \mun^{E-p_1}$.

We consider initially in detail the case with $p_i=p=E$ for all $i$ and 
$\G(u,v;y,\yz)$ is a symmetric polynomial in $y,\yz$ with degree $p$. Since it
must also be symmetric in $z,\bz$ \solid\ implies
\eqnn\Gsolt
$$\eqalignno{
\hG(u,v;y,\yz) = {}& -k + {(y-z)(\yz-\bz)\big ( f(z,\yz) + f(\bz,y) \big )
- (y-\bz)(\yz- z) \big ( f(z,y)+ f(\bz,\yz) \big )\over (z-\bz)(y-\yz)}\cr
&{} + (y-z)(y-\bz)(\yz-z)(\yz-\bz)\, \K(u,v;\lam,\mun) \, , & \Gsolt \cr}
$$
with $\K(u,v;\lam,\mun)={\hat \K}(u,v;y,\yz)$ defining an undetermined symmetric 
polynomial in $y,\yz$ of degree $p-2$. 
This term corresponds to the result \uncon\ described in the introduction.
To take account of the constraint \real\ we write
\eqn\ff{
f(z,y) = k + (y-z) \hf(z,y) \, ,
}
with $\hf(z,y)$ a free function, polynomial in $y$ of degree $p-1$.

The decomposition of $\hG(u,v;y,\yz)$ into the contributions for
different possible $SU(4)_R$ representations is given by \OPEf\
where  $a_{nm}$ are for this case the coefficients corresponding to
the representation with Dynkin labels $[n-m,2m,n-m]$. For this case
in \poly\ $P^{(0,0)}_n(y) = P_n(y)$, conventional Legendre polynomials.

We first consider the the contribution resulting from the constant $k$
in \Gsolt\ and \ff. It is easy to see that this gives only
\eqn\azero{
a^k_{00} = k \, .
}

To analyse the contributions arising from the function $\hf(z,y)$
this may be expanded as
\eqn\expf{
\hf(z,y) = \sum_{n=0}^{p-1} f_n(z)\, P_n(y) \, .
}
Using this in \ff\ and \Gsolt\ then $f_n$ gives rise to the following
contributions to $a_{nm}$ just for $m=0,1$,
\eqn\afour{\eqalign{
a^{f_n}_{n+1\,m} = {}&{(n+1)(n+2)\over (2n+1)(2n+3)}\, 
F_{nm}(z,\bz) \, , \qquad
a^{f_n}_{n-3\,m} =  {(n-1)n \over (2n-1)(2n+1)}\, F_{nm}(z,\bz) \, , \cr
a^{f_n}_{n\,m} = {}& -{n+1 \over 2n+1}\,  (z+\bz) F_{nm}(z,\bz)
\, , \qquad a^{f_n}_{n-2\,m} = -{n \over 2n+1}\, (z+\bz) F_{nm}(z,\bz) \, , \cr
a^{f_n}_{n-1\,m} = {}& \bigg ( z\bz + {1\over 2} 
+ {1 \over 2(2n-1)(2n+3)}\bigg )  F_{nm}(z,\bz) \, , \cr}
}
where
\eqn\FF{
F_{n1}(z,\bz) = - {f_n(z)-f_n(\bz) \over z-\bz} \, , \qquad
F_{n0}(z,\zz) = {zf_n(z)-\bz f_n(\bz) \over z-\bz} \, .
}
For low $n$ the results need to be modified but these can be obtained
from \afour\ by taking into account the symmetry relation in \poly. For 
$n=0$, $a^{f_0}_{11},a^{f_0}_{10}$ are as in \afour\ but for $a^{f_0}_{00}$
we need to take
\eqn\afz{
a^{f_0}_{00} - a^{f_0}_{-11} \to
a^{f_0}_{00} = - {(z^2-{1\over 3}) f_0(z) - (\bz^2-{1\over 3}) f_0(\bz)
\over z - \bz } \, ,
}
while for $n=1$, $a^{f_1}_{21},a^{f_1}_{20},a^{f_1}_{11},a^{f_1}_{10}$
are given by \afour\ but
\eqn\afone{
a^{f_1}_{00} = {\bz(z^2-{1\over 3}) f_1(z) - 
z (\bz^2-{1\over 3}) f_1(\bz) \over z - \bz } +{4\over 15}\, F_{11} \, . 
}

It remains to consider the contribution of the two variable function
$\K$ in \Gsolt\ which is expanded, for 
$P_{nm}(y,\yz) \equiv P^{(0,0)}_{nm}(y,\yz)$, as
\eqn\Kexp{
{\hat \K}(u,v;y,\yz) = \sum_{0\le m \le n \le p-2} \!\! A_{nm}(u,v) \, 
P_{nm}(y,\yz) \, , 
}
with $\half(p-1)p$ terms. In this case the Legendre recurrence relations
give
\eqnn\longr
$$\eqalignno{
a_{n{-2}\, m{-2}}^{A_{n m}}={}&{(m-1) m\, n (n+1)
\over(2m-1)(2m+1)(2n+1)(2n+3)}\, A_{n m} \, ,\cr
a_{n{-2}\, m{+2}}^{A_{n m}}={}& {(m+1)(m+2)n(n+1)
\over(2m+1)(2m+3)(2n+1)(2n+3)}\, A_{n m}  \, , \cr
a_{n{+2}\, m{-2}}^{A_{n m}}={}&{(m-1)m(n+2)(n+3)
\over(2m-1)(2m+1)(2n+3)(2n+5)}\, A_{n m} \, ,\cr
a_{n{+2}\, m{+2}}^{A_{n m}}={}&{(m+1)(m+2)(n+2)(n+3)
\over(2m+1)(2m+3)(2n+3)(2n+5)}\, A_{n m} \, , \cr
a_{n{-2}\, m{-1}}^{A_{n m}}={}& - {2 m n(n+1)\over(2m+1)(2n+1)(2n+3)}\,
{1-v\over u}\, A_{n m}  \, , \cr
a_{n{-2}\, m{+1}}^{A_{n m}}={}&-{2 (m+1) n(n+1)\over (2m+1)(2n+1)(2n+3)}\,
{1-v\over u}\, A_{n m}  \, , \cr
a_{n{-1}\, m{-2}}^{A_{n m}}={}&-{2 (m-1)m(n+1)\over(2m-1)(2m+1)(2n+3)} \,
{1-v\over u}\, A_{n m}  \, , \cr
a_{n{-1}\, m{+2}}^{A_{n m}}={}&-{2(m+1)(m+2)(n+1)\over(2m+1)(2m+3)(2n+3)}\,
{1-v\over u}\, A_{n m}  \, , \cr
a_{n{+2}\, m{-1}}^{A_{n m}}={}&-{2m(n+2)(n+3)\over(2m+1)(2n+3)(2n+5)}\,
{1-v\over u}\, A_{n m}  \, , \cr
a_{n{+2}\, m{+1}}^{A_{n m}}={}&-{2(m+1)(n+2)(n+3)\over(2m+1)(2n+3)(2n+5)}\,
{1-v\over u}\, A_{n m}  \, , \cr
a_{n{+1}\, m{-2}}^{A_{n m}}={}&-{2(m-1)m(n+2)\over(2m-1)(2m+1)(2n+3)}\,
{1-v\over u}\, A_{n m}  \, , \cr
a_{n{+1}\, m{+2}}^{A_{n m}}={}&-{2(m+1)(m+2)(n+2)\over(2m+1)(2m+3)(2n+3)}\,
{1-v\over u}\, A_{n m}  \, , \cr
a_{n{-1}\, m{-1}}^{A_{n m}}={}&{4 m(n+1)\over(2m+1)(2n+3)}\,
{(1-v)^2\over u^2}\, A_{n m}  \, , \cr
a_{n{-1}\, m{+1}}^{A_{n m}}={}& {4(m+1)(n+1)\over(2m+1)(2n+3)}\,
{(1-v)^2\over u^2}\, A_{n m}  \, , \cr
a_{n{+1}\, m{-1}}^{A_{n m}}={}& {4m(n+2)\over(2m+1)(2n+3)}\,
{(1-v)^2\over u^2}\, A_{n m}  \, , \cr
a_{n{+1}\, m{+1}}^{A_{n m}}={}& {4(m+1)(n+2)\over(2m+1)(2n+3)}\,
{(1-v)^2\over u^2}\, A_{n m}  \, , \cr
a_{n{-2}\, m}^{A_{n m}}={}& {2 n(n+1)\over (2n+1)(2n+3)} \, B_m\, A_{n m}  \, ,
\qquad\ \,  a_{n{+2}\, m}^{A_{n m}}={2(n+2)(n+3)\over (2n+3)(2n+5)}
\, B_m\, A_{n m}  \, , \cr
a_{n\, m{-2}}^{A_{n m}}={}& {2 (m-1)m \over(2m-1)(2m+1)} \, 
B_{n+1}\, A_{n m} \, ,
\quad\   a_{n\, m{+2}}^{A_{n m}}={2(m+1)(m+2)\over(2m+1)(2m+3)} \, 
B_{n+1}\, A_{n m}  \, ,\cr
a_{n{-1}\, m}^{A_{n m}}={}&-{4(n+1)\over 2n+3} \, B_m {1-v\over u}\, 
A_{n m}  \, , \qquad \ \, a_{n{+1}\, m}^{A_{n m}}=-{4(n+2)\over 2n+3}\, 
B_m {1-v\over u}\, A_{n m}  \, , \cr
a_{n\, m{-1}}^{A_{n m}}={}&-{4 m\over 2m+1} \, B_{n+1} {1-v\over u}\, 
A_{n m}  \, , \qquad a_{n\, m{+1}}^{A_{n m}}= -{4(m+1)\over 2m+1} 
\, B_{n+1} {1-v\over u}\, A_{n m}  \, , \cr
a_{n\, m}^{A_{n m}}={}& 4 B_m B_{n+1} \, A_{n m}  \, , & \longr \cr}
$$
where
\eqn\defBm{
B_m = {1+v \over u} - {m^2+m-1\over (2m-1)(2m+3)} \, .
}
For $m=n,n-1,n-2,n-3$, and also if $n=0,1,2$, \poly\ may be used 
to combine terms to ensure that we only have $a_{n'm'}^{A_{n m}}$ for
$0\le m' \le n'$. For $m=n=0$ this prescription gives
\eqn\Azero{\eqalign{
a_{22} = {}& {4\over 15}\, A_{00} \, , \qquad a_{21} = - {4\over 5} \, 
{1-v\over u} \,  A_{00} \, , \qquad a_{20} = {4\over 15} \Big (
3\, {1+v\over u}- 1 \Big ) \,  A_{00} \, , \cr
a_{11} = {}& {4\over 15} \Big ( 10\, {(1-v)^2\over u^2}
-5\, {1+v\over u} + 1 \Big ) \,  A_{00} \, , \quad
a_{10} = - {4\over 3} \Big ( 2\, {1+v\over u}- 1 \Big ){1-v\over u}\,
A_{00} \, , \cr
a_{00} = {}& {4\over 15} \Big ( 15\, {(1+v)^2\over u^2} - 
5\, {(1-v)^2\over u^2} - 8\, {1+v\over u} + 1 \Big ) \,  A_{00} \, , \cr}
}
which is equivalent to the results in  \scft. Similarly for $n=1,m=0,1$
the resulting $a_{n'm'}^{A_{n m}}$ correspond to those in  \ADHS.

The solution of the superconformal identities given by \azero, \afour\
and \longr\ may now be naturally interpreted in terms of the operator
product expansion. If in \longr\ we consider a single conformal partial
wave for $A_{nm}$ by letting
\eqn\AZG{
A_{nm} \to \G_{\Delta+4}^{(\ell)} \, ,
}
then, if $\A^\Delta_{[q,p,q],\ell}$ denotes a long superconformal
multiplet  whose lowest state has spin $\ell$, scale dimension
$\Delta$ and which belongs to a $SU(4)_R$ representation with Dynkin
labels $[q,p,q]$,  we obtain
\eqn\delL{
a_{n'm'}^{A_{n m}} \to a_{n'm'}^{\vphantom g}
\big (\A^\Delta_{nm,\ell} \big ) \, , \qquad \A^\Delta_{nm,\ell}
\equiv \A^\Delta_{[n-m,2m,n-m],\ell} \, .
}
The non zero results obtained from \longr\ with
\AZG\ may be conveniently expressed in the form
\eqn\Ared{
a_{n+i\,m+j}^{\vphantom g}\big (\A^\Delta_{nm,\ell} \big ) 
= N_{n+1,i} N_{m,j} \, {A}{}^{nm}_{|i|\, |j|} \, , \qquad 
i,j = \pm 2, \pm 1, 0 \, ,
}
for
\eqn\AA{\eqalign{
N_{m,2} = {}& {(m+1)(m+2)\over (2m+1)(2m+3)} \, , \quad
N_{m,1} = {m+1\over 2m+1}\, , \quad  N_{m,0} = 1 \, , \cr
N_{m,-1} = {}& {m\over 2m+1}\, , \qquad\quad
N_{m,-2} = {(m-1)m\over (2m-1)(2m+1)} \, . \cr}
}
and using \recurG\ we have
\eqn\expA{\eqalign{
& \hskip 3cm a_{n+i\,m+j}^{\vphantom g}\big (\A^\Delta_{nm,\ell} \big ) = 
\sum_{(\Delta';\ell')} b_{(\Delta';\ell')} \, \G^{(\ell')}_{\Delta'} \, ,\cr
& |i|=|j|=2 \, , \  (\Delta';\ell') = (\Delta+4; \ell) \, , \cr
& |i|=2, \, |j|=1 , \ |i|=1, \, |j|=2 \, , \  
(\Delta';\ell') = (\Delta+5,\Delta+3; \ell \pm 1 )\, ,\cr 
& |i|=|j|=1 \, , \   (\Delta';\ell') =  
(\Delta+6,\Delta+4,\Delta+2; \ell \pm 2,\ell)\, , \cr
& |i|=1, \, j=0 , \ i=0, \, |j|=1 \, , \  
(\Delta';\ell') = (\Delta+7,\Delta+1;\ell \pm 1 ), \, 
(\Delta+5,\Delta+3; \ell\pm 3, \ell \pm 1 ) \, , \cr
& i=j=0 \, , \   (\Delta';\ell') =  (\Delta+8,\Delta;\ell), \,
(\Delta+6,\Delta+2; \ell \pm 2,\ell), \,
(\Delta+4; \ell\pm 4,\ell \pm 2,\ell) \, .  \cr}
}
In consequence $a_{n'm'}^{\vphantom g}(\A^\Delta_{nm,\ell})$ 
corresponds to the contribution in the operator product expansion applied 
to the correlation function for all expected
operators belonging to $\A^\Delta_{nm,\ell}$. In \expA\
$b_{(\Delta';\ell')}  > 0$ if $\Delta > \ell+1$. 
If $m\le n \le m+3$ the results are modified since we then obtain 
from \longr\ contributions with $m'>n'$. In this case, for $n'\ge m'$ and 
$a_{m'-1\,n'+1}^{\vphantom g} \big (\A^\Delta_{nm,\ell} \big )$ non 
zero, we should take
\eqn\anm{
a_{n'\,m'}^{\vphantom g} \big (\A^\Delta_{nm,\ell} \big ) - 
a_{m'-1\, n'+1}^{\vphantom g} \big (\A^\Delta_{nm,\ell} \big ) \to 
a_{n'\,m'}^{\vphantom g} \big (\A^\Delta_{nm,\ell} \big ) \, .
}
Furthermore any contribution with $m'=n'+1$ should be dropped. 
Using this result and \anm\ we may then easily show that
\eqn\Azer{
a_{n'\,m'}^{\vphantom g} \big (\A^\Delta_{n\, n+1,\ell} \big ) = 0 \, ,
}
and for later reference we also note the symmetry relation
\eqn\Asym{
a_{n'\,m'}^{\vphantom g} \big (\A^\Delta_{n\, m,\ell} \big ) =
a_{m'-1\,n'+1}^{\vphantom g} \big (\A^\Delta_{m-1\, n+1,\ell} \big ) \, .
}

The unitarity condition for a long multiplet $\A^\Delta_{nm,\ell}$ requires 
\eqn\unitl{
\Delta \ge 2n + \ell + 2 \, , 
}
and so, as in \unit, using \AZG\ we must have for $u\sim 0$,
\eqn\unit{
A_{nm}(u,v) \sim u^{n+3+\ep} \, , \qquad \ep \ge 0 \, .
}

We now consider the operator product expansion interpretation of the remaining 
terms in the in the solution of the superconformal identities given by
\Gsolt\ and \ff. The constant $k$, whose contribution is just given by \azero,
clearly corresponds to the identity operator,
\eqn\Gident{
a_{nm}(\I) = \de_{n0} \de_{m0} \, .
}

To analyse the contribution of the single variable functions $f_n$ in \expf\
we use the result \ttwo\ for the conformal partial wave for twist
two operators as well as
\eqn\tzero{
\G_{\ell}^{(\ell)}(u,v)\big |_{\Delta_1=\Delta_2=\Delta_3=\Delta_4}
= \half \, { \bz\, g_{\ell}(x) - z \, g_{\ell} (\zz) \over z-\bz}\, , 
}
with $g_\ell$ as in \gell, for twist zero. Taking 
\eqn\fg{
f_{n+1}(z) \to \half g_{\ell+2}(x) \, , \qquad n=0,1,2,\dots \, ,
}
in \FF\ and \expf\ then leads to results corresponding to only 
twist zero and twist two operators. These operators can be interpreted as 
belonging to a multiplet $\D_{n\,0,\ell}$, where in general we denote by 
$\D_{n\,m ,\ell}\equiv \D_{[n-m,2m,n-m],\ell}$ the semi-short supermultiplet
in which the lowest dimension operator has $\Delta = 2m + \ell$, or
twist $2m$, and belongs to the $[n-m,2m,n-m]$ $SU(4)_R$ representation.
These non unitary super multiplets are discussed in appendix D.
For $\D_{n\,m ,\ell}$ the conformal partial waves may be expressed in general
in the form
\eqn\decD{
a_{n+i \, m+j} \big (\D_{n\,m,\ell}\big ) = N_{n+1,i}N_{m,j} \,
D^{nm}_{|i|\, j } \, , \qquad D^{nm}_{|i|\, 2} = 0 \, .
}
Corresponding to \fg\ we then have 
\eqn\single
{\eqalign{
D^{n0}_{21} = {}& \quar \, \G_{\ell+3}^{(\ell+1)}\, , \qquad \qquad
D^{n0}_{20} = \quar \big (\G_{\ell+2}^{(\ell)} +
a_{\ell+2} \, \G_{\ell+4}^{(\ell+2)} \big ) \, , \cr
D^{n0}_{11} = {}&
\G_{\ell+2}^{(\ell+2)} + \quar\big ( \G_{\ell+2}^{(\ell)} +
a_{\ell+2} \, \G_{\ell+4}^{(\ell+2)} \big )\, , \cr
D^{n0}_{10} = {}&
\G_{\ell+1}^{(\ell+1)} + a_{\ell+2}\, \G_{\ell+3}^{(\ell+3)} +
\quar \big ( \G_{\ell+1}^{(\ell-1)} +  b_\ell \,\G_{\ell+3}^{(\ell+1)} +
a_{\ell+2} a_{\ell+3}\, \G_{\ell+5}^{(\ell+3)} \big )  \, , \cr
D^{n0}_{01} = {}&
\G_{\ell+1}^{(\ell+1)} + a_{\ell+2} \, \G_{\ell+3}^{(\ell+3)}
+ \quar b_n \, \G_{\ell+3}^{(\ell+1)} \, ,  \cr
D^{n0}_{00} = {}& \G_{\ell}^{(\ell)} + b_\ell \, \G_{\ell+2}^{(\ell+2)}
+ a_{\ell+2} a_{\ell+3}\, \G_{\ell+4}^{(\ell+4)}
+ \quar b_n \big ( \G_{\ell+2}^{(\ell)}
+ a_{\ell+2} \, \G_{\ell+4}^{(\ell+2)} \big ) \, , \cr}
}
whereas $a_\ell$ is as in \defst\ and
\eqn\bell{
b_\ell = a_{\ell+2}+a_{\ell+1}=
{2\ell^2 + 6 \ell +3 \over (2\ell+1)(2\ell+5)} \, .
}
A list of relevant representations for differing dimensions contained in
$\D_{n\,0 ,\ell}\equiv \D_{[n,0,n],\ell}$ is listed in appendix D, the
twist zero and twist two representations correspond with those necessary for
\single.  For $f_0$ these results are modified. From \afz\ only twist two
contributions are required since, taking now $f_0(z) \to 2 g_{\ell+3}(x)$,
\eqn\ttwob{\eqalign{
a_{00}\big (\C_{00,\ell}\big ) = {}&
\G_{\ell+2}^{(\ell)} + {2(\ell+2)(\ell+3)\over 3(2\ell+3)(2\ell+7)} \,
\G_{\ell+4}^{(\ell+2)} + a_{\ell+3}a_{\ell+4} \, 
\G_{\ell+6}^{(\ell+4)} \, , \cr
a_{10}\big (\C_{00,\ell}\big ) = {}& {2\over 3} \Big (
\G_{\ell+3}^{(\ell+1)} + a_{\ell+3}  \,
\G_{\ell+5}^{(\ell+3)} \Big ) \, ,
\quad a_{11}\big (\C_{00,\ell}\big ) =
{2\over 3}  \, \G_{\ell+4}^{(\ell+2)}  \, .\cr}
}
Here we denote by $\C_{nm,\ell}\equiv \C_{[n-m,2m,n-m],\ell}$ the semi-short
supermultiplet in which the lowest dimension operator has $\Delta = 2n+\ell+2$
and belongs to the $[n-m,2m,n-m]$ $SU(4)_R$ representation.

The multiplets $\D_{[q,p,q],\ell}$  fail to satisfy the unitarity condition
\unitl\ on $\Delta$ and so their contributions as in \single\ must be
cancelled in a unitary theory.
This may be achieved by a corresponding long multiplet contribution.
When  $\Delta =  2m + \ell$ or $\Delta =  2n + \ell +2$ the long multiplet
$\A^\Delta_{nm,\ell}$ can be decomposed into semi-short multiplets resulting in
\eqn\DD{
a_{n'm'}^{\vphantom g}\big ( \A^{2m+\ell}_{n\, m,\ell} \big )
= 16 \, a_{n'm'}^{\vphantom g} \big ( \D_{n\, m,\ell} \big ){}
+ {4(m+1)\over 2m+1} \, a_{n'm'}^{\vphantom g}
\big ( \D_{n\,m{+1},\ell-1} \big ) \, ,
}
and, at the unitarity threshold \unitl,
\eqn\CC{
a_{n'm'}^{\vphantom g}\big ( \A^{2n+\ell+2}_{n\, m,\ell} \big )
= 16 \, a_{n'm'}^{\vphantom g} \big ( \C_{n\, m,\ell} \big ){}
+ {4(n+2)\over 2n+3} \, a_{n'm'}^{\vphantom g}
\big ( \C_{n{+1}\,m,\ell-1} \big ) \, .
}
When $n=m$ we have the special case
\eqn\CD{
a_{n'm'}^{\vphantom g}\big ( \A^{2n+\ell}_{n\,n,\ell} \big )
= 16 \, a_{n'm'}^{\vphantom g} \big ( \D_{n\,n,\ell} \big ){}
+ {(n+1)(n+2)\over (2n+1)(2n+3)} \,
a_{n'm'}^{\vphantom g} \big ( \C_{n+1\,n+1,\ell-2} \big ) \, .
}
The results \DD, \CC\ and \CD\ reflect a decomposition of long multiplets
at particular values of $\Delta$ as described in appendix D.
{}From \DD\ we may obtain $a_{n'm'}^{\vphantom g}\big ( \D_{n\, m,\ell} \big)$
iteratively starting from \single. With the  notation in \decD\ the results
are 
\eqnn\resD
$$\eqalignno{
D^{nm}_{2\,-2} = {}& {\ts{1\over 16}}\, \G^{(\ell)}_{2m+\ell+4} \, ,
\qquad\qquad D^{nm}_{2\,1} = \quar\, \G^{(\ell+1)}_{2m+\ell+3} \, , \cr
D^{nm}_{2\,-1} = {}& {\ts{1\over 16}}\big ( \G^{(\ell-1)}_{2m+\ell+3} +
4\, \G^{(\ell+1)}_{2m+\ell+3} + 
a_{m+\ell+2}\, \G^{(\ell+1)}_{2m+\ell+5} \big )\,, \cr
D^{nm}_{2\,0} = {}& {\ts{1\over 16}}\big ( 4\, \G^{(\ell)}_{2m+\ell+2} +
a_{m}\, \G^{(\ell)}_{2m+\ell+4} + 4a_{m+\ell+2}\, 
\G^{(\ell+2)}_{2m+\ell+4} \big )\,, \cr
D^{nm}_{1\,-2} = {}& {\ts{1\over 16}}\big ( \G^{(\ell-1)}_{2m+\ell+3} + 
4 \, \G^{(\ell+1)}_{2m+\ell+3} + \quar a_{m+1}\, \G^{(\ell-1)}_{2m+\ell+5} 
+ a_{m+\ell+2} \,\G^{(\ell+1)}_{2m+\ell+5}\big )\,,\cr
D^{nm}_{1\, 1} = {}& {\ts{1\over 4}}\big ( 
\G^{(\ell)}_{2m+\ell+2} + 4 \, \G^{(\ell+2)}_{2m+\ell+2} + \quar a_{m}\,
\G^{(\ell)}_{2m+\ell+4} + a_{m+\ell+2} \,\G^{(\ell+2)}_{2m+\ell+4}\big )\,,\cr
D^{nm}_{1\,-1} = {}& {\ts{1\over 16}}\big (
\G^{(\ell-2)}_{2m+\ell+2} + \quar a_{m+1}\, \G^{(\ell-2)}_{2m+\ell+4} \cr
&\quad {} + 8 \, \G^{(\ell)}_{2m+\ell+2} + ( b_{m+\ell} + a_m ) 
\G^{(\ell)}_{2m+\ell+4} 
+ \quar a_{m+1}a_{m+\ell+2}\, \G^{(\ell)}_{2m+\ell+6} \cr 
&\quad {} + 16 \, \G^{(\ell+2)}_{2m+\ell+2} + 8 a_{m+\ell+2}\,
\G^{(\ell+2)}_{2m+\ell+4} + a_{m+\ell+2}a_{m+\ell+3} \,
\G^{(\ell+2)}_{2m+\ell+6}\big ) \,, \cr
D^{nm}_{1\, 0} = {}& {\ts{1\over 16}}\big (
4\,\G^{(\ell-1)}_{2m+\ell+1} + 2 a_{m}\, \G^{(\ell-1)}_{2m+\ell+3} +
\quar a_{m}a_{m+1}\G^{(\ell-1)}_{2m+\ell+5}\cr
&\quad {} + 16\, \G^{(\ell+1)}_{2m+\ell+1} 
+ 4 ( b_{m+\ell} + a_m ) \G^{(\ell+1)}_{2m+\ell+3} 
+ 2 a_{m}a_{m+\ell+2}\, \G^{(\ell+1)}_{2m+\ell+5} \cr
&\quad {} + 16a_{m+\ell+2} \, \G^{(\ell+3)}_{2m+\ell+3} + 
4a_{m+\ell+2}a_{m+\ell+3} \, \G^{(\ell+3)}_{2m+\ell+5}\big ) \,, \cr
D^{nm}_{0\,-2} = {}& {\ts{1\over 16}}\big ( 4\, \G^{(\ell)}_{2m+\ell+2} +
\quar a_{m+1} \, \G^{(\ell-2)}_{2m+\ell+4} + b_n\, \G^{(\ell)}_{2m+\ell+4} \cr
&\quad {} + 4 a_{m+\ell+2}\, \G^{(\ell+2)}_{2m+\ell+4}
+ \quar a_{m+1} a_{m+\ell+2} \,\G^{(\ell)}_{2m+\ell+6}\big )\,,\cr
D^{nm}_{0\,1} = {}& {\ts{1\over 4}}\big (4\, \G^{(\ell+1)}_{2m+\ell+1} +
\quar a_m \, \G^{(\ell-1)}_{2m+\ell+3} + b_n \, \G^{(\ell+1)}_{2m+\ell+3} \cr
&\quad {} + 4 a_{m+\ell+2}\, \G^{(\ell+3)}_{2m+\ell+3} 
+ \quar a_m  a_{m+\ell+2} \,\G^{(\ell+1)}_{2m+\ell+5}\big )\, ,\cr
D^{nm}_{0\, -1} = {}& {\ts{1\over 16}}\big ( \quar a_{m+1}\, 
\G^{(\ell-3)}_{2m+\ell+3} + 4\,\G^{(\ell-1)}_{2m+\ell+1} + 
(a_{m} + b_n ) \, \G^{(\ell-1)}_{2m+\ell+3} 
+ \quar a_{m+1}b_{m+\ell}\, \G^{(\ell-1)}_{2m+\ell+5} \cr
&\quad {} + 16\, \G^{(\ell+1)}_{2m+\ell+1}
+ 4 ( b_{m+\ell} + b_n ) \G^{(\ell+1)}_{2m+\ell+3}
+ a_{m+\ell+2}(a_{m}+b_n) \, \G^{(\ell+1)}_{2m+\ell+5} \cr
&\quad {} + \quar a_{m+1} a_{m+\ell+2} a_{m+\ell+3}\, \G^{(\ell+1)}_{2m+\ell+7}\cr
&\quad {} + 16 a_{m+\ell+2}\, \G^{(\ell+3)}_{2m+\ell+3} +
4 a_{m+\ell+2}a_{m+\ell+3}\, \G^{(\ell+3)}_{2m+\ell+5} \big )  \, ,  \cr
D^{nm}_{0\, 0} = {}& {\ts{1\over 16}}\big (  a_{m}\,
\G^{(\ell-2)}_{2m+\ell+2} + \quar a_m a_{m+1} \,\G^{(\ell-2)}_{2m+\ell+4} +
16\, \G^{(\ell)}_{2m+\ell} + 4(a_{m} + b_n ) \, \G^{(\ell)}_{2m+\ell+2} \cr
&\quad {} +  a_{m}(b_{m+\ell}+b_n)\, \G^{(\ell)}_{2m+\ell+4}
+ \quar  a_{m} a_{m+1} a_{m+\ell+2}\, \G^{(\ell)}_{2m+\ell+6} \cr
&\quad {} + 16 b_{m+\ell} \, \G^{(\ell+2)}_{2m+\ell+2}
+ 4 a_{m+\ell+2}(a_{m}+b_n) \, \G^{(\ell+2)}_{2m+\ell+4} 
+  a_{m} a_{m+\ell+2} a_{m+\ell+3}\, \G^{(\ell+2)}_{2m+\ell+6}\cr
&\quad {} +
16 a_{m+\ell+2}a_{m+\ell+3}\, \G^{(\ell+4)}_{2m+\ell+4} \big )  \, . & \resD \cr
}
$$

The corresponding results for the semi-short multiplet $\C_{n\, m,\ell}$
may be obtained from those for $\D_{n\, m,\ell}$ given above by taking
\eqn\relDC{
a_{n'\, m'}^{\vphantom g} \big ( \C_{n\, m,\ell} \big ) =
a_{m'-1\, n'+1}^{\vphantom g} \big ( \D_{m-1\, n+1,\ell} \big ) \, .
}
Using \Asym\ then \CC\ easily follows from \DD. We may also verify
that \CD\ is satisfied.  Combining \DD\ for $m=n+1$ with \Azer\ we 
may then obtain
\eqn\CDe{\eqalign{
a_{n'\,m'}^{\vphantom g}\big (& \A^{2n+\ell}_{n\,n,\ell} \big ) 
- a_{m'-1\,n'+1}^{\vphantom g}\big ( \A^{2n+\ell}_{n\,n,\ell} \big )
=  16 \big ( a_{n'm'}^{\vphantom g} \big ( \D_{n\,n,\ell} \big ) 
- a_{m'-1\, n'+1 }^{\vphantom g} \big ( \D_{n\,n,\ell} \big ) \big ) \cr
&{} + {(n+1)(n+2)\over (2n+1)(2n+3)}\Big ( -
a_{m'-1\, n'+1}^{\vphantom g} \big ( \C_{n+1\,n+1,\ell-2} \big ) +
a_{n'm'}^{\vphantom g} \big ( \C_{n+1\,n+1,\ell-2} \big )\Big )  \, , \cr}
}
which for $n'\ge m'$, and noting the requirement \anm, gives exactly \CD.

In general the results from \relDC\ can be expressed as
\eqn\decC{
a_{n+i \, m+j} \big (\C_{n\,m,\ell}\big ) = N_{n+1,i}N_{m,j} \,
C^{nm}_{i\, j } \, , \qquad C^{nm}_{2\, j} = 0 \, .
}
For general $n,m$ the necessary operators are just those given in table 4
of \short. For $m=n$ the relation \relDC\ combined with \resD\ in this
case and applying the corresponding results to \anm\ gives
\eqnn\resC
$$\eqalignno{
C^{nn}_{1\,1} = {}& \G^{(\ell+2)}_{2n+\ell+4} \, , \cr
C^{nn}_{1\,0} = {}& \G^{(\ell+1)}_{2n+\ell+3} + \quar a_n \, 
\G^{(\ell+1)}_{2n+\ell+5} + a_{n+\ell+3}\, \G^{(\ell+3)}_{2n+\ell+5} \, , \cr
C^{nn}_{0\,0} = {}& \G^{(\ell)}_{2n+\ell+2} + \quar a_n \, 
\G^{(\ell)}_{2n+\ell+4} + (b_{n+\ell+1}-a_{n+1}) \, 
\G^{(\ell+2)}_{2n+\ell+4} \, , \cr
&{} +{\ts {1\over 16}} a_n a_{n+1}\, \G^{(\ell)}_{2n+\ell+6} 
+ \quar a_n a_{n+\ell+3}\, \G^{(\ell+2)}_{2n+\ell+6} 
+ a_{n+\ell+3}a_{n+\ell+4} \, \G^{(\ell+4)}_{2n+\ell+6} \, , \cr
C^{nn}_{1\,-1} = {}&\quar \, \G^{(\ell)}_{2n+\ell+4} + \G^{(\ell+2)}_{2n+\ell+4} 
+ \quar a_{n+\ell+3}\, \G^{(\ell+2)}_{2n+\ell+6} \, , \cr
C^{nn}_{0\,-1} = {}& \quar \, \G^{(\ell-1)}_{2n+\ell+3} +  
\G^{(\ell+1)}_{2n+\ell+3} + {\ts{1\over 16}} a_{n+1}\, 
\G^{(\ell-1)}_{2n+\ell+5} +  \quar b_{n+\ell+1}\, \G^{(\ell+1)}_{2n+\ell+5} \, , \cr
&{} +a_{n+\ell+3}\, \G^{(\ell+3)}_{2n+\ell+5} 
+ {\ts {1\over 16}} a_{n+1}a_{n+\ell+3} \, \G^{(\ell+1)}_{2n+\ell+7} 
+ \quar a_{n+\ell+3}a_{n+\ell+4} \, \G^{(\ell+3)}_{2n+\ell+7} \, , \cr
C^{nn}_{-1\,-1} = {}& {\ts {1\over 16}} \,\G^{(\ell-2)}_{2n+\ell+4} + \quar  \, 
\G^{(\ell)}_{2n+\ell+4} + \G^{(\ell+2)}_{2n+\ell+4} 
+ {\ts {1\over 16}} (b_{n+\ell+1}-a_{n+2}) \, \G^{(\ell)}_{2n+\ell+6} \cr
&{} + \quar a_{n+\ell+3} \, \G^{(\ell+2)}_{2n+\ell+6} 
+ {\ts {1\over 16}}a_{n+\ell+3}a_{n+\ell+4} \, \G^{(\ell+2)}_{2n+\ell+8} \, , \cr
C^{nn}_{1\,-2} = {}& \quar \, \G^{(\ell+1)}_{2n+\ell+5} \, , \cr
C^{nn}_{0\,-2} = {}& \quar \, \G^{(\ell)}_{2n+\ell+4} + 
{\ts{1\over 16}} a_{n+1}\, \G^{(\ell)}_{2n+\ell+6} 
+ \quar a_{n+\ell+3}\, \G^{(\ell+2)}_{2n+\ell+6} \, , \cr
C^{nn}_{-1\,-2} = {}& {\ts{1\over 16}}\,\G^{(\ell-1)}_{2n+\ell+5} + \quar \, 
\G^{(\ell+1)}_{2n+\ell+5} + {\ts {1\over 16}} a_{n+\ell+3}\, 
\G^{(\ell+1)}_{2n+\ell+7} \, , \cr
C^{nn}_{-2\,-2} = {}& {\ts {1\over 16}} \, \G^{(\ell)}_{2n+\ell+6} \, . &\resC \cr}
$$
The necessary operators correspond exactly to those listed in \short\ (see table
3) as present in the semi-short supermultiplet for this case.
For $n=0$ \resC\ reproduces \ttwo. We may also note that, since for $m\ge 1$,
$\quar < a_m \le {1\over 3}$ and $b_n > \half$, all coefficients
in \resC\ are positive as required by unitarity.

As in the $\N=2$ case the semi-short results also include the contributions
for short BPS multiplets when extended to negative $\ell$. Formally as shown
in \short\ $\C_{[q,p,q],-1} \simeq \B_{[q+1,p,q+1]}$ where $\B_{[q,p,q]}$
denotes the BPS supermultiplet whose lowest state has spin zero, $\Delta
= 2q+p$, and belongs to the $SU(4)_R$ $[q,p,q]$ representation. For $q>0$
the lowest state is annihilated by $\quar$ of the $Q$ and also $\bar Q$
supercharges whereas when $q=0$ we have a $\half$-BPS multiplet with $\half$
the $Q$ and $\bar Q$ supercharges annihilating the lowest state. As
earlier we identify, for $n\ge m$, $\B_{n\,m} \equiv \B_{[n-m,2m,n-m]}$ and 
we then have
\eqn\CBfour{
a_{n'm'}^{\vphantom g}\big ( \C_{n\, m,-1} \big ) = {n+1\over 2n+1} \,
a_{n'm'}^{\vphantom g}\big ( \B_{n+1\, m} \big ) \, ,
}
where
\eqn\decB{
a_{n+i \, m+j} \big (\B_{n\,m}\big ) = N_{n+1,i}N_{m,j} \,
B^{nm}_{i\, j } \, , \qquad B^{nm}_{2\, j} = B^{nm}_{1\, j} = 0 \, .
}
For general $n,m$ we have
\eqn\BB{
B^{nm}_{i\, j } = B^{nm}_{i\, |j| } \, ,
}
and
\eqn\resB{\eqalign{
B^{nm}_{0\,2} = {}& \quar \, \G^{(0)}_{2n+2} \, , \cr
B^{nm}_{-2\,2} = {}& {\ts {1\over 16}} \, \G^{(0)}_{2n+4}  \, , \cr
B^{nm}_{-1\,2} = {}& \quar \, \G^{(1)}_{2n+3} \, , \cr
B^{nm}_{0\,1} = {}&  \G^{(1)}_{2n+1} + \quar a_{n+1}\, \G^{(1)}_{2n+3} \, , \cr
B^{nn}_{-2\,1} = {}& \quar \, \G^{(1)}_{2n+3} + {\ts{1\over 16}} a_{n+2}\,
\G^{(1)}_{2n+5} \, , \cr
B^{nm}_{-1\,1} = {}& \quar  \,\G^{(0)}_{2n+2} + \G^{(2)}_{2n+2}+{\ts{1\over 16}}
a_{n} \,\G^{(0)}_{2n+4} + \quar a_{n+2} \, \G^{(2)}_{2n+4}\, , \cr
B^{nn}_{0\, 0} = {}& \G^{(0)}_{2n} + \quar(b_{m-1}-a_n ) \, \G^{(0)}_{2n+2}
+  a_{n+1}\, \G^{(2)}_{2n+2} + {\ts{1\over 16}} a_n  a_{n+1}\, 
\G^{(0)}_{2n+4} \, , \cr
B^{nm}_{-2\,0} = {}&\quar  \,\G^{(0)}_{2n+2} +  {\ts{1\over 16}}
(b_{m-1}-a_{n+1} )  \,\G^{(0)}_{2n+4} + \quar a_{n+2}\, \G^{(2)}_{2n+4} +
{\ts {1\over 64}} a_{n+1} a_{n+2}\, \G^{(0)}_{2n+6}  \, , \cr
B^{nm}_{-1\,0} = {}& \G^{(1)}_{2n+1} + \quar b_{m-1}  \,\G^{(1)}_{2n+3}
+ a_{n+2} \,  \G^{(3)}_{2n+3} + {\ts {1\over 16}} a_{n} a_{n+2} \,
\G^{(1)}_{2n+5} \, . \cr}
}
Again all coefficients are positive and the necessary operators are
exactly as expected for this supermultiplet (see table 2 in \short). For
$n=m+1$ the multiplet is truncated with, in \decB, the following non zero,
\eqn\resBm{\eqalign{
B^{m+1\,m}_{0\,1} = {}&  \G^{(1)}_{2m+3} \, , \cr
B^{m+1\,m}_{0\,0} = {}& \G^{(0)}_{2m+2} + \quar a_m \, \G^{(0)}_{2m+4}
+  a_{m+2}\, \G^{(2)}_{2m+4}  \, , \cr
B^{m+1\,m}_{-1\,0} = {}& \G^{(1)}_{2m+3} + \quar a_m \, \G^{(1)}_{2m+5}
+ a_{m+3} \,  \G^{(3)}_{2m+5} \, , \cr
B^{m+1\,m}_{0\,-1} = {}&  \G^{(1)}_{2m+3} + 
\quar a_{m+2}\, \G^{(1)}_{2m+5} \, , \cr
B^{m+1\,m}_{-1\,-1} = {}& \quar  \,\G^{(0)}_{2m+4} + 
\G^{(2)}_{2m+4}+{\ts{1\over 16}} a_{m+1} \,\G^{(0)}_{2m+6} + 
\quar a_{m+3} \, \G^{(2)}_{2m+4}\, , \cr
B^{m+1\,m}_{-2\,-1} = {}& \quar \, \G^{(1)}_{2m+5} + {\ts{1\over 16}} a_{m+3}\,
\G^{(1)}_{2m+7} \, , \cr
B^{m+1\, m}_{0\, -2} = {}& \quar  \, \G^{(0)}_{2m+4} \, , \qquad
B^{m+1\, m}_{-1\, -2} =  \quar  \, \G^{(1)}_{2m+5} \, , \qquad
B^{m+1\, m}_{-2\, -2} =  {\ts{1\over 16}}  \, \G^{(0)}_{2m+6} \, . \cr}
}
The necessary operators correlate again with those expected for this
$\quar$-BPS multiplet (see table 5 in \short).

If we consider the semi-short multiplet for $\ell=-2$ we get
\eqn\CBfs{
a_{n'm'}^{\vphantom g}\big ( \C_{n\, m,-2} \big ) = -4 \,
a_{n'm'}^{\vphantom g}\big ( \B_{n \, m} \big ) \, ,
}
which allows results for $a_{n'm'}^{\vphantom g}\big ( \B_{n \, m} \big )$
to be derived for $m=n$ in addition to $m<n$ as given by \CBfour.
However in this case there is a further decomposition into 
contributions corresponding to $\half$-BPS multiplets.
Such $\half$-BPS contributions are obtained in \decB\  by letting 
$\B_{nm} \to {\hat \B}_{nn}$ and $B^{nm}_{ij} \to {\hat B}{}^{nn}_{ij}$ where
\eqn\Bhalf{\eqalign{
{\hat B}{}^{nn}_{0\,0} = {}& \G^{(0)}_{2n} \, , \qquad\quad\quad \,
{\hat B}{}^{nn}_{0\,-1} = \G^{(1)}_{2n+1}\, , \qquad \quad 
{\hat B}{}^{nn}_{-1\,-1} = \G^{(2)}_{2n+2} \, , \cr
{\hat B}{}^{nn}_{0\,-2} = {}& \quar \, \G^{(0)}_{2n+2} \, , \qquad 
{\hat B}{}^{nn}_{-1\,-2} =  \quar \, \G^{(1)}_{2n+3} \, , \qquad
{\hat B}{}^{nn}_{-2\,-2} =  {\ts{1\over 16}} \, \G^{(0)}_{2n+4} \, , \cr}
}
(the relevant operators here correspond to table 1 in \short). With the
result given in \Bhalf\ we can then write in \CBfs
\eqn\CBh{
a_{n'm'}^{\vphantom g}\big ( \B_{n\, n} \big ) = 
a_{n'm'}^{\vphantom g}\big ( {\hat \B}_{n\, n} \big )
- {(n+1)(n+2)\over 4(2n+1)(2n+3)} \,
a_{n'm'}^{\vphantom g}\big ( {\hat \B}_{n+1\, n+1} \big ) \, .
}
{}From \Gzero\ and \Gident\ it is also easy to see that
\eqn\Id{
a_{nm}^{\vphantom g}\big ( {\hat \B}_{0\, 0} \big ) =
a_{nm}^{\vphantom g}\big ( \I \big ) \, .
}
Any $\half$-BPS contribution $a_{n'm'}
\big ({\hat \B}_{n\, n} \big )$ may then be isolated by considering 
appropriate linear combinations of 
$a_{n'm'}\big ( \C_{n\, n,-2} \big )$ together with
$a_{n'm'}\big ( \I \big )$.

We also consider the extremal and next-to-extremal cases. When $E=0$
$\G$ is independent of $y,\yz$ and so must also be the function $f$ in
\solid. From \real\ and \Gextr\ we then get the solution
\eqn\extp{
\G(u,v) = u^{{1\over 2}p_+} \, k  \, ,
}
where we define
\eqn\ppp{
p_\pm = p_2 \pm p_1 \, .
}
Noting that
\eqn\extPG{
P^{(p_1,p_2)}_{00}(y,\yz) = \half (p_+ +2) \, , \qquad
\G^{(0)}_{p_+}(u,v;p_-,p_+) = u^{{1\over 2}p_+} \, ,
}
it is clear that the only operator which is necessary in the operator product 
expansion  has $\Delta=p_+$ and is spinless belonging to the $[0,p_+,0]$
representation. This is of course may be identified with the contribution of
just the $\half$-BPS operator belonging to the short $\B_{[0,p_+,0]}$
supermultiplet so that for the extremal case, up to a constant factor,
\eqn\exts{
a_{nm}\big({\B}_{[0,p_+,0]} \big ) = \de_{n0} \de_{m0}\
\G^{(0)}_{p_+}\, . }
The correlation function in this case has the very simple form
\eqn\Fourext{\eqalign{
\l \vphi^{(p_1)}(x_1,t_1)&  \, \vphi^{(p_2)}(x_2,t_2)\,
\vphi^{(p_3)}(x_3,t_3)\, \vphi^{(p_4)}(x_4,t_4 )\r \big |_{p_4=p_1+p_2+p_3}\cr
&{} = {\big ( t_1 {\cdot t}_4 \big )^{p_1} \big ( t_2 {\cdot t}_4 \big )^{p_2} 
\big ( t_1 {\cdot t}_3 \big )^{p_3}  \over
r_{14}^{\, \, p_1}\ r_{24}^{\, \, p_2}\  r_{34}^{\, \, p_3 } }\, k \, . \cr}
}

For the next-to-extremal case, $E=1$, we have a similar solution to that
given by \Gsol\ and \ff, but with no arbitrary $\K$ term and $\hf$ a single
variable function of $z$,
\eqn\next{\eqalign{
\hG(u,v; y, \yz) = {}& u^{{1\over 2}p_+ -1} \Big ( k - {1\over z - \bz}
\big ( (y-z)(\yz - z) \hf(z) - (y- \bz)(\yz - \bz) \hf(\bz) \big ) \Big ) \cr
= {}& \sum_{0\le m \le n \le 1} \! \! a_{nm}(u,v) \, P^{(p_1-1,p_2-1)}_{nm}
(y,\yz) \, , \cr}
}
where we have expanded in terms of the different possible $SU(4)_R$ 
representations. From this we obtain
\eqn\deco{\eqalign{
{\ts {1\over 16}}p_+ (p_+ +1)& (p_+ +2)\, a_{11} = \ha_{11} = F_0 \, , \cr
{\ts {1\over 8}}(p_+ +1)& (p_+ +2)\, a_{10} = \ha_{10} = 
F_1 + {p_- \over p_+} \, F_0 \, ,\cr
\half p_+ \, a_{00} = \ha_{00} = {}& k \, u^{{1\over 2}p_+-1} + 
F_2 + {2p_-\over p_+ +2} \, 
F_1 + { p_-{\!\!}^2 - (p_+ +2)  \over (p_+ +1)(p_+ +2)} \, F_0 \, , \cr}
}
for
\eqn\Fn{
F_n(z,\bz) = - (-1)^n \, u^{{1\over 2}p_+ -1}\, { z^n \hf(z) - \bz^n \hf(\bz)
\over z - \bz } \, .
}

Keeping only the term in \deco\ involving $k$ we may easily from \extPG\ see 
that this represents the contribution of just the $\half$-BPS chiral
primary operator belonging to the $\B_{[0,p_+-2,0]}$ supermultiplet
so that in the next-to-extremal case we have
\eqn\azz{
\ha_{nm}\big ({\B}_{[0,p_+-2,0]} \big ) = 
\de_{n0} \de_{m0} \, \G^{(0)}_{p_+-2}\, .
}
If in \deco\ and \Fn\ we let $\hf(z) \to 2 g_{\ell+3}(x;p_1,p_2)$ and use the
definitions in \text\ we obtain the contributions for the semi-short
supermultiplet $\C_{[0,p_+-2,0],\ell}$,
\eqnn\semsh
$$\eqalignno{ 
\ha_{11}\big ( \C_{[0,p_+-2,0],\ell}\big ) = {}&
\G^{(\ell+2)}_{p_++\ell+2} \, , \cr
\ha_{10}\big (\C_{[0,p_+-2,0],\ell}\big ) = {}&
\G^{(\ell+1)}_{p_+ +\ell+1} + b_{\ell+2} \,\G^{(\ell+3)}_{p_++\ell+3} 
+ {4(\ell+2) p_- (p_+ + \ell+1) \over p_+ ( p_+ +2 \ell+ 2) ( p_+ +2 \ell+ 4)}\,
\G^{(\ell+2)}_{p_+ +\ell+2}  \, ,\cr
\ha_{00}\big (\C_{[0,p_+-2,0],\ell}\big ) 
= {}& 
\G^{(\ell)}_{p_+ +\ell}  + b_{\ell+2} b_{\ell+3} \,\G^{(\ell+4)}_{p_+ +\ell+4}
+ c_{\ell+2} \, \G^{(\ell+2)}_{p_++\ell+2} \cr
&{}+ {8(\ell+1)p_-(p_++\ell+1)\over (p_++2)(p_++2\ell)(p_++2\ell+4)}\,
\G^{(\ell+1)}_{p_+ +\ell+1} \cr
&{}+ {8(\ell+2)p_-(p_++\ell+2)\,b_{\ell+2}\over 
(p_++2)(p_++2\ell+2)(p_++2\ell+6)}\, \G^{(\ell+3)}_{p_+ +\ell+3} \, , & \semsh \cr}
$$
for
\eqn\bell{\eqalign{
b_\ell = {}& {4\,(\ell+1)(p_1+\ell)(p_2+\ell)(p_++\ell-1) \over
(p_+ + 2\ell -1)(p_+ + 2\ell)^2 (p_+ + 2\ell+1)} \, , \cr
c_\ell ={}& {2\, \ell(p_+ \! +\ell-1)\over(p_+ \! +1)(p_+ \! +2\ell-3)
(p_+ \! +2\ell+1)} \bigg (
p_+ \! -1 +{p_-{\!\!}^2 \big ( 8(\ell-1)(p_+ \! +\ell) - p_+(p_+ \! -1)\big ) \over
(p_+ \! +2)(p_+ \! +2\ell-2)(p_+ \! +2\ell) } \bigg ) \, . \cr}
}
The necessary operators required for \semsh\ correspond exactly with those in this 
semi-short supermultiplet (see table 3 in \short).

Just as previously we may extend these  \semsh\ to $\ell=-1,-2$ to obtain
results for short multiplets. Thus
\eqn\CBB{\eqalign{
\ha_{nm}\big ( \C_{[0,p_+-2,0],-1}\big ) = {}&
\ha_{nm}\big ( \B_{[1,p_+-2,1]}\big ) \, , \cr
\ha_{nm}\big ( \C_{[0,p_+-2,0],-2}\big ) = {}&
\ha_{nm}\big ( \B_{[0,p_+,0]}\big ) - 4 \, 
\ha_{nm}\big ( \B_{[0,p_+-2,0]}\big ) \, , \cr}
}
where, together with \azz,
\eqn\Bext{\eqalign{
\ha_{11}\big ( \B_{[1,p_+-2,1]}\big ) ={}& \G^{(1)}_{p_++1} \, , \cr
\ha_{10}\big ( \B_{[1,p_+-2,1]}\big ) ={}& \G^{(0)}_{p_+} 
+{4p_- \over p_+(p_++2)} \, \G^{(1)}_{p_++1} + b_1 \, \G^{(2)}_{p_++2} \, , \cr
\ha_{00}\big ( \B_{[1,p_+-2,1]}\big ) ={}& b_1 \bigg ( \G^{(1)}_{p_++1} 
+{8p_- (p_+ + 2) \over p_+(p_++2)(p_++4) } \, \G^{(2)}_{p_++2} + b_2 \,
\G^{(3)}_{p_++3} \bigg ) \, , \cr}
}
and
\eqn\Bex{
\ha_{11}\big ( \B_{[0,p_+,0]}\big ) = \G^{(0)}_{p_+} \, , \quad
\ha_{10}\big ( \B_{[0,p_+,0]}\big ) = b_0 \, \G^{(1)}_{p_++1} \, , \quad
\ha_{00}\big ( \B_{[0,p_+,0]}\big ) = b_0b_1 \, \G^{(2)}_{p_++2} \, .
}
The necessary operators here correspond to table 5 and table 1 in \short.

The results obtained above show that the operator product expansion for
$\half$-BPS operators can be decomposed into short, semi-short and long
supermultiplets. For $p_-=p_2-p_1 \ge 0$,
\eqn\decompB{\eqalign{
\B_{[0,p_1,0]} \otimes \B_{[0,p_2,0]} \simeq {}&
\bigoplus_{0 \le m \le n \le p_1} \!\! \B_{[n-m,p_-+ 2m,n-m]} \cr
&{} \oplus \bigoplus_{\ell\ge 0} 
\bigoplus_{0\le m \le n \le p_1-1} \!\! \C_{[n-m,p_-+ 2m,n-m],\ell} \cr
&{} \oplus  \bigoplus_{\ell\ge 0}  \bigoplus_{0 \le m \le n \le p_1-2} \!\!
\A^\Delta_{[n-m,p_-+ 2m,n-m],\ell}  \, , \cr}
}
in accordance with the results of Eden and Sokatchev \bpsN. In \decompB\
we identify $\B_{[0,0,0]} \simeq \I$, corresponding to the unit operator
in the operator product expansion. It immediately follows from \decompB\
that long supermultiplets, with non zero anomalous dimensions, cannot
contribute to extremal and next-to-extremal correlation functions.

\newsec{Crossing Symmetry}

The operator product expansion provides the strongest constraints when 
combined with crossing symmetry. For a correlation function for four identical
chiral primary operators the correlation function is invariant under
permutations of all $x_i, t_i$ for all $i=1,2,3,4$. Permutations 
of the form $(ij)(kl)$ act trivially so we may restrict to permutations 
leaving $x_4, t_4$ invariant so that crossing symmetry transformations  
correspond to the permutation group $\S_3$, which is of order 6. 
The action of each permutation on the essential conformal 
invariants $u,v$ or $x,\zz$ or $y, \bz$ and also on the $R$-symmetry
invariants $\lam,\mun$ or $\alpha,\bet$ or $y,\yz$ is given in table 1,
where the transformations of $\zz$ are identical to those of $x$, and
similarly for $\bz,\bet,\yz$.

\vskip 6pt
\vbox{\tabskip=0pt \offinterlineskip
\hrule
\halign{&\vrule# &\strut \ \hfil#\  \cr
height2pt&\omit&&\omit&&\omit&&\omit&&\omit&&\omit&&\omit&
&\omit&&\omit&&\omit&&\omit&&\omit&&\omit&\cr
& $~e$ && $~(12)$ && $~(13)$ && $~(23)$ && $(123)$ && $(132)$ &&&
& $~e$ && $~(12)$ && $~(13)$ && $~(23)$ && $(123)$ && $(132)$ &\cr
height2pt&\omit&&\omit&&\omit&&\omit&&\omit&&\omit&&\omit& 
&\omit&&\omit&&\omit&&\omit&&\omit&&\omit&\cr
\noalign{\hrule}
height4pt&\omit&&\omit&&\omit&&\omit&&\omit&&\omit&&\omit&
&\omit&&\omit&&\omit&&\omit&&\omit&&\omit&\cr
&\hfil$u$\hfil&
&$\hfil{u\over v}\hfil$ &&\hfil$v$\hfil &&\hfil${1\over u}\hfil$ &
&\hfil${v\over u}$\hfil  &&\hfil${1\over v}$\hfil&&&
&\hfil $\lam$\hfil &&\hfil$\mun$\hfil &&\hfil${\lam \over \mun}$\hfil &
&\hfil${1\over \lam}$\hfil &&\hfil${1\over \mun}$\hfil  &&
\hfil${\mun\over \lam}$\hfil  &\cr
height4pt&\omit&&\omit&&\omit&&\omit&&\omit&&\omit&&\omit&&\omit&&\omit&
&\omit&&\omit&&\omit&&\omit&\cr
\noalign{\hrule}
height4pt&\omit&&\omit&&\omit&&\omit&&\omit&&\omit&&\omit&&\omit&&\omit&
&\omit&&\omit&&\omit&&\omit&\cr
&\hfil $v$\hfil &&\hfil${1\over v}$\hfil &&\hfil$u$\hfil &&\hfil${v\over u}$\hfil&
&\hfil${1\over u}$\hfil &&\hfil${u\over v}$\hfil &&&
&\hfil $\mun$\hfil &&\hfil$\lam$\hfil &&\hfil${1\over \mun}$\hfil &
&\hfil${\mun\over \lam}$\hfil &&
\hfil${\lam\over \mun}$\hfil  &&\hfil${1\over \lam}$\hfil  &\cr
height4pt&\omit&&\omit&&\omit&&\omit&&\omit&&\omit&&\omit&
&\omit&&\omit&&\omit&&\omit&&\omit&&\omit&\cr
\noalign{\hrule}
height4pt&\omit&&\omit&&\omit&&\omit&&\omit&&\omit&&\omit&&\omit&&\omit&
&\omit&&\omit&&\omit&&\omit&\cr
&\hfil $x$\hfil &&\hfil${x\over x-1}$\hfil&&\hfil$1-x$\hfil&&\hfil${1\over x}$\hfil&
&\hfil${x-1\over x}$\hfil &&\hfil${1\over 1-x}$\hfil &&&
&\hfil $\alpha$\hfil&&\hfil$1-\alpha$\hfil &&\hfil${\alpha\over \alpha-1}$\hfil &
&\hfil${1\over \alpha}$\hfil &&
\hfil${1\over 1-\alpha}$\hfil  &&\hfil${\alpha-1\over \alpha}$\hfil  &\cr
height4pt&\omit&&\omit&&\omit&&\omit&&\omit&&\omit&&\omit&&\omit&&\omit&&\omit&
&\omit&&\omit&&\omit&\cr
\noalign{\hrule}
height4pt&\omit&&\omit&&\omit&&\omit&&\omit&&\omit&&\omit&&\omit&&\omit&
&\omit&&\omit&&\omit&&\omit&\cr
&\hfil $z$\hfil &&\hfil$-z$\hfil &&\hfil${z+3\over z-1}$\hfil&
&\hfil${3-z\over 1+z}$\hfil &
&\hfil${3+z\over 1-z}$\hfil &&\hfil${z-3\over z+1}$\hfil &&&
&\hfil $y$\hfil &&\hfil$-y$\hfil &&\hfil${y+3\over y-1}$\hfil &
&\hfil${3-y\over 1+y}$\hfil &&
\hfil${y+3\over y-1}$\hfil  &&\hfil${y-3\over y+1}$\hfil  &\cr
height4pt&\omit&&\omit&&\omit&&\omit&&\omit&&\omit&&\omit&&\omit&&\omit&
&\omit&&\omit&&\omit&&\omit&\cr}
\hrule}
Table 1. Symmetry transformations of variables under crossing.

For the $\N=4$ case with $p_i=p$ the crossing symmetry conditions on the
correlation function $\G(u,v;\lam,\mun)$ are generated by considering just
$(12)$ and $(13)$ which give
\eqn\cross{
\G(u,v;\lam,\mun) =  \G(u/v,1/v;\mun,\lam ) = 
\Big ( {u\over v}\, \mun \Big )^p \G(v,u;\lam/\mun,1/\mun) \, .
}
The general construction of such invariant correlation functions follows
by determining polynomials in $\lam,\mun$ which transform according
to the irreducible representations of $\S_3$. We first consider symmetric
polynomials satisfying
\eqn\polylm{
S_p(\lam,\mun) = S_p(\mun,\lam) = \mun^p S_p(\lam/\mun,1/\mun) \, .
}
As described by Heslop and Howe \Howe, for any given $p$, $\S_3$ acts on
the $\half(p+1)(p+2)$ monomials $\lam^r\mun^s, \, r+s\le p$, giving chains of 
length $6$ or $3$ or $1$ which may be added to give minimal polynomial 
solutions of \polylm. If the chain contains a monomial $(\lam\mun)^r$, for 
$0\le r \le [\half p]$, where $[x]$ denotes the integer part of $x$, then this 
term is invariant under the action of the permutation $(12)$ and the chain is 
of length 3, except if $p$ is divisible by 3 then $(\lam\mun)^{p/3}$ satisfies
\polylm\ by itself and so forms a chain of length 1. All other chains are of 
length 6. With this counting the number of independent such minimal symmetric 
polynomials is, 
\eqn\countN{
N_p = \cases{(n+1)3n+1 \, , \quad &\hbox{$p=6n\, ;$}  \cr
(n+1)(3n+q) \, , \quad &\hbox{$p=6n+q, \ q=1,2,3,4,5\, .$} \cr}
}

We list the first few non trivial cases in table 2, of course 
$S_0(\lam,\mun)=1$.

\vskip 6pt
\hbox{
\vbox{\tabskip=0pt \offinterlineskip
\hrule
\halign{&\vrule# &\strut \ \hfil#\  \cr
height2pt&\omit&&\omit&&\omit&\cr
&$p$&& polynomial\hfil &&$(i,j)$\hfil&\cr
height2pt&\omit&&\omit&&\omit&\cr
\noalign{\hrule}
height4pt&\omit&&\omit&&\omit&\cr
&$1$&&$\lam+\mun+1$\hfil&&$(0,0)$\hfil&\cr
height4pt&\omit&&\omit&&\omit&\cr
\noalign{\hrule}
height4pt&\omit&&\omit&&\omit&\cr
&$2$&&$\matrix{\scs\lam^2+\mun^2+1\hfill\cr\scs\lam\mun+\lam+\mun\hfill}$\hfil&
&${\scs(0,0),(1,0)}$\hfil&\cr
height4pt&\omit&&\omit&&\omit&\cr
\noalign{\hrule}
height4pt&\omit&&\omit&&\omit&\cr
&$3$&&$\matrix{\scs\lam^3+\mun^3+1\hfill\cr\scs\lam^2\mun+\lam\mun^2+\lam^2+
\mun^2+\lam+\mun\hfill\cr\scs\lam\mun\hfill}\hfil$&
&$\matrix{\scs(0,0),(1,0)\hfill\cr\scs(0,1)\hfill}$\hfil&\cr
height4pt&\omit&&\omit&&\omit&\cr
\noalign{\hrule}
height4pt&\omit&&\omit&&\omit&\cr
&$4$&&$\matrix{\scs\lam^4+\mun^4+1\hfill\cr\scs\lam^3\mun+\lam\mun^3+\lam^3+
\mun^3+\lam+\mun\hfill\cr
\scs\lam^2\mun^2+\lam^2+\mun^2\hfill\cr\scs\lam^2\mun+\lam\mun^2
+\lam\mun\hfill}$\hfil&&
$\matrix{\scs(0,0),(1,0),(2,0)\hfill\cr\scs(0,1)\hfill}$\hfil&\cr
height4pt&\omit&&\omit&&\omit&\cr
\noalign{\hrule}
height4pt&\omit&&\omit&&\omit&\cr
&$5$&&$\matrix{\scs\lam^5+\mun^5+1\hfill\cr\scs\lam^4\mun+\lam\mun^4+\lam^4+
\mun^4+\lam+\mun\hfill\cr\scs
\lam^3\mun^2+\lam^2\mun^3+\lam^3+\mun^3+\lam^2+\mun^2\hfill\cr
\scs\lam^3\mun+\lam\mun^3+\lam\mun\hfill\cr
\scs\lam^2\mun^2+\lam^2\mun+\lam\mun^2\hfill}$\hfil&&
$\matrix{\scs(0,0),(1,0),(2,0)\hfill\cr\scs(0,1),(1,1)\hfill}$\hfil&\cr
height4pt&\omit&&\omit&&\omit&\cr}
\hrule}
\vbox{\tabskip=0pt \offinterlineskip
\hrule
\halign{&\vrule# &\strut \ \hfil#\  \cr
height2pt&\omit&&\omit&&\omit&\cr
&$p$&& polynomial\hfil &&$(i,j)$\hfil&\cr
height2pt&\omit&&\omit&&\omit&\cr
\noalign{\hrule}
height4pt&\omit&&\omit&&\omit&\cr
&$6$&&$\matrix{\scs\lam^6+\mun^6+1\hfill\cr\scs\lam^5\mun+\lam\mun^5+\lam^5+
\mun^5+\lam+
\mun\hfill\cr\scs\lam^4\mun^2+\lam^2\mun^4+\lam^4+\mun^4+\lam^2+\mun^2\hfill\cr
\scs\lam^3\mun^3+\lam^3+\mun^3\hfill\cr
\scs\lam^4\mun+\lam\mun^4+\lam\mun\hfill\cr
\scs\lam^3\mun^2+\lam^2\mun^3+\lam^3\mun+\lam\mun^3+\lam^2\mun+\lam\mun^2\hfill\cr
\scs\lam^2\mun^2\hfill}$\hfil&&
$\matrix{\scs(0,0),(1,0),(2,0),(3,0)\hfill\cr\scs(0,1),(1,1)\hfill\cr
\scs(0,2)\hfill}$\hfil&\cr
height4pt&\omit&&\omit&&\omit&\cr
\noalign{\hrule}
height4pt&\omit&&\omit&&\omit&\cr
&$7$&&$\matrix{\scs\lam^7+\mun^7+1\hfill\cr\scs\lam^6\mun+\lam\mun^6+\lam^6+
\mun^6+\lam+\mun\hfill\cr\scs
\lam^5\mun^2+\lam^2\mun^5+\lam^5+\mun^5+\lam^2+\mun^2\hfill\cr
\scs\lam^4\mun^3+\lam^3\mun^4+\lam^4+\mun^4+\lam^3+\mun^3\hfill\cr
\scs\lam^5\mun+\lam\mun^5+\lam\mun\hfill\cr
\scs\lam^4\mun^2+\lam^2\mun^4+\lam^4\mun+\lam\mun^4+\lam^2\mun+\lam\mun^2\hfill\cr
\scs\lam^3\mun^3+\lam^3\mun+\lam\mun^3\hfill\cr
\scs\lam^3\mun^2+\lam^2\mun^3+\lam^2\mun^2\hfill}$\hfil&&
$\matrix{\scs(0,0),(1,0),(2,0),(3,0)\hfill\cr\scs(0,1),(1,1),(2,1)\hfill\cr
\scs(0,2)\hfill}$\hfil&\cr
height4pt&\omit&&\omit&&\omit&\cr}
\hrule\vskip21pt}
}

Table 2. Symmetric polynomials.

An alternative basis for $S_p$, valid for general $p$, may be obtained 
by constructing from $\lam,\mun$ two invariants $I_1,I_2$ under $\S_3$ 
and then introducing for any $p$ a factor
to ensure that \polylm\ holds. With suitable restrictions
the result becomes a polynomial expressible in the form
\eqn\Syy{\eqalign{
{S}_{p,(i,j)}(\lam,\mun) = {}& (\lam+\mun+1)^{p}\, I_1( \lam,\mun )^i
I_2( \lam,\mun )^j \, , \cr
I_1( \lam,\mun ) ={}& { \lam\mun+\lam+\mun \over (\lam+\mun+1)^2 } \, , 
\qquad I_2( \lam,\mun ) ={\lam\, \mun\over (\lam+\mun+1)^3} \, , \cr
i={}& 0,1,\dots ,[\half p] \, , \qquad
j = 0,1,\dots ,[{\ts{1\over 3}}(p-2i)] \, . \cr}
}
Lists of possible $(i,j)$ for $p$ up to 7 are given in table 2. 
This result may also be easily expressed as symmetric polynomial in $y,\yz$
by using
\eqn\lmyy{\eqalign{
\lam+\mun+1 = {}& \half ( y\yz+3 ) \, , \qquad \lam\mun = {\ts {1\over 16}}
(1-y^2)(1-\yz^2) \, , \cr
\Lambda = {}& (\lam+\mun+1)^2 - 4(\lam\mun+\lam+\mun)= 
{\ts {1\over 4}} ( y-\yz)^2 \, , \cr}
}
where $\Lambda$ is defined in \conlm.
Completeness of the basis provided by \Syy\ is straightforwardly demonstrated
by showing that it gives the same number of independent polynomials $N_p$
as given in \countN.

For the antisymmetric representation of $\S_3$ we require,
\eqn\repA{
a  \toinf{(12)} -a \, , \qquad a \toinf{(123)} a \, .
}
while the two-dimensional mixed symmetry representation of $\S_3$ is defined 
on a basis $(b,c)$ where
\eqn\mix{
\pmatrix{b\cr c} \toinf{(12)} \pmatrix{-1&0\cr0&1}\pmatrix{b\cr c} \, ,
\qquad
\pmatrix{b \cr c} \toinf{(123)} \pmatrix{-{1\over 2}&-{\sqrt 3 \over 2}\cr
{\sqrt 3 \over 2}&-{1\over 2}}  \pmatrix{b \cr c} \, .
}
It is easy to see that the tensor products formed by $a\,a'$ and 
$b \, b' + c\, c'$ are 
symmetric while $(b\, c' + c\, b', b\, b' + c\, c')$ is a basis for a
mixed symmetry representation and $b\, c' - c \, b'$ is antisymmetric.

For functions of $\lam,\mun$ \repA\  is satisfied by
\eqn\anti{
a(\lam,\mun) = {(\lam-\mun)(\lam-1)(\mun-1) \over (\lam+\mun+1)^3} \, .
} 
For $p\ge 3$, $a(\lam,\mun)S_{p,(i,j)}(\lam,\mun)$ is a polynomial if we allow
$i=0,1,\dots [\half(p-3)]$ and $j = 0,1,\dots ,[{\ts{1\over 3}}(p-2i-3)]$
giving $N_{p-3}$ antisymmetric polynomials.
For the mixed symmetry transformations in \mix\ there essentially two 
independent possibilities
\eqn\mixl{
b_1(\lam,\mun)= {\lam - \mun\over  \lam+\mun+1} \, ,
\qquad c_1(\lam,\mun) = {\lam+\mun - 2 \over \sqrt 3(\lam+\mun+1)} \, .
}
and
\eqn\mixlm{
b_2(\lam,\mun)= {\lam - \mun \over (\lam+\mun+1)^2} \, , 
\qquad c_2(\lam,\mun) = - {\lam+\mun - 2\lam\mun \over \sqrt 3(\lam+\mun+1)^2} \, .
}
By considering
$(b_r(\lam,\mun),c_r(\lam,\mun))S_{p,(i,j)}(\lam,\mun)$ for $p\ge r$, $r=1,2$,
for appropriate $i,j$ we obtain $N_{p-r}$ polynomial mixed symmetry
representations of $\S_3$. Together with the symmetric polynomials $S_{p,(i,j)}$
and $a S_{p,(i,j)}$ these provide a complete basis for two variable polynomials
in $\lam,\mun$ of order $p$ since 
$N_p + 2(N_{p-1}+N_{p-2})+N_{p-3}= \half(p+1)(p+2)$. We may also note that these 
polynomials form a closed set under multiplication since
\eqnn\mult
$$\eqalignno{
b_1{}^{\!2} + c_1{}^{\!2} = {}& {\ts{4\over 3}} - 4I_1 \, , \ \ \, \qquad
\sqrt 3 ( 2 b_1c_1 ,  b_1{}^{\!2} - c_1{}^{\!2}) 
= 2( b_1 - 3 b_2,  c_1 - 3 c_2 ) \, , \cr
b_1 b_2 + c_1 c_2 = {}& {\ts{2\over 3}}I_1 - 6I_2 \, , \qquad
\sqrt 3 ( b_1c_2 + c_1 b_2 ,  b_1 b_2  - c_1 c_2 ) 
= 2( I_1 b_1 - b_2,  I_1 c_1 - c_2 )  \, , \cr
b_2{}^{\!2} + c_2{}^{\!2} = {}&  {\ts{4\over 3}} I_1{\!}^2 - 4 I_2 \, , \qquad
\sqrt 3 ( 2 b_2c_2 ,  b_2{}^{\!2} - c_2{}^{\!2}) 
=  2( 3 I_2 b_1 - I_1 b_2,  3I_2 c_1 - I_1 c_2 ) \, , \cr
\sqrt 3(b_1c_2 - c_1 b_2) = {}& 2 a \, , \qquad\qquad\qquad
a^2 = I_1 {}^{\!2} - 4 I_1{}^{\!3} +18 I_1 I_2 - 4 I_2 - 
27 I_2{}^{\!2} \, . &\mult \cr}
$$
where $I_1,I_2$ are the invariants defined in \Syy.

For $\N=2$ there are further restrictions as a consequence of \conlm. Taking
$p\to 2n$ we construct, instead of \polylm\ since $\lam,\mun$ are expressible
in terms of just $\alpha$ by \defa, the single variable polynomials $f_n$ of
degree $2n$, satisfying under the action of $\S_3$
\eqn\Sy{\eqalign{
f_n(\alpha) = {}& f_n(1-\alpha) = (\alpha-1)^{2n} f_n \Big ( 
{\alpha\over \alpha-1} \Big ) = \alpha^{2n} f_n \Big ( {1\over \alpha } \Big ) \cr
= {}&  (\alpha-1)^{2n} f_n \Big ( {1 \over 1- \alpha} \Big ) =
(\alpha-1)^{2n} f_n \Big ( {\alpha -1 \over \alpha} \Big ) \, . \cr}
%{\hat S}_n(y) = {\hat S}_n(-y) = \big ( \half(1-y) \big )^{2n} {\hat S}_n
% \Big ( {y+3 \over 1-y} \Big ) \, . 
}
As shown by Heslop and Howe, \Howe, the sum of terms produced by the
action of $\S_3$ as given by \Sy\ starting from $\alpha^{r}$
generates a linearly independent set of polynomials
for $r=0,1,\dots [{\ts{1\over 3}}n]$, giving $[{\ts{1\over 3}}n]+1$ solutions
for $f_n$. Alternatively an equivalent basis is provided by
\eqn\Soly{
{S}_{n,j}(\alpha) = (\alpha^2-\alpha+1)^n s(\alpha)^j \, , 
\qquad j = 0,1, \dots , [{\ts{1\over 3}}n] \, ,
}
where $s(\alpha)$ is the $\S_3$ invariant
\eqn\Sinv{
s(\alpha) = {\alpha^2(1-\alpha)^2\over (\alpha^2-\alpha+1)^3} =
4\, {(1-y^2)^2 \over (y^2+3)^3}  \, .
}
The solutions given by \Soly\ correspond to \Syy\ for $i=0$ since $\Lambda=0$ 
in this case. A general polynomial solution of \Sy\ is then given by
\eqn\fgen{
f_n(\alpha) = (\alpha^2-\alpha+1)^n P(s(\alpha)) \, ,
}
with $P(s)$ a polynomial of degree $[{\ts{1\over 3}}n]$.

We may also consider othe representations of $\S_3$.
For the antisymmetric representation, as in \anti, we may define
\eqn\antiy{
a(\alpha) = (2\alpha-1)\,{(\alpha-2)(\alpha-1)\alpha(\alpha+1)\over
(\alpha^2-\alpha+1)^3} = 4 \, {y(y^2-1)(y^2-9)\over (y^2+3)^3 } \, ,
}
so that $a(\alpha){S}_{n,j}(\alpha)$ is then a polynomial for $n\ge 3$ and
$j = 0,1, \dots , [{\ts{1\over 3}}n]-1$. For the mixed symmetry representation 
there are two essential solutions which can be written in the form
\eqn\mixz{\eqalign{
b_1(\alpha) =  {}& {2\alpha-1\over\alpha^2-\alpha+1} = {4} \,  {y\over y^2+3} \, , \cr
c_1(\alpha) = {}& {1\over \sqrt 3}\,{2\alpha^2-2\alpha -1 \over \alpha^2-\alpha+1} =
{2\over \sqrt 3} \, {y^2 -3 \over y^2+3} \, , \cr}
}
and
\eqn\mixy{\eqalign{
b_2(\alpha) =  {}&  (2\alpha-1){\alpha(\alpha-1)\over
(\alpha^2-\alpha+1)^2} = {4} \, {y(y^2-1) \over (y^2+3)^2} \, , \cr
c_2(\alpha) = {}& \sqrt 3 \, {\alpha(\alpha-1)\over (\alpha^2-\alpha+1)^2} = 
4\sqrt 3 \, {y^2-1 \over (y^2+3)^2} \, . \cr}
}
It is easy to see that $(b_r(\alpha),c_r(\alpha)) {S}_{n,j}(\alpha)$
are polynomials for $j = 0,1, \dots , [{\ts{1\over 3}}(n-r)]$ if $n\ge r$
for $r=1,2$. The basis provided by ${S}_{n,j}(\alpha)$, 
$(b_r(\alpha),c_r(\alpha)) {S}_{n,j}(\alpha)$, $r=1,2$ and
$a(\alpha){S}_{n,j}(\alpha)$ is then complete in that it gives $2n+1$
linearly independent polynomials, allowing for the expansion of any
arbitrary polynomial of degree $2n$, $2\big([{\ts{1\over 3}}n]+
[{\ts{1\over 3}}(n-1)]+[{\ts{1\over 3}}(n-2)]\big)+5=2n+1$.

For $\N=4$ the superconformal Ward identities require
\eqn\scc{
\G(u,v;\lam,\mun)\big |_{\bet = {1\over \zz}} = f(x,\alpha) \, ,
}
so that \cross\ gives
\eqn\crossf{
f(x,\alpha) = f \Big ( {x\over x-1}, 1 - \alpha \Big ) =
\bigg ( {x(\alpha-1) \over 1-x} \bigg )^p f \Big ( 1-x , 
{\alpha \over \alpha-1} \Big ) =  ( x\alpha  )^p 
f\Big ( {1\over x}, {1\over \alpha } \Big ) \, .
}
To obtain an extension to a fully crossing symmetric correlation function
we may consider for any $S_p$ satisfying \polylm
\eqn\crossG{
\G(u,v;\lam,\mun) = S_p\Big ( u \, \lam , {u\over v} \,\mun \Big ) \, ,
}
which obeys \cross\ as a consequence of \polylm. From \scc\ we obtain
\eqn\fsym{
f(x,\alpha) = S_p \Big ( x \alpha, {x(1-\alpha) \over x-1} \Big ) \, ,
}
which automatically satisfies \crossf.

For the $\N=2$ case instead of \cross\ we have
\eqn\crosstwo{
\G(u,v;\lam,\mun) =  \G(u/v,1/v;\mun,\lam ) =
\Big ( {u^2\over v^2}\, \mun \Big )^n \G(v,u;\lam/\mun,1/\mun) \, ,
}
where, with $\lam,\mun$ constrained as in \defa, the superconformal Ward
identities are
\eqn\scct{
\G(u,v;\lam,\mun)\big |_{\alpha = {1\over \zz}} = f(x) = 
f \Big ( {x\over x-1} \Big ) = \bigg ( {x\over x-1} \bigg )^{\! 2n}\! f(1-x) 
= x^{2n} f \Big ( {1\over x} \Big ) \, ,
}
where we also exhibit the crossing symmetry relations for the single variable
function $f$. The corresponding solution to  \crossG\ is given by
\eqn\crossGt{
\G(u,v;\lam,\mun) = S_n\Big ( u^2 \, \lam , {u^2\over v^2} \,\mun \Big ) \, ,
}
which implies
\eqn\fSn{
f(x) = S_n \Big ( x^2 , {x^2\over (1-x)^2} \Big ) \, .
}
In this case if we consider the contribution of individual factors in the basis 
given by \Syy\ to $f(x)$ as expected from \crossGt\ and \fSn\ we have
\eqn\rela{ \eqalign{
P= {}& \Big ( u^2 \lam + {u^2\over v^2}\, \mun + 1 \Big )\Big |_{\alpha  
= {1\over \zz}} \!  =  p^2 \, , \quad
Q = {u^4\over v^2} \, \lam\mun \Big  |_{\alpha = {1\over \zz}} \! = q^2 \, , \cr
R = {}& \Big ( {u^4\over v^2}\,  \lam\mun + u^2 \lam + {u^2\over v^2}\, \mun 
\Big ) \Big  |_{\alpha = {1\over \zz}} \! = 2pq  \, ,  \cr}
}
where
\eqn\defpq{
p(x) = {x^2 -x +1 \over 1-x} \, , \qquad q(x) = {x^2\over 1-x} \, .
}
so that we have the relation $R^2=4PQ$. In consequence we may restrict in 
\Syy\ to those polynomials with $i=0,1$.

Conversely we may argue that for the $\N=2$ case all single variable 
functions $f(x)$ may be expressible in terms of $S_n$ as in \fSn\ 
and therefore may be extended to a fully crossing symmetric form for
$\G(u,v;\lam,\mun)$ as exhibited in \crossGt. To demonstrate this
we suppose all solutions of the crossing symmetry relations in \scct\ for
$f$ are solvable by writing
\eqn\fgr{
f(x) = p(x)^{2n} g\big ( s(x) \big ) \, , \qquad s(x) = {q(x) \over p(x)^3} \, ,
}
for some function $g$ of the crossing invariant $s$ given by $\Sinv$. 
Note that for $x\to 0, \, s \sim x^2$, $x \to 1 , \, s \sim (1-x)^2$ and for 
$x \to \infty, \, s \sim 1/x^2$. From the superconformal representation 
theory for the corresponding
contributions to the operator product expansion $f(x)$ should be analytic
in the neighbourhood of $x=0$ with singularities only at $x=1,\infty$. In
consequence $g(s)$ must be a polynomial which is then restricted to have
maximal degree $[{2\over 3}n]$ to avoid singularities when $x^2-x+1=0$. 
It is then easy to see that $f$
can be written as a polynomial in $P,Q$ with terms also linear in $R$, as 
defined in \rela, which is consistent with \fSn\ where $S_n$ has an  expansion
in terms of $S_{n,(i,j)}$ with $i=0,1$ and $j$ restricted as in \Syy.

A similar discussion is possible for $\N=4$. The function $f(x,\alpha)$
is required to be a general solution of the crossing symmetry conditions 
given by \crossf\ which is also a polynomial of degree $p$ in $\alpha$.
It is also analytic in $x$ in the neighbourhood of $x=0$ with singularities 
only at $x=1,\infty$. If we write
\eqn\Fnew{
f(x,\alpha) = P(x,\alpha)^p g(x,\alpha) \, , \qquad
P(x,\alpha) = {x^2\alpha - 2x\alpha + 2x  - 1 \over x-1 } \, ,
}
then $g$ is an invariant under the action of $\S_3$, as displayed in Table 1.
Determining a general form for $g$ is then reducible to finding a basis for
all possible independent invariants which may be formed from $x$ and 
$\alpha$. Since the action of $\S_3$ on any polynomial in $\alpha$ may be 
decomposed, up to functions of the invariant $s(\alpha)$, into contributions 
linear in $1$, $a(\alpha)$ and $(b_r(\alpha),c_r(\alpha)), \,  r=1,2$, as 
given in \antiy\ and \mixz, \mixy, then a basis for such invariants 
is obtained, in addition to the separate invariants $s(x), \, s(\alpha)$,
by combining these non trivial irreducible representations with corresponding
representations involving $x$ to give
\eqn\SSinv{
A ( x , \alpha) = a( x^{-1}) \, a ( \alpha ) \, ,
}
where $a( x^{-1}) = - a(x)$, and also
\eqn\Smix{\eqalign{
S_1 ( x,\alpha ) ={}&  b_1( x^{-1})\, b_1( \alpha ) + c_1( x^{-1})\, c_1( \alpha )\cr
= {}& {4\over 3} - {2(x \alpha  -1)^2 \over (\alpha^2-\alpha+1)
(x^2-x+1)} \, , \cr
S_2 ( x,\alpha ) ={}&  b_2( x^{-1})\, b_2( \alpha ) + c_2( x^{-1})\, c_2( \alpha )\cr
= {}&  {2 \alpha(1-\alpha) \, x(1-x) \over 
(\alpha^2-\alpha+1)^2(x^2-x+1)^2} \, ( x\alpha - 2\alpha - 2x + 1) \, ,\cr
S_3 ( x,\alpha ) ={}&  b_2( x^{-1})\, b_1( \alpha ) + c_2( x^{-1})\, c_1( \alpha )\cr
= {}& {2 x(1-x) \over (\alpha^2-\alpha+1)(x^2-x+1)^2} 
\, ( x\alpha^2 - 2x\alpha + 2\alpha  - 1) \, , \cr
S_4 ( x,\alpha ) ={}&  b_1( x^{-1})\, b_2( \alpha ) + c_1( x^{-1})\, c_2( \alpha )\cr
= {}& {2 \alpha(1-\alpha) \over (\alpha^2-\alpha+1)^2(x^2-x+1)}
\, ( x^2\alpha - 2x\alpha + 2x  - 1) \, .
\cr}
}
These are not independent since
\eqn\relAS{\eqalign{
& A ( x , \alpha) = {\ts {3\over 4}} \big ( S_1 ( x,\alpha )S_2 ( x,\alpha )
- S_3 ( x,\alpha ) S_4 ( x,\alpha ) \big ) \, , \cr
& S_2( x,\alpha) - \half \, S_1( x,\alpha)S_3(x,\alpha) - {\ts {1\over 3}}\,
S_3(x,\alpha) = - 2s(x) \, , \cr 
& S_2( x,\alpha) - \half \, S_1( x,\alpha)S_4(x,\alpha) - {\ts {1\over 3}}\,
S_4(x,\alpha) = - 2s(\alpha)  \, , \cr
& 2 \big ( S_3(x,\alpha) + S_4(x,\alpha) \big ) - 6 S_2( x,\alpha) +
S_1 ( x,\alpha )^2 - {\ts {2\over 3}} S_1 ( x,\alpha ) = {\ts {8\over 9}}\, .\cr}
}
A crucial further constraint arises from \real\
which here requires that $f(x,x^{-1})$ is a constant. Since $P(x,x^{-1})=3$
we also require that $g$ depends on invariants $s_r(x,\alpha)$ such that
$s_r(x,x^{-1})$ are constants. Taking account of the relations in \relAS\ 
there are then two independent solutions which we take as
\eqnn\solss
$$\eqalignno{
s_1(x,\alpha) = {}&  2 \, {S_3(x,\alpha)s(\alpha)\over 
S_4(x,\alpha)^2} = { R(x,\alpha) \over P(x,\alpha)^2} \, , \qquad
s_2(x,\alpha) = 8 \, {s(x)s(\alpha)^2\over S_4(x,\alpha)^3}
= { Q(x,\alpha) \over P(x,\alpha)^3} \, , \cr
R(x,\alpha) = {}&  {x(x\alpha^2 - 2x\alpha + 2\alpha  - 1) \over 1-x }  \, , 
\qquad\quad Q(x,\alpha) = {x^2 \alpha(1-\alpha) \over x-1} \, , & \solss \cr}
$$
where $R(x,x^{-1})=3, \, Q(x,x^{-1})=1$. It is then evident that $g$ in
\Fnew\ must be of the form
\eqn\gexp{
g = \sum_{{i,j\ge0\atop 2i+3j \le p}} c_{ij} s_1{}^{\! i}\,  s_2{}^{\! j} \, .
}
Noting that
\eqn\relb{ \eqalign{
P(x,\alpha) = {}& \Big ( u \,\lam + {u\over v}\, \mun + 1 \Big )\Big |_{\bet  
= {1\over \zz}} \, , \qquad
Q(x,\alpha)= {u^2\over v} \, \lam\mun \Big  |_{\bet = {1\over \zz}}  \, , \cr
R (x,\alpha)= {}& \Big ( {u^2\over v}\,  \lam\mun + u \,\lam + {u\over v}\, \mun 
\Big ) \Big  |_{\bet = {1\over \zz}}  \, ,  \cr}
}
it is easy to see, as a consequence of \Syy, that $I_r(u\lam, u \mun/v)
|_{\bet = 1/\zz} = s_r(x,\alpha)$ and hence the expression given by 
\Fnew\ and \gexp\ for the 
function $f$ may always be extended to a fully crossing symmetric result 
for the full correlation function $\G$ of the form \crossG\ with 
$S_p(\lam,\mun)= (\lam+\mun+1)^p \sum_{i,j} c_{ij} \, I_1(\lam,\mun)^i\, 
I_2(\lam,\mun)^j$ and where $f$ satisfies \fsym. With appropriate 
coefficients for the independent terms in $S_p$ \crossG\ corresponds to the 
results of free field theory. 
In general, using the formalism of harmonic superspace, the Intriligator
insertion technique \Intr\ demonstrates that only $\H$ as in \uncon\ or \uncont,
or $\K$ as in \Gsol\ or \Gsolt, can depend on the coupling $g$, and so are
dynamical. The functions $f(x)$ or $f(x,\alpha)$ are then identical with
the free theory, or $g=0$, results.

The remaining part of the correlation function may also be expressed
in terms of $\S_3$ representations. It is convenient  to define from \mixl\ 
and \mixlm\ $(b_r{\!}'(u,v),c_r{\!}'(u,v)) = (b_r(1/u,v/u),c_r(1/u,v/u))$.
For the $\N=2$ case we may then write for the factor which appears in the
solution of the superconformal identities in \uncont
\eqn\factwo{
(\alpha x - 1 ) (\alpha \zz - 1 ) = (\alpha^2-\alpha+1)(u+v+1) \big (
{\ts {1\over 3}} - \half ( b_1{\!}'(u,v) \, b_1(\alpha) + 
c_1{\!}'(u,v) \, c_1 (\alpha) ) \big ) \, .
} 
For $\N=4$ we may also note
\eqnn\factf
$$\eqalignno{
( \alpha x & -1  ) ( \alpha \zz - 1 ) ( \bet x -1  ) ( \bet \zz - 1 ) 
= {\ts {1\over 16}} u^2 (y-z)(y-\bz)(\yz-z)(\yz-\bz) \cr
={}& v + \lam^2 uv + \mun^2 u + \lam \, v(v-1-u) + \mun (u+v-1) + 
\lam \mun \, u(u-1-v)\cr
= {}& (\lam+\mun+1)^2(u+v+1)^2 \Big ({\ts {1\over 3}} I_1 (u,v) 
+ {\ts {1\over 3}} I_1 (\lam,\mun) - 2 I_1 (u,v) I_1 (\lam,\mun)  \cr
&\hskip 4.1cm {}- \half 
\big ( b_1{\!}'(u,v) \, b_2(\lam,\mun) +  c_1{\!}'(u,v) \, c_2(\lam,\mun) \cr 
& \hskip 4.5cm {}+ b_2{\!}'(u,v) \, b_1(\lam,\mun) 
+ c_2{\!}'(u,v) \, c_1(\lam,\mun) \cr
& \hskip 4.5cm {}  - 3 \, b_2{\!}'(u,v) \, b_2(\lam,\mun)
- 3\,  c_2{\!}'(u,v) \, c_2(\lam,\mun) \big ) \Big ) \, . & \factf \cr}  
$$
The function $\K$ in \Gsolt\ must then satisfy the crossing symmetry relations
\eqn\crossK{
\K(u,v;\lam,\mun) =  \K(u/v,1/v;\mun,\lam ) =
\Big ( {u\over v}\Big )^{p+2} \mun^{p-2}\, \K(v,u;\lam/\mun,1/\mun) \, .
}

It is also of interest to extend the considerations of crossing symmetry to
the next-to-extremal case when $p_1=p_2=p_3=p$, $p_4=3p-2$.  In this case
$\G$, defined by \GFp, must satisfy for the permutations $(12)$ and $(23)$
\eqn\Gcross{
\G(u,v;\lam,\mun) = v^{p-1} \G\Big ({u\over v},{1\over v};\mun, \lam\Big ) \, , 
\qquad \G(u,v;\lam,\mun) = u^{2p-1}\lam\, 
\G\Big ({1\over u},{v\over u};{1\over \lam}, {\mun \over \lam} \Big )  \, .
}
The solution \next\ can be rewritten as
\eqn\Gext{
\G(u,v;\lam,\mun) = u^{p -1} \bigg ( k + {x\zz\over x - \zz}
\Big ( \big (\alpha -1/x\big )\big (\bet  - 1/x\big ) f(x) - 
\big (\alpha  - 1/\zz\big )\big (\bet - 1/\zz \big ) f(\zz) \Big ) \bigg ) \, ,
}
and then \Gcross\ requires
\eqn\fcross{
f(x) = - f \Big ( {x \over x-1} \Big ) \, , \qquad f(x) = - x^2 f \Big ( {1\over x}
\Big ) + k \, x \, .
}
A particular solution of \fcross\ is given by
\eqn\fsol{
f(x) = {k\over 3}\, \Big ( x - {x \over x-1} \Big ) \, .
}
To obtain a general solution of \fcross\ it is then sufficient to seek the 
general solution $f_0(x)$ of \fcross\ with $k=0$. Using results obtained
above this is
\eqn\fgen{
f_0(x) = {(x-2)x(x+1)(2x-1)\over (x-1)(x^2-x+1)} \, h\big ( s(x) \big ) \, ,
}
where $s$ is the invariant defined by \fgr\ and \defpq. This introduces 
unphysical singularities for $x^2-x+1=0$ unless cancelled by $h$. However, for
compatibility with semi-short representations, $h(s)$ must be analytic in $s$
for $s\sim 0$ (if $h(s)=1/s$, which cancels the singularity at $x^2-x+1=0$, then
$f_0(x) \sim 1/x$ for $x\to 0$). Hence we conclude that there is no possible
solution of the form \fgen\ and hence we only have \fsol. In this case
\eqn\Gwex{
\G(u,v;\lam,\mun) =  {\ts {1\over 3}}k  \, u^{p-1} \Big ( 1 + \lam \, u + 
\mun \, {u\over v} \Big ) \, .
}

\newsec{Large $N$ Results}

In this paper we have endeavoured to work out the consequences of superconformal
symmetry for the four point correlation functions of BPS operators. As a result
our considerations are lacking in dynamical input since we do not consider
any details of $\N=2$ or $\N=4$ superconformal theories. The results of our
analysis demonstrate that the details of the dynamics resides in the 
function $\H$, which appears in \uncon\ and \uncont, or $\K$ as in \Gsolt.
In particular cases results have been obtained using perturbation theory \pert\
or with the AdS/CFT correspondence \refs{\Arut,\ADHS,\Degen}. We here summarise
some of the results obtained in \refs{\Arut,\ADHS,\Degen} in the context
of this paper.

For $p_i=p$ in the large $N$ limit the leading result is, with a suitable 
normalisation, simply obtained from free field theory
\eqn\GlN{
\G_0(u,v;\lam,\mun) = 1 + (\lam u)^p + \Big ( \mun \, {u\over v} \Big )^p \, .
}
The definitions \solid\ and \real\ then give
\eqn\flN{
f_0(z,y) = 1 + \Big ( {1+y \over 1+z} \Big )^p + \Big ( {1-y \over 1-z} \Big )^p \, ,
\qquad k=3 \, .
}
Using \Gsolt\ we can then determine, assuming
$\K(u,v;\lam,\mun) = {1\over 16}u^2 \H(u,v;\lam,\mun)$, for $p=2$
\eqn\Hztwo{
\H_0(u,v;\lam,\mun)= 1 + {1\over v^2} \, ,
}
and for $p=3$
\eqnn\Hzthree
$$\eqalignno{ 
\H_0(u,v;\lam,\mun)= {1\over v^3} \Big (& \half(\lam+\mun) u (1+v^3)
+ \half(\lam- \mun) \big ( -3u(1-v^3) + 2(1-v)(1+v^3) \big ) \cr 
&{}+ u (1+v^3) - 1 + 2v+2v^3 - v^4 \Big ) \, , & \Hzthree \cr}
$$
while for $p=4$
\eqnn\Hzfour
$$\eqalignno{
\H_0(u,v;\lam,\mun)= {1\over v^4}&  \Big ( \lam\mun \, u^2(1+v^4) \cr
&{}  + \half (\lam-\mun)^2
\big ( 2(1-v)^2(1+v^4) - 5u (1+v^5) + 3uv (1+v^3) + 4u^2(1+v^4) \big ) \cr
&{}+ \half (\lam^2 - \mun^2) \big ( u(1-v)(1+v^4) - 2u^2 (1-v^4) \big ) \cr
&{} + \half (\lam+\mun) \big ( - (1-v)^2 + u (1+v) \big ) (1+v^4)  \cr
&{} + \half (\lam-\mun) \big ( (1-v) (- 3(1+v)+7u) (1+v^4) + 8v(1-v)(1+v^3)
- 4u^2 (1-v^4) \big ) \cr
&{}+ 1+v^6 - 3v(1-v)(1-v^3) - 2u (1-v)(1-v^4) + u^2(1+v^4) 
\Big ) \, . & \Hzfour \cr}
$$
In each case the crossing symmetry relation $\H_0(u,v;\lam,\mun) =
\H_0(u/v,1/v;\mun,\lam)/v^2$ is satisfied but the corresponding one for
$u\leftrightarrow v$ is not since it is necessary to take account of the
function $f_0(z,y)$ then as well.

The large $N$ results obtained through the AdS/CFT correspondence are expressible 
in terms of functions $\oD_{n_1n_2n_3n_4}(u,v)$
which satisfy various identities listed in \scft\ and \ADHS.
When $p=2$
\eqn\Htwo{
\H(u,v;\lam,\mun) = - {4\over N^2} \, u^2 \oD_{2422}(u,v) \, ,
}
and for $p=3$,
\eqn\Hthree{
\H(u,v;\lam,\mun) = - {9\over N^2} \, u^3 \big ( ( 1+\lam+\mun) \oD_{3533}
+ \oD_{3522} + \lam \, \oD_{2523} + \mun \, \oD_{2532} \big ) \, .
}
For $p=4$ the results obtained in \Degen\ can be rewritten as
\eqnn\Hfour
$$\eqalignno{
\H(u,v;\lam,\mun) = - {4\over N^2} \, u^4 & \Big ( (1+\lam^2 + \mun^2
+ 4 \lam + 4 \mun + 4 \, \lam\mun ) \oD_{4644} \cr 
&{}+ 2 ( \oD_{4633} + \oD_{4622} ) + 2\lam^2 ( \oD_{3634} + \oD_{2624} )
+2 \mun^2 ( \oD_{3643} + \oD_{2642} ) \cr
&{} - 4\, \lam( \oD_{4624} -2 \oD_{3623} ) - 4 \, \mun( \oD_{4642} -2 \oD_{3632} ) \cr
&{} - 4\, \lam\mun( \oD_{2644} -2 \oD_{2533} ) \Big ) \, . & \Hfour \cr}
$$
Since $\K(u,v;\lam,\mun) = {1\over 16}u^2 \H(u,v;\lam,\mun)$ it is easy to
verify both the crossing symmetry conditions \crossK\ using $\oD$ identities.
Furthermore the results given by \Htwo, \Hthree\ and \Hfour, in which overall
factors of $u^p$ are present, are manifestly
compatible with the unitarity conditions flowing from \Kexp\ and \unit\ 
since the leading log. term $\oD_{n_1n_2n_3n_4}(u,v)$ is $\log u$ itself.
When expressed in terms of conformal partial waves $\G^{(\ell)}_{\Delta+4}$
it is easy to see in each case that only contributions with minimum twist 
$\Delta - \ell = 2p$ are required. Hence \Htwo, \Hthree\ and \Hfour\ require the
presence of operators belonging to long multiplets which have anomalous dimensions 
with twist, at zeroth order in $1/N$, $\Delta - \ell = 2(p+t), \ t=0,1,2,\dots $
for the lowest scale dimension operators in each multiplet. The condition
$\Delta - \ell = 2p$ is stronger than that required by unitarity \unit, with
$n\le p-2$, which shows that for any representation some low twist multiplets
decouple (thus for the singlet case twist 2 is absent as it disappears in the large $N$
limit but twist 4 multiplets, which are necessary in the $p=2$ correlation function, 
decouple from the correlation functions for $p=3,4$).  

To obtain the anomalous scale dimensions in detail it is necessary to decompose 
both \Htwo, \Hthree, \Hfour\ and \Hztwo, \Hzthree, \Hzfour\ in terms of different 
representations, as in \Kexp, and then to expand each term in conformal partial waves.
The expressions \Hztwo, \Hzthree\ and \Hzfour\ require 
contributions with twist zero and above but the corresponding low twist operators
in long supermultiplets, for which there are no anomalous dimensions, 
are cancelled by semi-short multiplets which are required by the expansion
of $f_0(z,y)$. For $p=2$ and $p=3$ a detailed discussion is contained in \scft\ and
\ADHS\ (although some details are different the analysis is equivalent to the
the results that would be obtained by expanding $\H$ as given by \Hztwo\ and \Hzthree).

\newsec{Conclusion}

In this paper we have derived requirements arising from $\N=2$ and $\N=4$
superconformal symmetry for the four point correlation functions of BPS
operators. The derived conditions are clearly necessary but not manifestly
sufficient in that it is possible to imagine that there are further constraints
arising from superconformal transformations involving higher levels, although
we have no reason to suppose that there any such additional conditions. A related
question is whether the four point correlations functions of all descendant
operators are determined uniquely from the basic correlation function for
the superconformal primary BPS operators. In the simplest three point function case
the correlation function for various descendants, including the energy momentum
tensor three point function, was calculated by hand in \scft. In a superspace
formalism these questions are straightforward to address, the question of
uniqueness depends on whether there are any nilpotent superconformal invariants
which are formed from the anti-commuting $\theta$ coordinates. Nevertheless
it would be nice to show directly that the correlation function of all
descendant operators could in principle be obtained by the action of 
differential operators acting on the basic correlation function, subject
to the conditions for superconformal invariance derived here.

Another area of possible future investigation is whether the requirements 
of crossing symmetry and superconformal invariance might be extended further
using constraints arising from factorisation and the operator product
expansion, as in the classic bootstrap framework. In the above 
we showed how there were conditions on the single variable functions that
arise in the solution of superconformal identities which dictate that they
are essentially of free field form. Our arguments are restricted to the case
where all but one of the operators in the correlation function are identical.

Finally we may mention that the use of null vectors $t$ to conveniently
express arbitary rank traceless symmetric tensor fields in the form
$\vphi^{(p)}(x,t)$, homogeneous of degree $p$ in $t$, may for conformal
fields be also written in terms of homogeneous coordinates $\eta^A$ on the null
cone $\eta^2=0$ \Dirac\ such that $\phi{(p)}(\lambda \eta, t) ) =
\lambda^{-\Delta} \phi^{(p)} (\eta,t)$. For $\N=4$ both $\eta$ and $t$ 
are 6-vectors. The expansion in terms of harmonic polynomials as discussed
in appendix B has a direct analogue to the conformal partial wave expansion
which was explored in \CFT.

\bigskip
\noindent
{\bf Acknowledgements}
\medskip
We are particularly grateful to Francis Dolan for much help and collaboration
during various stages of this investigation. HO would also like to thank
Gleb Arutyunov and Emeri Sokatchev for many very useful conversations and emails and
for showing us details of their work at an early stage.
MN would like to thank PPARC, the Cambridge European Trust and the Isaac Newton
Trust for financial assistance.

\vfill\eject

\appendix{A}{Results for Null Vectors}

We discuss here some results for null vectors $t_r$ which are useful
in the text. For generality we allow $t$ to be $d$-dimensional. As
a consequence of \nul\ differentiation requires some care but for any
null vector $a$ we may define as usual
\eqn\diffo{
{\pr \over \pr t} (a{\cdot t})^n = n\, (a{\cdot t})^{n-1} a \, .
}
More generally for a set of null vectors $a_1,a_2,\dots, a_p $ we have
\eqn\diffm{\eqalign{
\! {\pr \over \pr t} \prod_{i=1}^p (a_i{\cdot t})^{n_i} = {}&
\sum_{i=1} ^p n_i\, (a_1{\cdot t})^{n_1} \dots (a_i{\cdot t})^{n_i-1}\dots
(a_p{\cdot t})^{n_p} a_i \cr
&{}- R \!\! \sum_{1\le i < j \le p} n_i n_j \, a_i{\cdot a_j} \, 
(a_1{\cdot t})^{n_1} \dots (a_i{\cdot t})^{n_i-1}\dots 
(a_j{\cdot t})^{n_j-1} \dots (a_p{\cdot t})^{n_p} t \, , \cr}
}
where
\eqn\defR{
R = {2\over 2N+d-4} \, , \qquad N = \sum_{i=1}^p n_i \, .
}
The right hand side of \diffm, with \defR, may be represented in the form
\eqn\tdiff{
\bigg ( {\pr\over \pr t } - t \, {1\over 2t{\cdot \pr} +d } \, \pr^2 \bigg )
\prod_{i=1}^p (a_i{\cdot t})^{n_i} \, ,
}
where the action of the derivatives is as usual, without regard to the 
constraint $t^2=0$. The resulting operator  is equivalent to a definition given in 
\Barg\ for an interior differential operator on the complex null cone.

{}From \diffm\ we may readily find
\eqn\Lap{
\bigg [ {\pr \over \pr t_r}, {\pr \over \pr t_s} \bigg ] 
\prod_{i=1}^p (a_i{\cdot t})^{n_i} = 0 \, , \quad
{\pr \over \pr t}{\cdot {\pr \over \pr t}} \prod_{i=1}^p (a_i{\cdot t})^{n_i}
= 0 \, , \quad t {\cdot {\pr \over \pr t}} \prod_{i=1}^p (a_i{\cdot t})^{n_i}
= N \, \prod_{i=1}^p (a_i{\cdot t})^{n_i} \, .
}
We also have
\eqn\dr{
\bigg [ {\pr \over \pr t_r}, \,  t_s \bigg ]
\prod_{i=1}^p (a_i{\cdot t})^{n_i} = \bigg ( \de_{rs} - {2\over 2N+d-2}
\, t_r {\pr \over \pr t_s} \bigg ) \prod_{i=1}^p (a_i{\cdot t})^{n_i} \, ,
}
which implies
\eqn\div{
{\pr \over \pr t_r} \bigg (  t_r \prod_{i=1}^p (a_i{\cdot t})^{n_i} \bigg )
= {(2N+d)(N+d-2)\over 2N+d-2} \,  \prod_{i=1}^p (a_i{\cdot t})^{n_i} \, .
}

Defining the generators of $SO(d)$ by 
\eqn\gend{
L_{rs} = t_r \pr_s - t_s \pr_r \, ,
}
then the above results give
\eqn\LLe{
L_{ru}L_{su} \prod_{i=1}^p (a_i{\cdot t})^{n_i} = - \big ( (N+d-3)\,
t_r \pr_s + (N-1)\, t_s \pr_r + N \, \de_{rs} \big ) 
\prod_{i=1}^p (a_i{\cdot t})^{n_i} \, ,
}
and 
\eqn\eiSO{
\half L_{rs} L_{rs} \prod_{i=1}^p (a_i{\cdot t})^{n_i} = - 
N (N+d-2) \prod_{i=1}^p (a_i{\cdot t})^{n_i}  \, ,
}
which reproduces the appropriate eigenvalue of the Casimir operator for the
representation formed by traceless rank $N$ tensors.

If $V_r(t)$ is homogeneous of degree $N$ then in general
\eqn\Vd{
V_r = {\hat V}_r + {1\over N+1}{\pr \over \pr t_r } \big ( t_s V_s \big ) \, ,
\qquad t_r {\hat V}_r = 0 \, ,
}
as used in \VU\ and \VUp. If $V_r$ also satisfies
\eqn\VD{
{\pr \over \pr t_r } V_s - {\pr \over \pr t_s } V_r = 0 \, , \qquad
{\pr \over \pr t_r } V_r = 0 \, ,
}
then, by contracting with $t_s$ and using \Lap, \dr, we easily see that
${\hat V}_r=0$.
As a further corollary if $V_r = \pr_s U_{rs}$, $U_{rs}=-U_{sr}$, 
$\pr_{[r} U_{su]} =0$ then $(N+1) V_r = \pr_r ( \half L_{su} U_{su})$ with
$L_{su}$ as in \gend. In general we have the decomposition
\eqn\Vdiv{\eqalign{
V_r = {}& {2N+d-4\over (2N+d-2)(N+d-3)}\, t_r \, \pr{\cdot V} \cr
&{} -{1\over (2N+d)(N+d-3)} \big ( (2N+d-2) \pr_s (t_r V_s - t_s V_r)
+ 2\, \pr_r (t{\cdot V}) \big ) \, . \cr}
}

\appendix{B}{Two Variable Harmonic Polynomials}

For the expansion of four point functions in terms of $R$-symmetry
representations we consider here the eigenfunctions
of the $SO(d)$ Casimir operator
\eqn\Ls{
L^2 = \half L_{rs}L_{rs} \, ,
}
where the generators  are
\eqn\gen{
L_{rs} = t_{1r} \pr_{1s} - t_{1s} \pr_{1r} +  
t_{2r} \pr_{2s} - t_{2s} \pr_{2r} \, ,
}
formed by homogeneous functions of the null vectors $t_1,t_2,t_3,t_4$.
Obviously $L_{rs} t_1{\cdot t_2} = 0$ and hence $L^2 (t_1 {\cdot t_2})^k
(t_3 {\cdot t_4})^l f(\lam,\mun) = (t_1 {\cdot t_2})^k (t_3 {\cdot t_4})^l
L^2 f(\lam,\mun)$, where $\lam,\mun$ are given by \deflu. 
We therefore first consider eigenfunctions which are polynomials in $\lam,\mun$
\eqn\polyY{
Y(\lam,\mun) = \sum_{t\ge 0} \sum_{q=0}^t  c_{t,q} \, \lam^{t-q} \mun^q \, ,
}
satisfying
\eqn\eigL{
L^2 Y(\lam,\mun) = - 2C Y(\lam,\mun) \, .
}

With the aid of the given in  appendix A we may easily calculate the 
action of $L^2$ on a monomial formed from $\lam,\mun$,
\eqn\Llu{
L^2( \lam^p \mun^q) = - 2\big ( (d-2)(p+q) + 4 pq\big ) \lam^p \mun^q
+ 2(1-\lam-\mun)
\big ( p^2 \lam^{p-1} \mun^q + q^2 \lam^p \mun^{q-1} \big ) \, , 
}
or 
\eqn\LDd{
\half L^2 \to \D_d = (1-\lam-\mun)\Big ( {\pr \over \pr \lam} \lam
{\pr \over \pr \lam} + {\pr \over \pr \mun} \mun {\pr \over \pr \mun} \Big )
- 4 \lam\mun{\pr^2 \over \pr \lam\pr\mun} - (d-2) 
\Big ( \lam {\pr \over \pr \lam} +  \mun {\pr \over \pr \mun} \Big ) \, .  
}
Alternatively $\D_d$ may be written in the form
\eqn\Dad{
\D_d = {1\over w} \, {\underline \pr}^T w {\underline G}\, {\underline \pr}\, ,
\qquad {\underline G} = \pmatrix { \lam(1-\lam - \mun) & - 2\lam\mun\cr
- 2\lam\mun & \mun(1-\lam - \mun)} \, , \quad {\underline \pr} =
\pmatrix { \pr_\lam\cr \pr_\mun} \, ,
}
where, with $\Lambda=(\sqrt \lam + \sqrt \mun + 1) (\sqrt \lam + \sqrt \mun - 1)
(\sqrt \lam - \sqrt \mun + 1)(\sqrt \lam - \sqrt \mun - 1)$ as in \conlm,
\eqn\defw{
w = \Lambda^{{1\over 2}(d-5)} \, .
}
In general, for a polynomial as in \polyY\ with $t_{\rm max} = n$, we must have that 
$c_{n,q}$ forms an eigenvector for an $(n{+1}) \times (n{+1})$ 
matrix $M_n$,
\eqn\eigc{
M_{n,pq} c_{n,q} = C c_{n,p} \, , \quad M_{n,pq} = \de_{p\, q}
\big ( n(n+d-2)+ 2 p (n-p) \big ) + \de_{p \, q{-1}} \, q^2 +
\de_{p \, q{+1}} (n-q)^2 \, .
}
The coefficients $c_{t,q}$ with $t<n$ may then be obtained by solving
recurrence relations. For given $n$ there are $n+1$ eigenvectors 
solving \eigc\ and the corresponding eigenfunctions are
\eqn\eig{
Y_{nm}(\lam,\mun) \, , \quad C_{nm} =  n(n+d-3) + m(m+1) \, , \quad
n=0,1,2, \dots , \  m=0,\dots n \, .
}
As a consequence of \Dad\ and \defw\ the polynomials are orthogonal for
$d>5$ with respest to integration over $\lam,\mun \ge 0, \, 
\sqrt \lam + \sqrt \mun \le 1$ with weight $w$ (for a general discussion
of such two variable orthogonal polynomials see \refs{\Koo,\ortho}).

The polynomials $Y_{nm}$ are also eigenfunctions for higher order Casimir
invariants. Letting
\eqn\defQ{ 
\quar \, L_{rs}L_{st}L_{tu}L_{ur} - \half \, L^2 L^2 + \quar (d-2)(d-3)\, L^2  \to \Q
}
then acting on any $Y(\lam,\mun)$ we may express $\Q$ in a form similar to \Dad,
\eqn\Qact{\eqalign{
\Q ={}& - {1\over \Lambda^{{1\over 2}(d-5)}} 
\pmatrix { \pr_\lam{\!}^2 &  \pr_\mun{\!}^2} \Lambda^{{1\over 2}(d-3)}
\pmatrix { \lam^2 & - \lam\mun\cr - \lam\mun & \mun^2}
\pmatrix { \pr_\lam{\!}^2 \cr  \pr_\mun{\!}^2} \cr
&{} + (d-3) \,  {1\over \Lambda^{{1\over 2}(d-5)}} 
\pmatrix { \pr_\lam &  \pr_\mun} \Lambda^{{1\over 2}(d-5)}
\pmatrix { 2\lam & \lam+\mun-1\cr \lam+\mun-1 & 2\mun} 
\pmatrix { \pr_\lam\cr \pr_\mun} \cr 
&{} + (d-2) \,  {1\over \Lambda^{{1\over 2}(d-5)}} \big ( \pr_\lam
\, \Lambda^{{1\over 2}(d-3)} \, \pr_\mun + 
\pr_\mun \,\Lambda^{{1\over 2}(d-3)}\, \pr_\lam \big ) \, . \cr}
}
The harmonic polynomials then satisfy
\eqn\Qeig{
\Q \, Y_{nm} = - (n-m)(n+m+1)(n+m+d-3)(n-m+d-4) \, Y_{nm} \, .
}

Using \Llu\ it is  straightforward to construct the first few eigenfunctions 
satisfying \eigL\ by hand. With an arbitrary normalisation, we find for $n=0,1,2,3$,
\eqn\Ynm{\eqalign{
Y_{00}(\lam,\mun) = {}& 1  \, , \cr
Y_{10}(\lam,\mun) = {}& \lam - \mun \, , \cr
Y_{11}(\lam,\mun) = {}& \lam + \mun - {\ts {2\over d}}\, , \cr
Y_{20}(\lam,\mun) = {}& \lam^2 + \mun^2 - 2\lam\mun
- {\ts {2\over d-2}}(\lam + \mun) + {\ts {2\over (d-2)(d-1)}} \, , \cr
Y_{21}(\lam,\mun) = {}& \lam^2 - \mun^2 - {\ts {4\over d+2}}
(\lam - \mun) \, , \cr
Y_{22}(\lam,\mun) = {}& \lam^2 + \mun^2 + 4\lam\mun 
- {\ts {8\over d+4}}(\lam + \mun) + {\ts {8\over (d+2)(d+4)}} \, ,\cr
Y_{30}(\lam,\mun) = {}& \lam^3 - 3\lam^2\mun + 3\lam\mun^2 - \mun^3
- {\ts {6\over d}}(\lam^2 - \mun^2) + {\ts {12\over d(d+1)}}(\lam - \mun)
 \, , \cr
Y_{31}(\lam,\mun) = {}& \lam^3 - \lam^2\mun - \lam\mun^2 + \mun^3
- {\ts {8(d-1)\over (d+4)(d-2)}}(\lam^2 + \mun^2) +
{\ts {8(d-6)\over (d+4)(d-2)}}\, \lam\mun \cr
&{}+{\ts {4(3d+2)\over (d+1)(d+4)(d-2)}}(\lam + \mun) - 
{\ts {8\over (d+1)(d+4)(d-2)}} \, , \cr
Y_{32}(\lam,\mun) = {}& \lam^3 + 3\lam^2\mun - 3\lam\mun^2 - \mun^3
- {\ts {12\over d+6}}(\lam^2 - \mun^2) + 
{\ts {24\over (d+4)(d+6)}}(\lam - \mun)
 \, , \cr
Y_{33}(\lam,\mun) = {}& \lam^3 +9\lam^2\mun  + 9 \lam\mun^2 + \mun^3
- {\ts {18\over d+8}}(\lam^2 + \mun^2)- {\ts {72\over d+8}} \,\lam\mun \cr
&{}+{\ts {72\over (d+6)(d+8)}}(\lam + \mun) 
- {\ts {48\over (d+4)(d+6)(d+8)}} \, . \cr}
}
Up to an overall normalisation for $d=6$ each term may be identified
with terms in the projection operators constructed in \ADHS\ where $Y_{nm}$
corresponds to the $SU(4)\simeq SO(6)$ representation with Dynkin labels
$[n-m,2m,n-m]$. For $m=n$ in \eig\ we have $c_{n,q} = {n \choose q}^2$ 
and the recurrence relations may be easily solved giving
\eqn\Ynn{
Y_{nn} (\lam,\mun) = A_n F_4 (-n, n+\half d -1 ; 1,1; \lam,\mun) \, ,
}
where $F_4$ is one of Appell's generalised hypergeometric functions\foot{
$$F_4(a,b;c,c';x,y) = \sum_{m,n} {(a)_{m+n} (b)_{m+n} \over
(c)_m (c')_n m! \, n!} \, x^m y^n \, .
$$} and $A_n$ is some overall constant.

To obtain more general forms (see \Vretare) we used the variables $\alpha,\bet$
defined in \defab. Acting on $Y(\lam,\mun)= \P(\alpha,\bet)=
\P(\bet,\alpha)$
\eqn\LD{
\half L^2 \P(\alpha,\bet) = \cD_d \P(\alpha,\bet) \, ,
}
where, using \Llu\ or \LDd, we now have
\eqn\Dop{
\cD_d = {\pr \over \pr \alpha} \alpha(1-\alpha) 
{\pr \over \pr \alpha} + {\pr \over \pr \bet} \bet(1-\bet)
{\pr \over \pr \bet} + (d-4) {1\over \alpha- \bet} \Big (
\alpha(1-\alpha) {\pr \over \pr \alpha} - \bet(1-\bet)
{\pr \over \pr \bet} \Big ) \, .
}
Corresponding to \eigL\ and \eig\ we have
\eqn\eigD{
\cD_d \P_{nm}(\alpha,\bet) = - \big ( n(n+d-3) + m(m+1) \big ) 
\P_{nm}(\alpha,\bet) \, ,
}
where $\P_{nm}(\alpha,\bet)$ are generalised symmetric Jacobi polynomials.
For particular $d$ simplified formulae may be found in terms of well
know single variable Legendre polynomials $P_n$.  When $d=4$ it
is clear from \Dop\ that $\cD_4$ is just the sum of two independent 
Legendre differential operators  so that
\eqn\eigfour{
\P_{nm}(\alpha,\bet) = \half \big ( P_n(y) P_m(\yz) +
P_m(y) P_n(\yz) \big ) \, , \quad n\ge m \, , 
}
with $y,\yz$ defined in \yy.
For $d=6$ we may use the result
\eqn\DF{
\cD_6  {1\over \alpha -\bet } = {1\over \alpha -\bet } ( \cD_4 + 2 )  \, ,
}
to see that we can take the eigenfunctions to be of the form
\eqn\eigsix{
\P_{nm}(\alpha,\bet) =  p_{n{+1}m}(y,\yz) \, , \quad n\ge m \, , 
}
where
\eqn\defpnm{
p_{nm}(y,\yz) = - p_{mn}(y,\yz) = 
{ P_{n}(y) P_m(\yz) - P_m(y) P_{n} (\yz) \over y - \yz} \, .
}

It is also of interest to consider $d=8$ when we take
\eqn\PFe{
\P (\alpha,\bet) = {F(\alpha,\bet) \over (\alpha-\bet)^2}\, ,
}
and  the eigenvalue equation becomes
\eqnn\DFe
$$\eqalignno{
\cD_6 F(\alpha,\bet) &{} - {2\over (\alpha-\bet)^2} \Big ( \alpha(1-\alpha)
{\pr \over \pr \alpha} \big ( (\alpha-\bet)   F(\alpha,\bet) \big )
- \bet( 1- \bet ) {\pr \over \pr \bet} \big ( (\alpha-\bet) 
F(\alpha,\bet) \big ) \Big ) \cr
&{} = - (C+4) F(\alpha,\bet) \, . &  \DFe \cr}
$$
If we assume
\eqn\FPnm{
F(\alpha,\bet) = \sum_{n,m} a_{nm}\,  p_{nm}(y,\yz)\, ,
}
and use, from standard identities for Legendre polynomials,
\eqnn\recur
$$\eqalignno{ 
&{1\over y-\yz} \Big ( (1-y^2){\pr \over \pr y} - (1-\yz^2)
{\pr \over \pr \yz}\Big )\big ( (y-\yz) p_{nm}(y,\yz)\big ) \cr 
&\quad {} = {m(m+1)\over 2m+1} 
\big ( p_{nm{+1}}(y,\yz)- p_{nm{-1}}(y,\yz)\big ) -
{n(n+1)\over 2n+1} \big ( p_{n{+1}m}(y,\yz) - p_{n{-1}m}(y,\yz) \big )\cr 
&(y-\yz)  p_{nm}(y,\yz) = {1\over 2n+1} \big ( (n+1) p_{n{+1}m}(y,\yz) + n\,
p_{n{-1}m}(y,\yz) \big ) \cr 
& \qquad \qquad \qquad \qquad {}- {1\over 2m+1}\big ( (m+1) p_{nm{+1}}(y,\yz)
 + m \, p_{nm{-1}}(y,\yz) \big ) \, , & \recur \cr}
$$
then we may set up recurrence relations for $a_{nm}$ which for the
appropriate value of $C$ have just four terms.  For $C=n(n+1)+m(m+1)-6$
\FPnm\  gives a solution
\eqn\defq{\eqalign{
q_{nm}(y,\yz) = {1\over (y-\yz)^2} \bigg \{ & {n+1\over 2n+1} (n+m)(n-m-1)
p_{n{+1}m}(y,\yz) \cr &{} + {n\over 2n+1} (n+m+2)(n-m+1) p_{n{-1}m}(y,\yz) \cr
&{}- {m+1\over 2m+1} (n+m)(n-m+1) p_{nm{+1}}(y,\yz) \cr &
{}- {m\over 2m+1} (n+m+2)(n-m-1) p_{nm{-1}}(y,\yz) \bigg \} \, , \cr}
}
where $q_{nm}(y,\yz) = - q_{mn}(y,\yz), \ q_{nn}(y,\yz) = 
q_{n{+1}\, n}(y,\yz)=0$. Hence we can take
\eqn\eigeight{
\P_{nm}(\alpha,\bet) =  q_{n{+2}\, m}(y,\yz) \, . 
}

The above results for harmonic polynomials in $\lam,\mun$ are relevant
for discussing four point functions when each field belongs to the same
$SO(d)$ representation. For the more general case we also consider instead
of \eigL,
\eqn\eigLab{
L^2 \big ( (t_1{\cdot t_4})^a (t_2{\cdot t_4})^b 
Y^{(a,b)}(\lam,\mun)\big ) = -2C 
\big ( (t_1{\cdot t_4})^a (t_2{\cdot t_4})^b Y^{(a,b)}(\lam,\mun) \big )\, ,
}
where now the action of $L^2$ is determined by
\eqn\Lab{ \eqalign{
L^2 & \big ( (t_1{\cdot t_4})^a (t_2{\cdot t_4})^b  \lam^p \mun^q \big )\cr
& {}= (t_1{\cdot t_4})^a (t_2{\cdot t_4})^b \Big ( 
\big ( 2\D_d -(a+b)(a+b+d-2) - 4a p - 4bq \big ) ( \lam^p \mun^q ) \cr
&\qquad \qquad \qquad \qquad \ {}+ 2(1-\lam - \mun) \big ( bp \, 
\lam^{p-1} \mun^q  + a q \, \lam^p \mun^{q-1} \big ) \Big ) \, , \cr}
}
or
\eqn\defDab{
\half L^2 \big ( (t_1{\cdot t_4})^a (t_2{\cdot t_4})^b f(\lam,\mun) \big )
= (t_1{\cdot t_4})^a (t_2{\cdot t_4})^b \big ( \D^{(a,b)}_d
 - \half (a+b) (a+ b + d-2) \big ) f (\lam,\mun) \, ,
}
where
\eqn\LDa{
\D^{(a,b)}_d = \D_d + (1-\lam-\mun) \Big ( a \, {\pr \over \pr \mun} +
b\,{\pr \over \pr \lam} \Big ) -2a \, \lam {\pr \over \pr \lam} - 2b \,
\mun {\pr \over \pr \mun} \, .
}
This may also be written in the form \Dad\ with 
$w = \lam^b \mun^a \Lambda^{{1\over 2}(d-5)}$.
The possible eigenvalues for polynomial eigenfunctions with maximum 
power $p+q=n$ are then determined by the matrix
\eqnn\matM
$$\eqalignno{
M_{n,pq} = {}& \de_{p\, q}
\big ( n(n+d-2+a+b)+ \half (a+b)(a+b+d-2) + 2 p (n-p) + a(n-p) + b p\big ) \cr
&{} + \de_{p \, q{-1}} \, q(q+a)  + \de_{p \, q{+1}} (n-q)(n-q+b) \, . 
& \matM \cr}
$$
The eigenfunctions $Y^{(a,b)}_{nm}(\lam,\mun)$ for $m=0,1,\dots, n$
then have eigenvalues
\eqn\eigab{
C_{nm} =  \big ( n + \half (a+b)\big )\big (n+ \half (a+b) + d-3 \big ) + 
\big (m + \half (a+b)\big )\big (m + \half (a+b) + 1 \big ) \, .
}
For $d=6$ $Y^{(a,b)}_{nm}$ corresponds to the representation
$[n-m,a+b+2m,n-m]$. The simplest non trivial examples are
\eqn\Yab{\eqalign{
Y^{(a,b)}_{10}(\lam,\mun) = {}& \lam - \mun + {a-b \over a+b+d-2} \,, \cr
Y^{(a,b)}_{11}(\lam,\mun) = {}& {1\over b+1} \, \lam +
{1\over a+1} \, \mun - { 1\over a+b+\half d} \, . \cr}
}
Corresponding to \Ynn\ we have in general
\eqn\Ynnab{
Y^{(a,b)}_{nn} (\lam,\mun) = A_n 
F_4 (-n, n+ a+b+ \half d -1 ;b+1,a+1; \lam,\mun) \, .
}

Again more explicit results can be obtained by using the variables
$\alpha,\bet$. In \LDa\ the differential operator now becomes
\eqn\Dopab{
\cD^{(a,b)}_d = \cD_d^{\vphantom g} - \big (a\, \alpha - b(1-\alpha)\big ) 
{\pr\over \pr \alpha}- \big (a \, \bet - b(1-\bet )\big ) 
{\pr\over \pr \bet } \, ,
}
with $\cD_d$ given in \Dop.
Denoting the eigenfunctions of $\cD^{(a,b)}_d$ by $\P^{(a,b)}_{nm}
(\alpha,\bet)$ then previous results for $d=4,6$ for the eigenfunctions 
can be extended by using Jacobi polynomials $P_n^{(a,b)}$. For $d=4$
$\cD^{(a,b)}_4 = D^{(a,b)}_\alpha + D^{(a,b)}_\bet$ where $D^{(a,b)}$
is the ordinary differential operator defined by
\eqn\Dthreeab{
D^{(a,b)}_\alpha = {\d \over \d \alpha} \alpha(1-\alpha) {\d \over \d \alpha}
- a \, \alpha {\d \over \d \alpha} + b (1-\alpha) {\d \over \d \alpha} \, .
}
The eigenfunctions of $D^{(a,b)}_\alpha$ are just
$P^{(a,b)}_n(y)$, where $ y = 2\alpha-1$ and the eigenvalues are
$-n(n+a+b+1)$.
For $d=6$ the generalisation of \eigsix\ and \defpnm\ is then
\eqn\eigsixab{
\P^{(a,b)}_{nm}(\alpha,\bet) =  
{ P^{(a,b)}_{n+1}(y) P^{(a,b)}_m(\yz) - P^{(a,b)}_m(y) P^{(a,b)}_{n+1} (\yz) 
\over y - \yz} \, .
}

When $d=3$ the above results need to be considered separately since 
$\lam, \mun$ are not independent and satisfy the constraint \conlm.
The eigenfunctions $Y^{(a,b)}_{nm}(\lam,\mun)$ are also restricted
since $Y^{(a,b)}_{nm}(\lam,\mun) = 0$ for $m<n-1$ as a consequence
of \conlm. To obtain eigenfunctions of $L^2$ in general we make use
of the solution \defa\ which amounts to setting $\alpha= \bet$ in the
above, so that we are restricted just to single variable functions. 
Instead of \defDab\ we have
\eqn\Lab{ 
L^2 \big ( (t_1{\cdot t_4})^a (t_2{\cdot t_4})^b f(\alpha) \big )
= (t_1{\cdot t_4})^a (t_2{\cdot t_4})^b 
\big ( D^{(2a,2b)}_\alpha - (a+b)(a+b+1) \big )  f(\alpha) \, ,
}
using the definition \Dthreeab. In consequence
\eqn\Leig{
L^2 \big ( (t_1{\cdot t_4})^a (t_2{\cdot t_4})^b P^{(2a,2b)}_n(y) \big ) 
= - (n+a+b)(n+a+b+1) \, (t_1{\cdot t_4})^a (t_2{\cdot t_4})^b 
P^{(2a,2b)}_n(y) \, ,
}
corresponding to the $(n{+a}{+b})$-representation for $SU(2)\simeq SO(3)$. 
Hence for $d=3$ we may then take
\eqn\eigthreeab{
\P^{(a,b)}_{nn}(\alpha,\alpha) = P^{(2a,2b)}_{2n}( y ) \, , \qquad
\P^{(a,b)}_{n\, n{-1}}(\alpha,\alpha) = P^{(2a,2b)}_{2n-1}( y ) \, ,
}
with $\P^{(a,b)}_{nm}(\alpha,\alpha)=0$ for $m<n-1$.

For $d=3$ there are also eigenfunctions involving cross products. To
consider these we first define
\eqn\defT{
T_1 = t_1 \, {\cdot \, t_3 \times t_4} \, (t_1{\cdot t_4})^{a-1} 
(t_2{\cdot t_4})^b \, , \quad
T_2 = t_2 \, {\cdot \, t_3 \times t_4} \, (t_1{\cdot t_4})^{a}
(t_2{\cdot t_4})^{b-1} \, ,
}
and consider eigenfunctions of the form $T_1 f_1(\alpha) + T_2
f_2(\alpha)$. The action of $L^2$ on such functions is given by
\eqn\Lcross{\eqalign{
\big (L^2 &{} + (a+b-1)(a+b) \big )\, ( T_1 f_1 + T_2 f_2 ) \cr
&{}= \pmatrix { T_1 & T_2} \pmatrix { D^{(2a-1,2b+1)} - 2a & -2a \cr
- 2b &  D^{(2a+1,2b-1)} - 2b} \pmatrix {f_1 \cr f_2} \, . \cr}
}
However for $t_i$ three dimensional null vectors the basis given
by \defT\ is not independent since we have from \idthree
\eqn\TTid{
T_1 (1-\alpha ) + T_2 \alpha = 0 \, ,
}
so that $f_1, f_2$ are not unique. If we use this freedom to set $f_2=0$ 
the eigenvalue equation for $L^2$ reduces to
\eqn\eigf{\eqalign{
\Big ( &  D^{(2a-1,2b+1)} -2a + 2b \, {1-\alpha \over \alpha} - 
(a+b-1)(a+b) \Big ) f_1 \cr
&{}= {1\over \alpha} \Big (  D^{(2a-1,2b-1)} - (a+b-1)(a+b) \Big )
(\alpha f_1) = - C f_1 \, ,\cr}
}
which has solutions proportional to Jacobi polynomials,
\eqn\solcr{
f_1(\alpha) = {1\over \alpha}\, P_n^{(2a-1,2b-1)}(y) \, , \qquad
C = (n+a+b-1)(n+a+b) \, .
}
For $n\ge 1$ the apparent singularity for $\alpha \to 0$ may be removed
by using \TTid\ to give an appropriate non zero  $f_2$. 
The eigenfunctions for the solution in \solcr\  correspond to the $SU(2)$
$(n{+a}{+b}{-1})$-representation. Alternatively we may set $f_1=0$
and obtain the corresponding equation
\eqn\eigftwo{
{1\over 1-\alpha} \Big (  D^{(2a-1,2b-1)} - (a+b-1)(a+b) \Big )
\big ((1-\alpha) f_2 \big ) = - C f_2 \, .
}

\appendix{C}{Calculation of Differential Operators}

A non trivial aspect in the derivation of the superconformal identities
is the determination of the differential operators \DDD\ which appear in
\crit. To sketch how these were obtained we first obtain, for any
dimension $d$ and arbitrary $f(\lam,\mun)$,
\eqnn\Ld
$$\eqalignno{
\half (k + & a + \half d - 2)\, L_{2[rs}\pr_{1u]} 
\big ( (t_1{\cdot t_2})^k 
(t_3{\cdot t_4})^l (t_1{\cdot t_4})^a (t_2{\cdot t_4})^b f \big ) \cr
&{}= - (t_1{\cdot t_2})^{k-2} (t_3{\cdot t_4})^{l-1} 
(t_1{\cdot t_4})^a (t_2{\cdot t_4})^b 
\big ( t_{1[r}t_{2s}^{\vphantom g}t_{3u]} \, t_2{\cdot t_4} \, \D_1 
+  t_{1[r}t_{4s}^{\vphantom g}t_{2u]} \, t_2{\cdot t_3}\, \D_2  \cr
&\qquad\qquad\qquad\qquad {}+ (k + a + \half d - 2) \, 
t_{2[r}t_{3s}^{\vphantom g}t_{4u]} \, 
t_1{\cdot t_2}\, ( \D_\lam - \D_\mun ) \big ) f \, , & \Ld \cr}
$$
where
\eqn\defDD{\eqalign{
\D_1 = {}& {\pr \over \pr \lam} \D_\mun +
\Big (\lam {\pr \over \pr \lam} + \mun {\pr \over \pr \mun} + 1 - k \Big )
\big (\D_\lam - \D_\mun \big ) \, , \cr
\D_2 = {}& - \Big ( {\pr \over \pr \mun}+ {a\over \mun} \Big ) \D_\lam
+ \Big (\lam {\pr \over \pr \lam} + \mun {\pr \over \pr \mun} +1 - k
\Big ) \big (\D_\lam - \D_\mun \big ) \, , \cr}
}
for
\eqn\DD{\eqalign{
\D_\lam = {}& \lam(1-\lam) {\pr^2 \over \pr \lam^2} - \mun^2
{\pr^2 \over \pr \mun^2} - 2\lam\mun\, {\pr^2 \over \pr \lam\pr\mun}\cr
&{}+ (b+1) {\pr \over \pr \lam} -  (a+b+\half d) 
\Big (\lam {\pr \over \pr \lam} + \mun {\pr \over \pr \mun} \Big )
+ k ( k+a+b+\half d -1 ) \, , \cr
\D_\mun = {}& \mun(1-\mun) {\pr^2 \over \pr \mun^2} - \lam^2
{\pr^2 \over \pr \lam^2} - 2\lam\mun \,{\pr^2 \over\pr\lam\pr\mun}\cr
&{}+ (a+1) {\pr \over \pr \mun} -  (a+b+\half d)
\Big (\lam{\pr \over \pr \lam} + \mun {\pr \over \pr \mun} \Big )
+ k ( k+a+b+\half d -1 ) \, . \cr}
}
In terms of \LDa\ and \LDd\ we have
\eqn\DDD{
\Delta^{(a,b)}_d \equiv \D_\lam + \D_\mun + ( \lam - \mun ) ( \D_\lam - \D_\mun ) 
= \D^{(a,b)}_d +  2k ( k+a+b+\half d -1 ) \, .
}
The operators in \defDD\ satisfy the identity
\eqn\simD{\eqalign{
& 2(\lam-\mun + 1 ) \D_1 + 2(\mun -\lam + 1 )\D_2 \cr
&\ = D_2 \, \Delta^{(a,b)}_d - \Lambda 
\bigg ( {\pr \over \pr \lam} + {\pr \over \pr \mun} + {a \over \mun} \bigg )
\big (\D_\lam - \D_\mun \big ) - (2k+2a-1)(\D_\lam - \D_\mun ) \, , \cr}
}
with $\Lambda$ as in \conlm\ and, as well as \DDD, defining
\eqn\Dtwo{
D_2 = ( \lam-\mun+1){\pr \over \pr \lam} + ( \lam-\mun - 1)\Big (
{\pr \over \pr \mun} + {a \over \mun} \Big ) \, .
}

When $d=3$, and $\Lambda =0$,  this result with \idthree\ leads to 
the simplified form for \Ld
\eqn\Dlth{\eqalign{
& 4(k+a-\half) \,{\pr \over \pr t_1}\cdot L_2  
\big ((t_1{\cdot t_2})^k (t_3{\cdot t_4})^l
(t_1{\cdot t_4})^a (t_2{\cdot t_4})^b {\hat f} \, \big ) \cr
&{}=  - t_2 \, {\cdot \, t_3 \times t_4} \,
(t_1{\cdot t_2})^{k-1} (t_3{\cdot t_4})^{l-1}(t_1{\cdot t_4})^a 
(t_2{\cdot t_4})^b  \,
\hD_2 \big ( D^{(2a,2b)} +2k(2k+2a+2b+1) \big ) {\hat f} \, , \cr}
}
letting $f(\lam,\mun) = {\hat f}(\alpha)$ and $D_2 \to \hD_2$ for
\eqn\Dtw{
\hD_2 = {\d \over \d \alpha} - {2a\over 1-\alpha} \, .
}
{}From \Lab\ the operator $D^{(2a,2b)} +2k(2k+2a+2b+1)$ acting on 
$ {\hat f}(\alpha)$ corresponds to  $L^2 + (2k+a+b)(2k+a+b+1)$. 
We may also note that
\eqn\Dinv{
\hD_2 D^{(2a,2b)} = {1\over 1-\alpha} \, D^{(2a-1,2b+1)} (1-\alpha) \hD_2 \, ,
}
and in \Dlth\ from \Lcross
\eqnn\LDD
$$\eqalignno{
& t_2 \, {\cdot \, t_3 \times t_4} \,(t_1{\cdot t_2})^{k-1} 
(t_3{\cdot t_4})^{l-1}(t_1{\cdot t_4})^a (t_2{\cdot t_4})^b \cr
\noalign{\vskip -4pt}
& \ {}\times {1\over 1- \alpha} \big (  D^{(2a-1,2b+1)} 
+2k(2k+2a+2b+1) \big ) \big ((1-\alpha) f \big ) & \LDD \cr
&{} = \big ( L^2 + (2k+a+b)(2k+a+b+1) \big )
t_2 \, {\cdot \, t_3 \times t_4} \,(t_1{\cdot t_2})^{k-1} 
(t_3{\cdot t_4})^{l-1}(t_1{\cdot t_4})^a (t_2{\cdot t_4})^b \, f \, . \cr}
$$

The equivalent results to \Dlth\ for $L_2 \to L_3$ and $L_2 \to L_4$
can be found by using the permutations $2\to 3 \to 4 \to 2$, along
with $a \to a'= k-l, \ b\to -a, \ k \to a+l$, ${l \to a+b+l}$ and
$\alpha \to \alpha' = -(1-\alpha)/\alpha$, and also $2\to 4 \to 3 \to 2$, 
along with in this case $a \to a''= -b, \, b\to l-k, \, k \to a+b+l, \, 
l \to b+k$ and
$\alpha \to \alpha'' = 1/(1-\alpha)$. From  \Dlth\ we then find
\eqn\Dltf{\eqalign{
\hD_3 = {}& \alpha^{2(a+l-1)}(1-\alpha)^{-2a}\hD'{}_{\!2} \,
\alpha^{-2(a+l)}(1-\alpha)^{2a} = {\d \over \d \alpha} + {2a\over 1-\alpha}
- {2(k+a) \over \alpha} \, , \cr
\hD_4 = {}& \alpha^{-2b}(1-\alpha)^{2(b+k-1)}\hD''{}_{\!\!2}  \,
\alpha^{2b}(1-\alpha)^{-2(b+k)} = {\d \over \d \alpha} + {2k\over 1-\alpha}\, . 
\cr}
}
Together with \Dtw, \Dltf\ is equivalent to \DDD.

For the analysis of the $\N=4$ superconformal identities a particular solution of
the constraints \consa\ is obtained by expressing $T_i$ in terms of scalar 
functions $Y_i(u,v;t)$ 
\eqn\TY{
T_i = - \, \bga{\cdot {\pr \over \pr t_1}} \, Y_i 
\ \gamma {\cdot {\overleftarrow{\pr \over \pr t_i}}}  \, . 
}
\TV\ and \VUp\ then give
\eqn\solY{\eqalign{
U_i ={}& \big ( L_{1,rs} L_{i,rs} + p_1 p_i \big ) Y_i \, , \qquad
W_{i,rsu} = 3 \big ( \pr_{1,[r}  L_{i,su]} Y_i \big )_{\rm sd} \, , \cr
\hV_{i,r} = {}& \pr_{1s} L_{i,rs} Y_i - {1\over p_1}\, \pr_{1r}\big (
\half L_{1,su} L_{i,su} Y_i \big ) \, . \cr}
}
Writing
\eqn\exYp{
Y_i(u,v;t) = \big ( t_1 {\cdot t}_4 \big )^{p_1-E} \big ( t_2 {\cdot t}_4
\big )^{p_2-E} \big ( t_1 {\cdot t}_2 \big )^{E}
\big ( t_3 {\cdot t}_4 \big )^{p_3} \, \Y_i(u,v;\lam,\mun) \, ,
}
then for $i=2$, using \Ld\ with $k=E, \, k+a=p_1, \, k+b=p_2$, we find
\eqn\Utwo{
\U_2 = \Delta^{(p_1-E,p_2-E)}_6 \Y_2 \, , \qquad 
\W_2 = 6 (\D_\lam - \D_\mun ) \Y_2 \, .
}
$\A_2$ and $\B_2$ are then given by \Ld\ and \defDD\ in terms of $\U_2, \W_2$ in 
accord with \relABC. The other results may be obtained by cyclic permutations. For
$2\to 3 \to 4 \to 2$, when $\lam \to \mun/\lam, \, \mun \to 1/\lam$ and $E\to
E - p_4-p_2$, then $\U_2 \to \mun^{p_1-E} \lam^{p_2-p_4-E} \U_3$, 
$\W_2 \to \mun^{p_1-E} \lam^{p_2-p_4-E+1} \W_3$,
$\A_2 \to \mun^{p_1-E} \lam^{p_2-p_4-E+2} \C_3$ and
$\B_2 \to \mun^{p_1-E} \lam^{p_2-p_4-E+2} \A_3$. For $2\to 4 \to 3 \to 2$, so that
$\lam \to 1/\mun, \, \mun \to \lam/\mun$ and $E\to - E +p_1+p_2$, in this case 
$\U_2 \to \mun^{-p_2}\, \lam^{p_2 -E} \U_4$, $\W_2 \to \mun^{1-p_2}\, \lam^{p_2 -E}\W_4$,
$\A_2 \to \mun^{2-p_2}\, \lam^{p_2 -E} \B_4$ and
$\B_2 \to \mun^{2-p_2}\, \lam^{p_2 -E} \C_4$.

However the representation \TY\ is not valid in general since it excludes
contributions involving the $\vep$-tensor. Nevertheless equivalent results
may be obtained by use of \WUp. With the expansion
\eqn\UVW{\eqalign{
& t_{2[r} \pr_{1s]} U_2 + p_1 \, t_{2[r} \hV_{2,s]} + p_1 \, W_{2,rsu} t_{2u} \cr
&{} = \big ( t_1 {\cdot t}_4 \big )^{p_1-E-1}
\big ( t_2 {\cdot t}_4 \big )^{p_2-E} \big ( t_1 {\cdot t}_2 \big )^{E-1}
\big ( t_3 {\cdot t}_4 \big )^{p_3-1} \cr
& \quad {} \times \bigg ( t_{2[r} t_{3s]} \, t_1 {\cdot t}_4 \, t_2 {\cdot t}_4 
\Big ( \pr_\lam \U_2 + p_1 \, \J_2 - {\ts {1\over 6}} p_1 ( \A_2 + \W_2 ) \Big ) \cr
& \qquad\quad {}+  t_{2[r} t_{4s]} \, t_1 {\cdot t}_4 \, t_2 {\cdot t}_3 
\Big ( \pr'{\!}_\mun \U_2 - p_1 {1\over \mun} ( \I_2 + \lam  \J_2) 
+ {\ts {1\over 6}} p_1 ( \B_2 + \W_2 ) \Big ) \cr
& \quad\qquad {}+ t_{1[r} t_{2s]} \, t_2 {\cdot t}_4 \, t_3 {\cdot t}_4
\Big ( {\tau \over p_1+1} \big ( ( \pr_\lam + \pr'{\!}_\mun ) ( E - \lam \pr_\lam
- \mun \pr_\mun ) + \pr_\lam \pr'{\!}_\mun \big ) \U_2 \cr
\noalign{\vskip-8pt}
&\hskip 7.4cm {}  - p_1 \, \V_2 
- {\ts {1\over 6}} p_1 \tau  ( \A_2 - \B_2 ) \Big )  \bigg ) \, ,  \cr }
}
where $\pr'{\!}_\mun = \pr_\mun + (p_1-E)/\mun$, then \WUp\ requires
\eqn\UUU{\eqalign{
6 \, \pr_\lam \U_2 = {}& - 6 p_1 \, \J_2 + p_1 ( \A_2 + \W_2 ) = 2 (p_1+1)\A_2
- (\O_\lam - p_1 ) \W_2 \, , \cr
6 \, \pr'{\!}_\mun \U_2 = {}& 6 p_1 {1\over \mun} ( \I_2 + \lam  \J_2) -
p_1 ( \B_2 + \W_2 ) = - 2 (p_1+1)\B_2 + (\O_\lam - p_1 ) \W_2 \, , \cr}
}
using \solIJ\ for $i=2$, which gives the first two equations in \relABC.
The remaining results in \relABC\ can be obtained by using permutations. 
In addition with \VABC\ we also obtain
\eqn\req{\eqalign{
(p_1+1) \big ( ( \O_\lam -p_1 +1 )& \B_2 - ( \O_\mun -p_1 +1 ) \A_2 \big ) \cr
= {}& {6} \big ( ( E -1 - \lam \pr_\lam - \mun \pr_\mun )
( \pr_\lam + \pr'{\!}_\mun ) + \pr_\lam \pr'{\!}_\mun \big ) \U_2 \cr
= {}& - {3} \big ( ( \O_\lam -p_1 +1  ) \pr'{\!}_\mun +
 ( \O_\mun -p_1 +1  ) \pr_\lam \big ) \U_2 \, . \cr} 
}
It is then straightforward to see that \req\ follows from \UUU\ using
$[\O_\lam, \O_\mun] = \O_\lam - \O_\mun$.

\appendix{D}{Non Unitary Semi-short Representations}

In section 5 the analysis of the operator product expansion in general potentially
required contributions below the unitarity threshold on the scale dimension
$\Delta$. We show here how such truncations of the full representation
space arise for the superconformal algebra $PSU(2,2|4)$, following the
approach in \short\foot{This reproduces an analysis in \phd.}

The essential results are found by considering the chiral subalgebra $SU(2|4)$
(although no hermeticity conditions are imposed)
which has generators $Q^i{}_{\! \alpha}, \,S_i{}^{\!\alpha}$, $\alpha=1,2, \,
i = 1,\dots 4$, where
\eqn\algQS{
\big \{ Q^i{}_{\! \alpha} ,  S_j{}^{\!\beta} \big \} =   4 \big (
\de^i{}_{\! j} ( M_\alpha{}^{\! \beta} + \half  \,
\de_\alpha{}^{\! \beta} \hD ) - \de_\alpha{}^{\! \beta} R^i{}_{\! j} \big )   \, ,
}
as well as $\{ Q^i{}_{\! \alpha} , Q^j{}_{\! \beta} \} =
\{ S_i{}^{\!\alpha} , S_j{}^{\!\beta} \} = 0$. 
In \algQS\ $M_\alpha{}^{\! \beta}$ are generators of $SU(2)$ and
$R^i{}_{\! j}$ are generators of $SU(4)$,
$ \sum_i R^i{}_{\! i}=0$, with standard commutation relations.
$\hD$ is the dilation operator, with eigenvalues the scale dimension.
The commutators with $Q^i{}_{\! \alpha}$ and $S_i{}^{\!\alpha}$ are then
\eqn\MQS{\eqalign{
[M_\alpha{}^{\! \beta} , Q^i{}_{\! \gamma} ] = {}& \de_\gamma{}^{\! \beta}
Q^i{}_{\! \alpha} - \half \de_\alpha{}^{\! \beta} Q^i{}_{\! \gamma} \, , 
\qquad [M_\alpha{}^{\! \beta} ,  S_i{}^{\!\gamma} ] = - \de_\alpha{}^{\! \gamma}
S_i{}^{\!\beta} + \half  \de_\alpha{}^{\! \beta}  S_i{}^{\!\gamma} \, , \cr
[ R^i{}_{\! j} ,  Q^k{}_{\! \alpha} ] = {}&\de^k{}_{\! j}  Q^i{}_{\! \alpha}
- \quar  \de^i{}_{\! j}  Q^k{}_{\! \alpha} \, , \qquad \ \
[ R^i{}_{\! j} ,  S_k{}^{\!\alpha} ] =  - \de^i{}_{\! k} S_j{}^{\!\alpha}
+ \quar  \de^i{}_{\! j}   S_k{}^{\!\alpha} \, , \cr
[ \hD , Q^i{}_{\! \alpha}] ={}& \half  Q^i{}_{\! \alpha} \, , \hskip 3.05cm
[ \hD , S_i{}^{\!\alpha} ]  = - \half S_i{}^{\!\alpha} \, . \cr}
}
In terms of the usual $J_3, J_\pm$
\eqn\MJ{
\big [M_\alpha{}^{\! \beta}\big ] = \pmatrix{ J_3 & J_+ \cr J_- & -J_3 } \, ,
}
and it clear then that $( Q^i{}_{\! 1}, Q^i{}_{\! 2})$ and
$(S_i{}^{\!2},-S_i{}^{\! 1})$ form $j=\half$ doublets.
In terms of a standard Chevalley basis 
$E_r{}^{\!\pm}, \, H_r$, $r=1,2,3$, where $H_r{\!}^\dagger
=H_r, \, E_r{}^{\!+}{}^\dagger = E_r{}^{\!-}$ with commutators $[H_r,H_s]=0, \, 
[ E_r{}^{\! +}, E_s{}^{\! -} ] = \de_{rs} H_s, \, 
[H_r ,  E_s{}^{\! \pm}] = \pm K_{sr}  E_s{}^{\! \pm}$, for $[K_{rs}]$ the
$SU(4)$ Cartan matrix, 
\eqn\Car{
[K_{rs}] = \pmatrix{2& -1 & 0\cr -1& 2 &-1\cr 0&-1&2} \, ,
}
then we may take $R^1{}_{\! 2} = E_1{}^{\! +}, \,
R^2{}_{\! 3} = E_2{}^{\! +}, \, R^3{}_{\! 4} = E_3{}^{\! +}$ and 
$R^i{}_{\! i} = \quar (3H_1 + 2H_2 + H_3) - \sum_{r=1}^{i-1} H_r$.

For $SU(4)\otimes SU(2)$ highest weight states
$|p_1,p_2,p_3;j\rangle^{\rm hw} \equiv |p;j\rangle^{\rm hw}$ we have
\eqn\hw{\eqalign{
H_r |p;j\rangle^{\rm hw} = {}& p_r |p;j\rangle^{\rm hw} \, , \ \
J_3 |p;j\rangle^{\rm hw} = j |p;j\rangle^{\rm hw} \, , \ \
J_+ |p;j\rangle^{\rm hw} = E_r{}^{\!+} |p;j\rangle^{\rm hw} =0 \, , \cr}
}
from which states defining a representation with Dynkin labels $[p_1,p_2,p_3]j$
are constructed by the action of $E_r{}^{\! -},J_-$.
The representations of $SU(2|4)$ may then be formed from a highest weight state
which is also superconformal primary,
\eqn\DSw{
\hD |p;j\rangle^{\rm hw} = \Delta |p;j\rangle^{\rm hw} \, , \qquad
S_i{}^{\!\alpha} |p;j\rangle^{\rm hw} = 0 \, . 
}
The states of a generic supermultiplet, labelled
by $a^\Delta_{[p_1,p_2,p_3]j}$, are obtained by the action of the
supercharges, giving 
$\prod_{i,\alpha} \big ( Q^i{}_{\! \alpha} \big )^{n_{i\alpha}}
|p;j\rangle^{\rm hw}$ with $n_{i\alpha}=0,1$, together with the lowering
operators $E_r{}^{\!-}$. The possible $SU(4)\otimes SU(2)$ representations
$[p_1{\!}',p_2{\!}',p_3{\!}']j'$, with scale dimension $\Delta'$, forming  
the supermultiplet $a^\Delta_{[p_1,p_2,p_3]j}$ are obtained by adding the
$SU(4),SU(2)$ weights with $n_{i\alpha} =0,1$ so that
\eqn\weight{
p_r{\!}'=p_r+\sum_\alpha (n_{r\alpha}-n_{r{+1}\,\alpha}) \, , \quad
j{\,}'=j+\half \sum_i ( n_{i\,1} - n_{i\, 2} ) \, , \quad \Delta' = \Delta + \half
\sum_{i,\alpha} n_{i\alpha} \, ,
}
It is easy to see that ${\rm dim}\, a^\Delta_{[p_1,p_2,p_3]j} = 2^8
d(p_1,p_2,p_3)(2j+1)$, where $d(p_1,p_2,p_3)$ is the dimension of the $SU(4)$
representation with Dynkin labels $[p_1,p_2,p_3]$. If in \weight\ any
$p_r{\!}'$ or $j'$ are negative the Racah-Speiser algorithm, described in
\short, provides a precise prescription for removing such
$[p_1{\!}',p_2{\!}',p_3{\!}']j'$.

Shortening conditions arise for suitable $\Delta$ when descendant representations
satisfy the conditions \DSw\ to be superconformal primary. Since all 
$S_i{}^{\!\alpha}$ are obtained by commutators of $S_1{}^{\! 1}$ with $E_r{}^{\!+}$
and $J_+$ it is sufficient to impose only that $S_1{}^{\! 1}$ annihilates the
highest weight state of the representation. In such cases we may impose that
the appropriate combinations of $Q^i{}_{\! \alpha}$  
annihilate $|p;j\rangle^{\rm hw}$.
For application here it is convenient to define, acting on states $|\psi \rangle$ 
such that $J_3 |\psi \rangle = j |\psi \rangle$,
\eqn\Qti{
\tQ^i = Q^i{}_{\! 2} - {1\over 2j} \, Q^i{}_{\! 1} J_-   \, .
}
If $J_+ |\psi \rangle=0$ then $J_+ \tQ^i |\psi \rangle=0$ and
$J_3 \tQ^i |\psi \rangle = (j-\half) \tQ^i |\psi \rangle$. From \algQS\ we have
\eqn\sQ{
\half j \big \{ S_1{}^{\!1} , \tQ^ 1\big \} = 
\big ( 2j - J_3 - \half \hD + \quar (3H_1+2H_2+H_3) \big ) J_- \, , \quad
\half j \big \{ S_1{}^{\!1} , \tQ^2 \big \} = E_{1}{}^{\! -} J_- \, .
}
It is straightforward to show that $\tQ^1 |p;j\rangle^{\rm hw} \sim
|p_1+1,p_2,p_3;j-\half \rangle $. The shortening conditions considered in \short 
and previously are obtained by imposing
\eqn\shQ{
\tQ^i |p;j\rangle^{\rm hw} = 0 \ \cases  {\ i=1 \, , \cr \ i=1,2 &
if $\ p_1 =0 \, ,$\cr \ i=1,2,3 & if $\ p_1 = p_2 = 0 
\, , $\cr \ i=1,2,3,4 & if $\  p_1 = p_2 = p_3 =0 \, . $\cr}
}
In each case there is a consistency condition on $\Delta$ which can be found
by using \sQ\ and \DSw,
\eqn\sDel{
\Delta = 2 + 2j + \half (3p_1+2p_2+p_3) \, .
}
The corresponding supermultiplet is here denoted by $c_{[p_1,p_2,p_3]j}$. Detailed
results were given in \short, the $SU(4)\times SU(2)$ representations
present may be calculated as in \weight\ with the restriction $n_{i2} =0$
for those $i$ listed in \shQ\ for each case.

There are also additional shortening conditions of the form
\eqn\snew{
\Big ( \tQ^2 - {1\over p_1} \, \tQ^1 E_{1}{}^{\! -} \Big ) |p;j\rangle^{\rm hw} 
= 0 \, , \ p_1>0 \, , \quad
\tQ^1 \tQ^2 |0,p_2,p_3;j\rangle^{\rm hw} = 0 \, ,
}
where the left hand sides correspond to highest weight states 
$|p_1-1,p_2+1,p_3;j-\half \rangle$ and $|0,p_2+1,p_3;j- 1 \rangle$ respectively. 
Using \sQ\ these conditions then require
\eqn\sDelid{
\Delta = 2j + \half ( - p_1+2p_2+p_3) \, .
}
For $p_2=0$ the condition \snew\ extends also to
$\big ( \tQ^3 - {1\over p_1} \, \tQ^1 [E_{2}{}^{\! -}, E_{1}{}^{\! -}] \big ) 
|p;j\rangle^{\rm hw} = 0$.
The supermultiplet in each  case is denoted by $d_{[p_1,p_2,p_3]j}$.
The representations are obtained as in \weight\ with $n_{22}=0$, or if
$p_2=0$ then $n_{22}=n_{32}=0$.
For $p_1=0$ it is sufficient  to exclude $n_{12}=n_{22}=1$.

These semi-short representations lead to decompositions of the generic
multiplet,
\eqn\dem{\eqalign{
a^{2+2j+{1\over 2}(3p_1+2p_2+p_3)}_{[p_1,p_2,p_3]j}
\simeq {}& c_{[p_1,p_2,p_3]j} \oplus c_{[p_1+1,p_2,p_3]j{-{1\over 2}}} \, ,\cr
a^{2j+{1\over 2}(-p_1+2p_2+p_3)}_{[p_1,p_2,p_3]j}
\simeq {}& d_{[p_1,p_2,p_3]j} \oplus d_{[p_1-1,p_2+1,p_3]j{-{1\over 2}}} \, ,\cr
a^{2j+{1\over 2}(2p_2+p_3)}_{[0,p_2,p_3]j}
\simeq {}& d_{[0,p_2,p_3]j} \oplus c_{[0,p_2+1,p_3]j{-1}} \, .\cr}
}
Formally, as discussed in \short, we have
\eqn\bc{
c_{[p_1,p_2,p_3]-{1\over 2}} \simeq b_{[p_1+1,p_2,p_3]} \, , \qquad
c_{[p_1,p_2,p_3]-1} \simeq - b_{[p_1,p_2,p_3]} \, ,
}
where $b_{[p_1,p_2,p_3]}$ is the short supermultiplet formed by imposing
$Q^1{}_{\! \alpha}|p;0\rangle^{\rm hw}=0$ where we require $\Delta =
\half (3p_1+2p_2+p_3)$. Just as in \bc\ $d_{[p_1,p_2,p_3]-{1\over 2}}$
may be identified with a multiplet obtained from the highest weight
state $|p_1-1,p_2+1,p_3;0\rangle^{\rm hw}$, with $\Delta = \half (-p_1 + 2p_2 +p_3
+1)$, annihilated by 
$Q^2{}_{\! \alpha} - {1\over p_1}Q^1{}_{\! \alpha}E_1{}^{\! -}$.
Formally we have
\eqn\dcel{
d_{[p_1,p_2,p_3]j} \simeq c_{[-p_1-2,p_1+p_2+p_3+2,-p_3-2]j} \, ,
}
where, in accord with the Racah-Speiser algorithm described in \short, we may 
identify $SU(4)$ representations $[p_1,p_2,p_3] \simeq
[-p_1-2,p_1+p_2+p_3+2,-p_3-2]$ which are related by an even element of the Weyl 
group. This allows the detailed representation content and dimension for 
$d_{[p_1,p_2,p_3]j}$ to be determined from the results given in \short.

The generators of the superconformal group $PSU(2,2|4)$ are obtained by
extending those of $SU(2|4)$ to include the hermitian conjugates,
$\bQ_{i\dal} = Q^i{}_{\! \alpha}{}^\dagger, \, \bS{}^{i\dal} =
S_i{}^{\!\alpha}{}^\dagger , \, {\bar M}{}^\dbe{}_{\!\smash{\dal}} =
(M_\alpha{}^{\! \beta}){}^\dagger$, with an algebra obtained by conjugation
of that for $SU(2|4)$, assuming $\hD^\dagger = - \hD$ and
$(R^i{}_{\! j})^\dagger = R^j{}_{\! i}$. In addition
$\{ S_i{}^{\!\alpha} ,  \bQ_{j\dal} \} = \{ Q^i{}_{\! \alpha} , \bS{}^{j\dal}\}
= 0$ and $\{ Q^i{}_{\! \alpha} ,  \bQ_{j\dal} \} =  2\de^i{}_{\! j} 
(\si^a)_{\alpha\dal}P_a, \, 
\{ \bS{}^{i\dal} , S_j{}^{\!\alpha} \} = 2 \de^i{}_{\! j}
(\bsi^a)^{\dal\alpha} K_a$ where $P_a$ is the momentum operator and $K_a$ the
generator of special conformal transformations. The supermultiplets
are generated from highest weight superconformal primary states
$|p_1,p_2,p_3;j,\bj\rangle^{\rm hw}$, where $\bj$ is the $SU(2)$ 
quantum number for $\bJ_\pm, \bJ_3$ obtained from 
${\bar M}{}^\dbe{}_{\!\smash{\dal}}$, which is annihilated by $S_i{}^{\!\alpha},
\bS{}^{i\dal}$ and $K_a$. The representations are of course infinite 
dimensional, since they are generated by arbitrary powers of $P_a$, but
they are formed by a finite set of conformal primary representations,
annihilated by $K_a$, which are straightforwardly constructed from $SU(2|4)$
supermultiplet representations, as described above, combined with 
with their conjugates formed by the action of $\bQ_{i\dal}$. For these
$j\to \bj$ and $p_1 \leftrightarrow p_3$. The generic supermultiplet is
denoted $\A^\Delta_{[p_1,p_2,p_3](j,\bj)}$ where $[p_1,p_2,p_3](j,\bj)$
are the labels for the representation with lowest scale dimension $\Delta$.
The conformal primary states  form representations labelled by
$[p_1{\!}',p_2{\!}',p_3{\!}'](j{\,}',\bj{\,}')$, with scale dimension $\Delta'$,
which are given by
\eqn\weightL{\eqalign{
p_r{\!}'= {}& p_r +\sum_\alpha (n_{r\alpha}-n_{r{+1}\,\alpha}) +
\sum_{\dal} ( {\bar n}_{r{+1}\, \dal} - {\bar n}_{r \dal} ) \, , 
\cr
j{\,}'={}& j+\half \sum_i ( n_{i\, 1} - n_{i\, 2} ) \, , \qquad 
\bj{\,}'= \bj+\half \sum_i ( {\bar n}_{i\, 2} - {\bar n}_{i\, 1} ) \, , \cr 
\Delta' = {}& \Delta + \half
\sum_{i,\alpha} n_{i\alpha} + \half \sum_{i,\dal} {\bar n}_{i\dal}
\, , \qquad n_{i\alpha}, {\bar n}_{i\dal} = 0,1 \, . \cr}
}
The total dimension is $2^{16}d(p_1,p_2,p_3)(2j+1)(2\bj+1)$.

We here consider the case when shortening conditions are imposed for
both the $Q$ and $\bQ$ charges. Requiring \shQ\ with \sDel\ together with
its conjugate we have the semi-short multiplets,
\eqn\semiC{
\C_{[p_1,p_2,p_3](j,\bj)}\, , \qquad p_1-p_3 = 2(\bj-j)\, ,
\qquad \ \Delta = 2 + j+ \bj + p_1 + p_2 + p_3 \, ,
}
where we impose in \weightL\ $n_{12} = {\bar n}_{41}=0$, 
with further restrictions if $p_1$ or $p_3$ are zero. Requiring
\snew\ and \sDelid\ in both cases gives
\eqn\semiD{
\D_{[p_1,p_2,p_3](j,\bj)}\, , \qquad p_1-p_3 = 2(j-\bj)\, ,
\qquad \ \Delta = j+ \bj + p_2  \, ,
}
and we require in \weightL\ $n_{22}={\bar n}_{31}=0$. If $p_2=1$ then we
exclude $n_{32}={\bar n}_{21}=1$ while if $p_2=0$ then we require 
$n_{32}={\bar n}_{21}=0$ as well.
Corresponding to \dcel\ we have
\eqn\DCrel{
\D_{[p_1,p_2,p_3](j,\bj)} \simeq 
\C_{[-p_1-2,p_1+p_2+p_3+2,-p_3-2](j,\bj)} \, .
}
This result has essentially been used in \relDC.
We may also impose  \shQ\ with \sDel\ for the $Q$ charges and \snew\ and \sDelid\
for $\bQ$ giving
\eqn\semiE{
\E_{[p_1,p_2,p_3](j,\bj)}\, , \qquad p_1+p_3 = 2(\bj-j-1)\, ,
\qquad \ \Delta = 2+ j+ \bj + p_1 + p_2  \, ,
}
and we may also obtain a conjugate ${\bar \E}_{[p_1,p_2,p_3](j,\bj)}$.
Only for \semiC, where $\Delta$ is at the unitarity threshold, is there
a unitary representation. 

For relevance in section 5 we list the self-conjugate representations, when
$p_1=p_2, \, j=\bj$, arising in 
$\D_{[q,0,q](j,j)}$, obtained  by the action of equal powers of the $Q$
and $\bQ$ supercharges for each $\Delta$ 

\vskip 12pt
\hskip -1.7cm
\vbox{\tabskip=0pt \offinterlineskip
\hrule
\halign{&\vrule# &\strut \ \hfil#\  \cr
height2pt&\omit&&\omit&\cr
& $\ell$\hfil && $\ell+1$\hfil  &\cr
height2pt&\omit&&\omit&\cr
\noalign{\hrule}
height4pt&\omit&&\omit&\cr
& $~[q,0,q]_\ell$ &
&  $\matrix{\scs [q-1,0,q-1]_{\ell+1},[q-1,2,q-1]_{\ell+1}, 
3[q,0,q]_{\ell+1},[q+1,0,q+1]_{\ell+1}\cr
\scs [q-1,0,q-1]_{\ell-1},2[q,0,q]_{\ell-1},
[q+1,0,q+1]_{\ell-1}}$  &\cr
height4pt&\omit&&\omit&\cr}
\hrule}

\vskip -5pt
\hskip -1.7cm
\vbox{\tabskip=0pt \offinterlineskip
\hrule
\halign{&\vrule# &\strut \ \hfil#\  \cr
height2pt&\omit&\cr
& $\ell+2$\hfil &\cr
height2pt&\omit&\cr
\noalign{\hrule}
height4pt&\omit&\cr
&  $\matrix{\scs [q-2,2,q-2]_{\ell+2},
[q-1,0,q-1]_{\ell+2},2[q-1,2,q-1]_{\ell+2},4[q,0,q]_{\ell+2},[q,2,q]_{\ell+2},
[q+1,0,q+1]_{\ell+2}\cr
\scs [q-2,0,q-2]_\ell,[q-2,2,q-2]_\ell, 5[q-1,0,q-1]_\ell,2[q-1,2,q-1]_\ell,
8[q,0,q]_\ell,
[q,2,q]_\ell, 5[q+1,0,q+1]_\ell,[q+2,0,q+2]_\ell\cr
\scs [q,0,q]_{\ell-2}} $ &\cr
height4pt&\omit&\cr}
\hrule}

\vskip -5pt
\hskip -1.7cm
\vbox{\tabskip=0pt \offinterlineskip
\hrule
\halign{&\vrule# &\strut \ \hfil#\  \cr
height2pt&\omit&\cr
&$\ell+3$\hfil &\cr
height2pt&\omit&\cr
\noalign{\hrule}
height4pt&\omit&\cr
&$\matrix{\scs [q-1,0,q-1]_{\ell+3},[q-1,2,q-1]_{\ell+3},3[q,0,q]_{\ell+3},
[q+1,0,q+1]_{\ell+3}\cr
\scs [q-3,2,q-3]_{\ell+1},[q-2,0,q-2]_{\ell+1},4[q-2,2,q-2]_{\ell+1},
6[q-1,0,q-1]_{\ell+1},6[q-1,2,q-1]_{\ell+1}\cr
\scs 10[q,0,q]_{\ell+1},4[q,2,q]_{\ell+1},
6[q+1,0,q+1]_{\ell+1},[q+1,2,q+1]_{\ell+1},[q+2,0,q+2]_{\ell+1}\cr
\scs [q-1,0,q-1]_{\ell-1},[q-1,2,q-1]_{\ell-1},3[q,0,q]_{\ell-1}, 
[q+1,0,q+1]_{\ell-1}\cr}$&
\cr height4pt&\omit&\cr}
\hrule}

\vskip -5pt
\hfuzz = 10pt{
\hskip -1.7cm
\vbox{\tabskip=0pt \offinterlineskip
\hrule
\halign{&\vrule# &\strut \ \hfil#\  \cr
height2pt&\omit&\cr
& $\ell+4$ \hfil &\cr
height2pt&\omit&\cr
\noalign{\hrule}
height4pt&\omit&\cr
& $\matrix{\scs [q,0,q]_{\ell+4}\cr
\scs \!\! [q-2,0,q-2]_{\ell+2},[q-2,2,q-2]_{\ell+2},5[q-1,0,q-1]_{\ell+2},
2[q-1,2,q-1]_{\ell+2},8[q,0,q]_{\ell+2},[q,2,q]_{\ell+2},
5[q+1,0,q+1]_{\ell+2},[q+2,0,q+2]_{\ell+2}\!\!\! \cr
\scs [q-2,2,q-2]_{\ell},[q-1,0,q-1]_{\ell},2[q-1,2,q-1]_{\ell},
4[q,0,q]_{\ell},[q,2,q]_{\ell},[q+1,0,q+1]_{\ell}}$ & \cr
height4pt&\omit&\cr}
\hrule}
}

\vskip -18pt
\hskip -1.7cm
\vbox{\tabskip=0pt \offinterlineskip
\hrule
\halign{&\vrule# &\strut \ \hfil#\  \cr
height2pt&\omit&&\omit&\cr
& $\ell+5$\hfil &&$\ell+6$\hfil &\cr
height2pt&\omit&&\omit&\cr
\noalign{\hrule}
height4pt&\omit&&\omit&\cr
& $\matrix{\scs [q-1,0,q-1]_{\ell+3},2[q,0,q]_{\ell+3},[q+1,0,q+1]_{\ell+3}\cr
\scs[q-1,0,q-1]_{\ell+1},[q-1,2,q-1]_{\ell+1},3[q,0,q]_{\ell+1},
[q+1,0,q+1]_{\ell+1}}$&
& $[q,0,q]_{\ell+2}$ & \cr  
height4pt&\omit&&\omit&\cr}
\hrule}

\vskip -3pt
\noindent
Table 3. Diagonal representations for each $\Delta$ in
$\D_{[q,0,q]({1\over 2}\ell,{1\over 2}\ell)}$.

For application in the text we have the decompositions of self conjugate 
multiplets
\eqn\deCDE{\eqalign{
\A^{2+2j+p+2q}_{[q,p,q](j,j)}
\simeq {}& \C_{[q,p,q](j,j)} \oplus \C_{[q+1,p,q](j-{1\over 2},j)} \oplus
\C_{[q,p,q+1](j,j-{1\over 2})} \cr
&{} \oplus \C_{[q+1,p,q+1](j-{1\over 2},j-{1\over 2})}\, , \cr
\A^{2j+p}_{[q,p,q](j,j)}
\simeq {}& \D_{[q,p,q](j,j)} \oplus \D_{[q-1,p+1,q](j-{1\over 2},j)} \oplus
\D_{[q,p+1,q-1](j,j-{1\over 2})} \cr
&{} \oplus \D_{[q-1,p+2,q-1](j-{1\over 2},j-{1\over 2})}\, , \cr
\A^{2j+p}_{[0,p,0](j,j)}
\simeq {}& \D_{[0,p,0](j,j)} \oplus \E_{[0,p+1,0](j-1,j)} \oplus
{\bar \E}_{[0,p+1,0](j,j-1)} \cr
&{} \oplus \C_{[0,p+2,0](j-1,j-1)}\, . \cr }
}
The first case represents the decomposition of a long multiplet into semi-short
multiplets at the unitarity threshold, the second plays a crucial role in section
5 in relating the solution of the superconformal Ward identities to the operator
product expansion.

\listrefs
\bye